\documentclass[numberedappendix]{emulateapj}

\usepackage{graphicx}
\usepackage[]{natbib}
\usepackage{placeins}
\usepackage{lineno}
\usepackage{ulem}
\usepackage{rotating}
\usepackage{adjustbox}

\usepackage[caption=false]{subfig}    

\usepackage{color}

\pdfoutput=1

\newcommand{\degree}{\ensuremath{^\circ}}

\newcommand{\ltsima} {$\; \buildrel < \over \sim \;$}
\newcommand{\fermi}{{\it Fermi}}
\newcommand{\gtsima} {$\; \buildrel > \over \sim \;$}
\newcommand{\lta} {\lower.5ex\hbox{\ltsima}}
\newcommand{\gta} {\lower.5ex\hbox{\gtsima}}

\newcommand{\grba} {GRB~$080916$C~}
\newcommand{\grbnosa} {GRB~$080916$C}

\newcommand{\grbb} {GRB~$090926$A~}
\newcommand{\grbnosb} {GRB~$090926$A}

\shorttitle{A New Model for GRB Prompt Emission Spectra}
\shortauthors{S.Guiriec}

\begin{document}

\title{Towards a Better Understanding of the GRB Phenomenon: a New Model for GRB Prompt Emission and its effects on the New L$_\mathrm{\lowercase{i}}^\mathrm{NT}$--E$_{peak,\mathrm{\lowercase{i}}}^\mathrm{rest,NT}$ relation}

\author{S. Guiriec\altaffilmark{1,2,3,4}, C. Kouveliotou\altaffilmark{5}, F. Daigne\altaffilmark{6}, B. Zhang\altaffilmark{7}, R. Hasco\"et\altaffilmark{8}, R. S. Nemmen\altaffilmark{9}, D. J. Thompson\altaffilmark{1}, P. N. Bhat\altaffilmark{10}, N. Gehrels\altaffilmark{1}, M. M. Gonzalez\altaffilmark{11}, Y. Kaneko\altaffilmark{12}, J. McEnery\altaffilmark{1,2}, R. Mochkovitch\altaffilmark{6}, J. L. Racusin\altaffilmark{1}, F. Ryde\altaffilmark{13,14}, J. R. Sacahui\altaffilmark{15} \& A. M. \"{U}nsal\altaffilmark{12}}


\altaffiltext{1}{NASA Goddard Space Flight Center, Greenbelt, MD 20771, USA}
\altaffiltext{2}{Department of Physics and Department of Astronomy, University of Maryland, College Park, MD 20742, USA}
\altaffiltext{3}{Center for Research and Exploration in Space Science and Technology (CRESST)}
\altaffiltext{4}{NASA Postdoctoral Program}
\altaffiltext{5}{Office of Science and Technology, ZP12, NASA/Marshall Space Flight Center, Huntsville, AL 35812, USA}
\altaffiltext{6}{UPMC-CNRS, UMR7095, Institut d'Astrophysique de Paris, F-75014 Paris, France}
\altaffiltext{7}{Department of Physics and Astronomy, University of Nevada, Las Vegas, NV 89012, USA}
\altaffiltext{8}{Physics Department and Columbia Astrophysics Laboratory, Columbia University, 538 West 120th Street, New York, NY 10027, USA}
\altaffiltext{9}{Instituto de Astronomia, Geof\'{\i}sica e Ci\^encias Atmosf\'ericas, Universidade de S\~ao Paulo, S\~ao Paulo, SP 05508-090, Brazil}
\altaffiltext{10}{University of Alabama in Huntsville, NSSTC, 320 Sparkman Drive, Huntsville, AL 35805, USA}
\altaffiltext{11}{Instituto de Astronom'a, UNAM, MŽxico 04510, Mexico}
\altaffiltext{12}{Sabanc\i~University, Faculty of Engineering and Natural Sciences, Orhanl\i-Tuzla 34956 Istanbul, Turkey}
\altaffiltext{13}{Department of Physics, Royal Institute of Technology, AlbaNova, SE-106 91 Stockholm, Sweden}
\altaffiltext{14}{The Oskar Klein Centre for Cosmo Particle Physics, AlbaNova, SE-106 91 Stockholm, Sweden}
\altaffiltext{15}{Instituto Nacional de Pesquisas Espaciais -- INPE, Avenida dos Astronautas 1758, 12227-010, S‹o JosŽ dos Campos-SP, Brazil}


\email{sylvain.guiriec@nasa.gov}


\begin{abstract}

Gamma Ray Burst (GRB) prompt emission spectra in the keV-MeV energy range are usually considered as adequately fitted with the empirical Band function. 
Recent observations with the {\it Fermi Gamma-Ray Space Telescope} ({\it Fermi}) revealed deviations from the Band function, sometimes in the form of an additional black-body (BB) component, while on other occasions in the form of an additional power law (PL) component extending to high energies.  
In this article we investigate the possibility that the three components may be present simultaneously in the prompt emission spectra of two very bright GRBs (080916C and 090926A) observed with {\it Fermi}, and how the three components may affect the overall shape of the spectra. While the two GRBs are very different when fitted with a single Band function, they look like ``twins'' in the three-component scenario. Through fine-time spectroscopy down to the 100 ms time scale, we follow the evolution of the various components. We succeed in reducing the number of free parameters in the three-component model, which results in a new semi-empirical model---but with physical motivations---to be competitive with the Band function in terms of number of degrees of freedom. From this analysis using multiple components, the Band function is globally the most intense component, although the additional PL can overpower the others in sharp time structures. The Band function and the BB component are the most intense at early times and globally fade across the burst duration. The additional PL is the most intense component at late time and may be correlated with the extended high-energy emission observed thousands of seconds after the burst with {\it Fermi}/Large Area Telescope (LAT). Unexpectedly, this analysis also shows that the additional PL may be present from the very beginning of the burst, where it may even overpower the other components at low energy. We investigate the effect of the three components on the new time-resolved luminosity-hardness relation in both the observer and rest frames and show that a strong correlation exists between the flux of the non-thermal Band function and its E$_\mathrm{peak}$ only when the three components are fitted simultaneously to the data (i.e., F$_\mathrm{i}^\mathrm{NT}$--E$_\mathrm{peak,i}^\mathrm{NT}$ relation). In addition, this result points toward a universal relation between those two quantities when transposed to the central engine rest frame for all GRBs (i.e., L$_\mathrm{i}^\mathrm{NT}$--E$_\mathrm{peak,i}^\mathrm{rest,NT}$ relation).
We discuss theoretical implications of the three spectral components within this new empirical model. The BB component can be interpreted as the photosphere emission of a magnetized relativistic outflow. The Band component can be interpreted as synchrotron radiation in an optically thin region above the photosphere, either from internal shocks or magnetic field dissipation. The extra power law component extending to high energies likely has an inverse Compton origin of some sort, even though its extension to a much lower energy remains a mystery. 

\end{abstract}


\section{Introduction}

The fireball model remains the most popular scenario for the Gamma Ray Burst (GRB) phenomenon~\citep{Cavallo:1978,Paczynski:1986,Goodman:1986,Shemi:1990,Rees:1992,Meszaros:1993,Rees:1994}. In this model, the GRB central engine is a stellar-mass black hole or a rapidly spinning and highly-magnetized neutron star formed by either the collapse of a supermassive star~\citep[collapsar;][]{Woosley:1993,MacFadyen:1999,Woosley:2006} or the merger of two compact objects~\citep{Paczynski:1986,Fryer:1999,Rosswog:2003}. In both cases, the original explosion creates a bipolar collimated jet composed mainly of photons, electrons, positrons and a small fraction of baryons. The relativistic explosion ejecta within the jet are not homogeneous---they form multiple high density layers, which propagate at various velocities. When the fastest layers catch up with the slowest, the charged particles contained in the layers are accelerated through mildly relativistic collisionless shocks ~\citep[internal shocks;][]{Rees:1994,Kobayashi:1997,Daigne:1998}. The particles subsequently cool via emission processes such as synchrotron, Synchrotron Self Compton (SSC), and Inverse Compton (IC). The internal shock phase is usually associated to the so-called GRB prompt emission, mainly observed in the keV$-$MeV energy range~\citep[see e.g., the spectral catalogs by][]{vonKienlin:2014, Gruber:2014} and usually lasting from a few ms up to several tens to hundreds of seconds. 
As the ejecta interact with the interstellar medium they slow down via relativistic collisionless shocks~\citep[external shocks;][]{Rees:1992,Meszaros:1993} accelerating charged particles, which then emit non-thermal synchrotron photons. This external shock phase is usually associated to the so-called GRB afterglow emission observed at radio wavelengths up to X-rays and in some cases even up to the GeV regime hours after the prompt phase, and days and even years for the lowest frequencies. The detailed origin of the gamma-ray emission, however, is not fully understood and many theoretical difficulties remain, such as the composition of the jet, the energy dissipation mechanisms, as well as the radiation mechanisms~\citep[e.g.,][]{Zhang:2011}.

Another prediction of the fireball model is the existence of intense, thermal-like emission from the jet photosphere, expected to be observed simultaneously with the non-thermal prompt emission~\citep{Goodman:1986,Meszaros:2002,Daigne:2002,Rees:2005}. Indeed, the high density of the outflow makes it optically thick to Thomson scattering close to the source, but as the jet expands its density decreases and radiation can escape resulting in a thermal-like component more or less affected by sub-photospheric processes and jet-curvature effects.

GRB prompt emission spectra in the keV--MeV energy regime were predominantly fitted by the so-called Band function ~\citep{Band:1993,Greiner:1995} prior to the launch of the {\it Fermi Gamma-ray Space Telescope} (hereafter {\it Fermi}) in 2008. This empirical function is a smoothly broken power law whose shape is defined with four parameters: two indices $\alpha$ and $\beta$ corresponding to the spectral slopes of the low- and high-energy power laws (PL), respectively; a break energy parametrized to correspond to the maximum of the $\nu$F$_\nu$ spectrum E$_\mathrm{peak}$~\citep{Gehrels:1997}; and a normalization factor. The derived Band spectra usually suggested a non-thermal origin of the radiation leading to the natural proposition that they were produced by synchrotron mechanisms. However, the high values obtained for $\alpha$ were often in conflict with the synchrotron scenario predictions in both the slow and fast electron cooling regimes~\citep{Crider:1997,Preece:1998,Ghisellini:2000}. With the aim to identify emission from the fireball model's jet photosphere, \citet{Ghirlanda:2003} and \citet{Ryde:2004} fitted a thermal spectral shape---using a pure black-body component (BB), and a combination of a BB and a PL, respectively---to the GRB prompt emission observed with the Burst And Transient Source Experiment (BATSE) on board the {\it Compton Gamma Ray Observatory} ({\it CGRO}). Despite the good fits obtained in a few cases, it was often difficult to assess whether thermal spectra were better than the non-thermal Band ones. At the same time, \citet{Gonzalez:2003} fitted simultaneously BATSE and {\it CGRO}/Energetic Gamma Ray Experiment Telescope (EGRET) data of GRB~$941017$, finding a significant deviation from the Band function at high energies ($>1$ MeV); these fits improved with the addition of a PL to account for the high-energy deviation.

The Gamma-ray Burst Monitor (GBM) and the Large Area Telescope (LAT) on board {\it Fermi} have significantly improved GRB prompt emission observations after 2008. Although the Band function remains a good model to describe GBM spectra~\citep[e.g.,][]{Gruber:2014}, the joint spectral analysis of some GRBs using both GBM and LAT confirmed that in some cases an additional PL is required to improve the fit quality~\citep[e.g.,][]{Abdo:2009:GRB090902B,Guiriec:2010,Ackermann:2010:GRB090510,Ackermann:2011:GRB090926A,Ackermann:2013}. Moreover, \fermi~results show that this additional PL can also account for deviations from the Band function below a few tens of keV. Indeed, \citet{Ackermann:2011:GRB090926A} reported the existence of an intense peak in the prompt emission light curves (LCs) of GRB~$090926$A, from the lowest to the highest energies, associated to this additional PL. Similar associations of sharp structures in GRB LCs with an additional PL were also reported in~\citet{Guiriec:2011b} and \citet{Gonzalez:2012}, the latter using {\it CGRO} data. In GRB~$090926$A, the additional PL exhibited a spectral break around 1.4 GeV, which was interpreted as resulting from $\gamma$--$\gamma$ opacity making possible an estimate of the bulk Lorentz factor, $\Gamma$, between 200 and 700.~\citet{Guiriec:2010} reported that a similar PL deviation from the Band function could also be identified in GBM-only data of some GRBs, suggesting a spectral turnover well below the 1.4 GeV of GRB~$090926$A, since little or no emission was observed in the LAT data for these events. The origin of this spectral component remains challenging.

The quest for GRB prompt photospheric emission continued in the \fermi~Era. Within the framework of the fireball model, the expected photosphere component should outshine the non-thermal Band component \citep{Zhang:2009}, but this photosphere component remained undetected. Such a surprising observational result triggered a heated debate on the GRB prompt emission mechanism. A highly-magnetized jet was suggested to suppress the photosphere emission component \citep{Daigne:2002,Nakar:2005,Zhang:2009}. Alternatively, it was suggested that the Band component is produced from a dissipative photosphere \citep[e.g.,][]{Beloborodov:2010,Lazzati:2011,Rees:2005,Thompson:2006}. \citet{Ryde:2010} and \citet{Peer:2012} fitted simultaneously a thermal component with a PL to the data of GRB~$090902$B and suggested that the main keV--MeV prompt emission could be of thermal origin and, therefore, the signature of the jet's photosphere. Then, \citet{Guiriec:2011a} reported for the first time that the prompt emission $\nu$F$_\nu$ spectra of GRB~$100724$B were best fitted with a double curvature model (BB+Band) rather than the Band function alone. Contrary to the conclusions of~\citet{Ryde:2010} and \citet{Peer:2012}, the photospheric component (BB) identified in~\citet{Guiriec:2011a} was subdominant compared to the non-thermal component (Band function). Such a subdominant BB is not expected from the pure fireball model, but is more consistent with a highly-magnetized outflow, as suggested by \citet{Daigne:2002}, \citet{Nakar:2005}, and \citet{Zhang:2009}. 
\citet{Guiriec:2011a} suggested that to produce such a subdominant BB component, the outflow must be highly-magnetized close to the source but the jet must have a low magnetization at large radii for the internal shocks to be efficient. Alternatively, \citet{Zhang:2011b} suggested that efficient energy dissipation is still possible, even if the magnetization parameter $\sigma$ is still greater than unity. They envisaged a collision-induced magnetic reconnection and turbulence (ICMART) to effectively dissipate the magnetic energy in the outflow \citep[see also][for a simulation of GRB lightcurves within such a model.]{Zhang:2014} A similar component has now been reported in a handful of other GRBs~\citep{Guiriec:2011a,Guiriec:2011b,Guiriec:2013a,Burgess:2011,Axelsson:2012:GRB110721A} observed with \fermi. Interestingly, as initially proposed in~\citet{Guiriec:2011a} and confirmed in \citet{Guiriec:2013a}, fits using a Band function with a BB result in Band function shapes that are much more compatible with the predictions of the synchrotron emission origin. Moreover, as shown by~\citet{Guiriec:2013a}, in those GRBs in which an intense but still subdominant thermal-like emission is detected, a strong correlation appears between the energy fluxes and the $\nu$F$_\nu$ spectral peak energy of the non-thermal component only (hereafter F$_\mathrm{i}^\mathrm{NT}$--E$_\mathrm{peak,i}^\mathrm{NT}$ relation\footnote{NT stands for non-thermal component; in the context of the multi-component model that includes thermal and non-thermal components, this non-thermal component is usually adequately fitted with a Band function or a cutoff power law.}.) For GRBs in which the thermal-like contribution affects very little the shape of the non-thermal one, the fit of a Band function alone to the data leads to a similar F--E$_\mathrm{peak}$ relation~\citep{Golenetskii:1983,Borgonovo:2001,Liang:2004,Guiriec:2010,Ghirlanda:2010b,Ghirlanda:2011a,Ghirlanda:2011b,Lu:2012}. The F$_\mathrm{i}^\mathrm{NT}$--E$_\mathrm{peak,i}^\mathrm{NT}$ relations have similar slopes for all GRBs, indicating a universal-like mechanism for the non-thermal prompt emission, and, when corrected for the redshift, the fits between the luminosity and E$_\mathrm{peak}^\mathrm{rest}$ of the non-thermal component (hereafter L$_\mathrm{i}^\mathrm{NT}$--E$_\mathrm{peak,i}^\mathrm{rest,NT}$) align perfectly for all tested GRBs~\citep{Guiriec:2013a}. 

In summary, until now GRB prompt emission spectra have been found to be composed of: 1) a Band function alone; 2) a combination of a Band function and a PL; 3) a combination of a BB and a PL; or 4) a combination of a Band function and a BB \citep{Abdo:2009:GRB080916C,Abdo:2009:GRB090902B,Guiriec:2010,Ackermann:2010:GRB090510,Ackermann:2011:GRB090926A,Guiriec:2010,Guiriec:2011a,Guiriec:2011b,Guiriec:2013a,Peer:2012,Ryde:2010,Zhang:2011c}. The non-thermal Band function usually overpowers the spectra, the quasi-thermal photosphere component usually carries less than a few tens of percent of the total energy, and the additional PL component extends from low to high energies.

Here we reanalyze two of the brightest \fermi~GRBs, \grba and~\grbnosb, in the context of the multiple components described  above and discuss the possible simultaneous existence of such separate components by comparing spectral analyses at various time scales from time-integrated to very fine-time intervals. Through the evolution of these components with time, we trace back the emission mechanisms which produced them, following the jet physical mechanisms back to the central engine energy reservoir. In Section~\ref{section:observation}, we describe the data selection as well as the observations, and in Section~\ref{sec:technique} we present the procedures we followed for the analysis. Sections~\ref{sec:tisa} and \ref{sec:ctr} are dedicated to the results of the time-integrated and the coarse-time analyses, respectively. Section~\ref{sec:ftr} reports the fine-time spectral analysis results, which are the main topic of this article. In Section~\ref{sec:flux-epeak}, we investigate the impact of the multiple spectral components on the relations between the energy-flux of the non-thermal component and its observed $\nu$F$_\nu$ spectral-peak energy (namely F$_\mathrm{i}^\mathrm{NT}$--E$_\mathrm{peak,i}^\mathrm{NT}$) and between the luminosity of the non-thermal component and its intrinsic $\nu$F$_\nu$ spectral-peak energy (namely L$_\mathrm{i}^\mathrm{NT}$--E$^\mathrm{rest,NT}_\mathrm{peak,i}$.) Section~\ref{sec:newmodel} introduces our suggested model for GRB prompt emission. Section~\ref{section:interpretation} discusses an interpretation of our observational results. Finally, in Section~\ref{sec:partialConclusion}, we summarize the most important observational results and comment on their future impact.

\section{Data selection and Observations}
\label{section:observation}

The GBM consists of 12 Sodium Iodide (NaI) detectors covering energies between 8 keV and 1 MeV. GBM also includes two bismuth germanate (BGO) detectors (b0, b1) covering energies between 200 keV and 40 MeV, located on opposite sides of the spacecraft. The LAT is a pair-conversion telescope sensitive to photons with energies from 20 MeV to $>$300 GeV. When a photon enters the LAT, it is converted into an electron-positron pair when interacting into the conversion tungsten foils located between the tracker silicon strip detector planes. Detailed information about the GBM and the LAT can be found in~\citet{Meegan:2009} and~\citet{Atwood:2009}, respectively\footnote{See also http://fermi.gsfc.nasa.gov for additional and up-to-date information about \it{Fermi}.}.

\begin{figure*}
\begin{center}
\includegraphics[totalheight=0.498\textheight, clip]{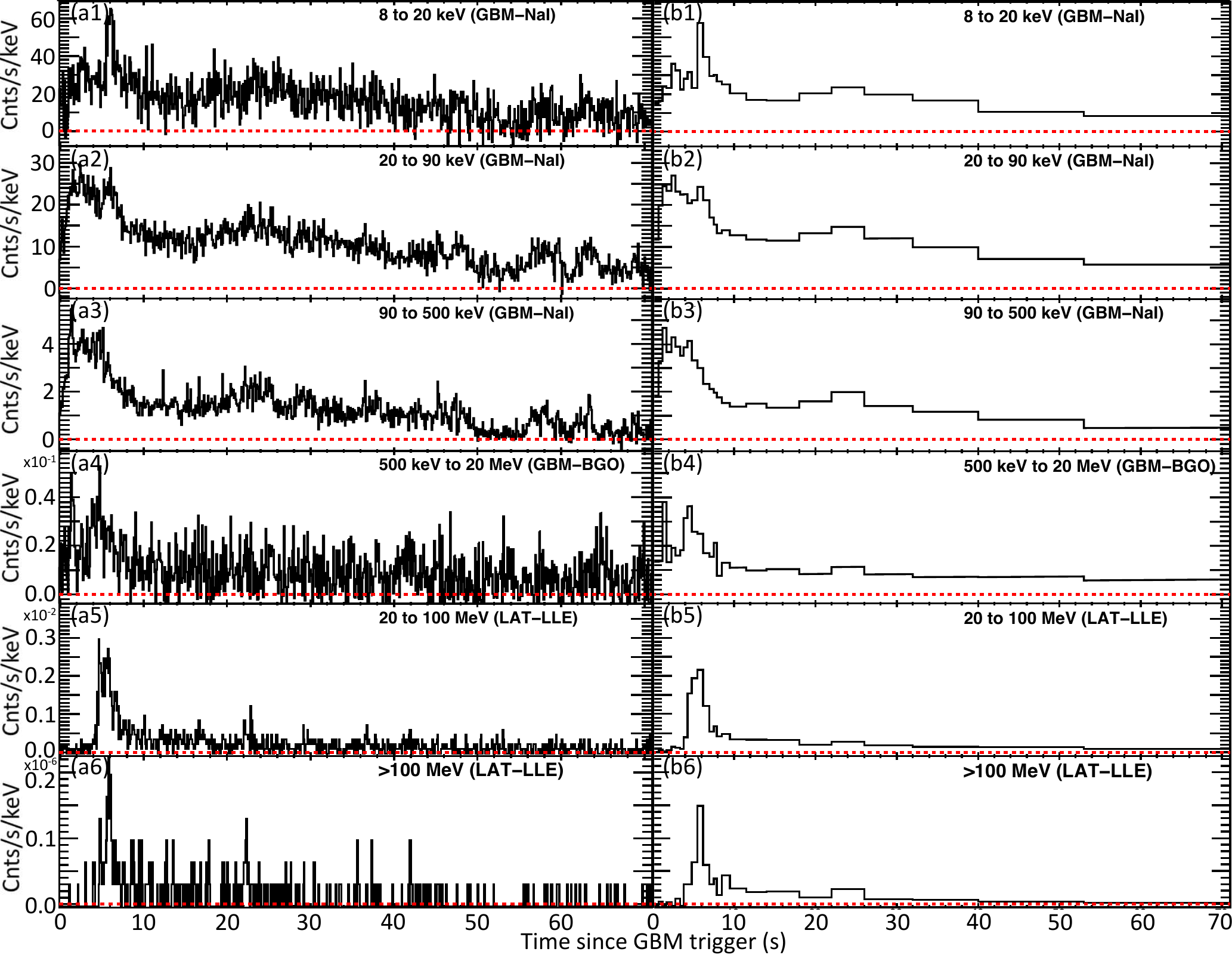}
\caption{\label{fig01}GRB~080916C : Light curves as observed with GBM NaI (8--500 keV) and BGO (500 keV--20 MeV) detectors and with LAT/LLE ($>$20 MeV). (a1--6)  0.1 s time-resolved light curves. (b1--6) Light curves with the binning used for the fine-time analysis presented in Section~\ref{sec:ftr}.}
\end{center}
\end{figure*}

\begin{table*}
\caption{\label{tab07}List of GRB properties.}
\begin{center}
\begin{tabular}{|c|c|c|c|c|}
\hline
GRB Name\footnote{See GBM GRB catalog at http://fermi.gsfc.nasa.gov/ssc.} & Location & Redshift z & T$_\mathrm{90} (s)\footnote{Duration computed between 50 and 300 keV~\citep{Kouveliotou:1993}}$ & Observed Highest Energy Photon (GeV)\\
\hline
GRB~080916C & (RA,Dec)=(119.84717\degree,-56.63833\degree) ($\pm$0.5'')\footnote{The Gamma-Ray burst Optical/Near-Infrared Detector \citep[GROND --][]{Clemens:2008}} & 4.24$\pm$0.26\footnote{\citet{Greiner:2009}} & 63$\pm$1 & 33 \\
GRB~090926A & (RA,Dec)=(353.40070\degree,-66.32390\degree) ($\pm$1.5'')\footnote{Swift/UVOT\citep{Vetere:2009}} & 2.1062\footnote{Very Large Telescope (VLT) \citep{Malesani:2009}}                & 20$\pm$2 & 19.6 \\
\hline
\end{tabular}
\end{center}
\end{table*}

\begin{figure*}
\begin{center}
\includegraphics[totalheight=0.498\textheight, clip]{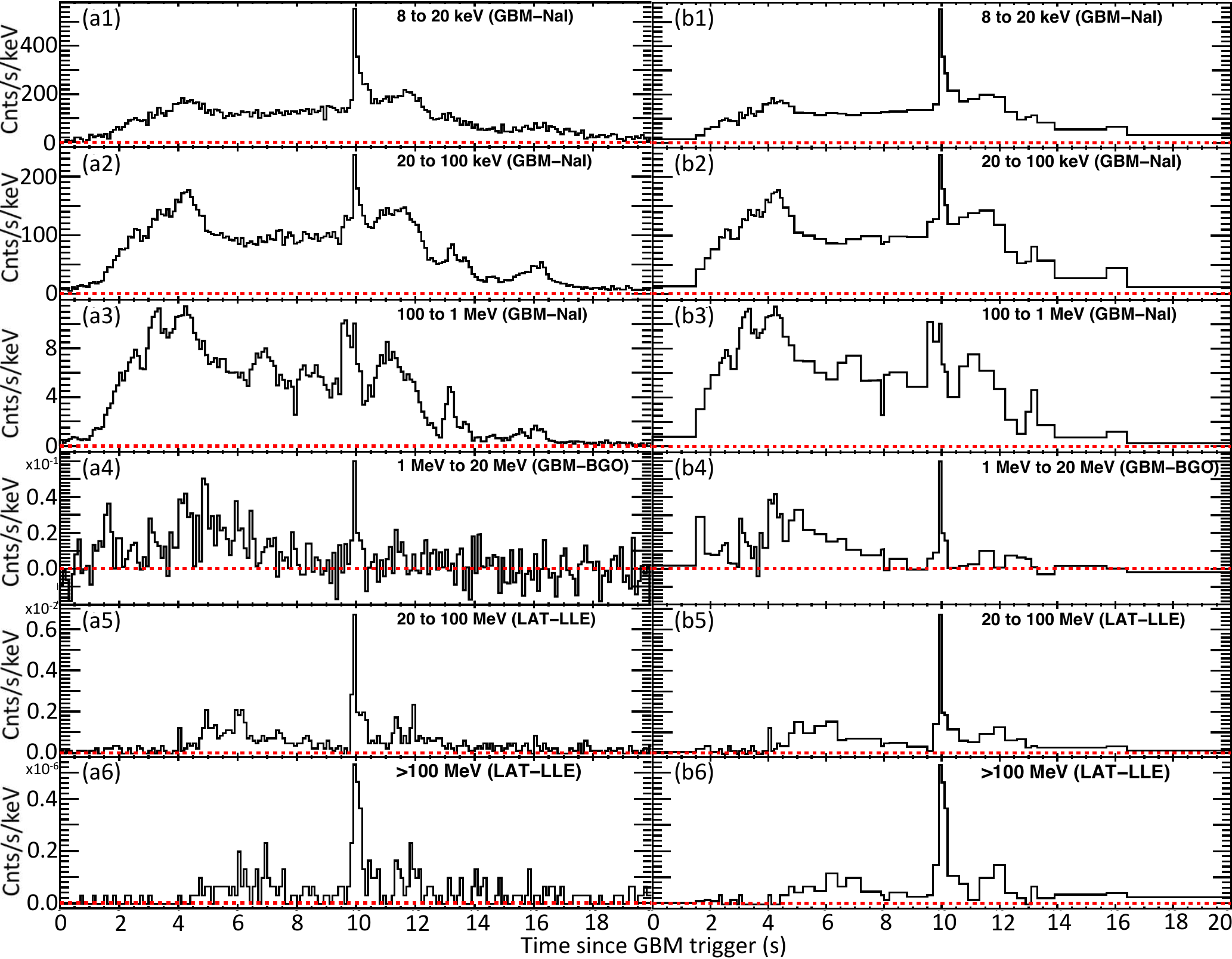}
\caption{\label{fig02}GRB~090926A : Light curves as observed with GBM NaI (8 keV--1 MeV) and BGO (1--20 MeV) detectors and with LAT/LLE ($>$20 MeV). (a1--6) 0.1 s time-resolved light curves. (b1--6) Light curves with the binning used for the fine-time analysis presented in Section~\ref{sec:ftr}.}
\end{center}
\end{figure*}

GBM detected~\grba and~\grbb on 2008 September 8 at T$_\mathrm{0}$=00:12:45 UT~\citep{Goldstein:2008} and on 2009 September 26 at T$_\mathrm{0}$=04:20:26.99 UT~\citep{Bissaldi:2009}, respectively. These bursts were also detected with the LAT~(\citet{Tajima:2008,Abdo:2009:GRB080916C} for \grbnosa, and~\citet{Uehara:2009,Ackermann:2011:GRB090926A} for \grbnosb.) The properties of the two GRBs are reported in Table~\ref{tab07}. We show the light curves of  \grba and \grbb in several energy ranges in Figures~\ref{fig01} and  \ref{fig02}, respectively.


For the analysis we only selected the optimal NaI detectors with angles to the source smaller than 30$\degree$ based on the best source locations and with no blockage by another part of the spacecraft (e.g., LAT, radiators, solar panels) as well as with no shadowing by another GBM module. According to these criteria, we selected detectors n0, n3 and n4 for GRB~080916C and n3, n6 and n7 for GRB~090926A\footnote{The NaIs are named nx with x varying from 0 to 9 for the first 10 detectors, and ``a'' and ``b'' for detectors 11 and 12, respectively}.

Usually, the source is only in the field of view of one BGO detector, and only this BGO detector is retained for the analyses. Therefore, for GRB~080916C, we used b0. However, in the case of~\grbnosb, the burst occurred in the median plane of the spacecraft, which is parallel to the sides to which the BGO detectors are attached. Therefore, the two BGO detectors can be simultaneously used in this specific case. Because of the peculiar location of GRB~090926A in the spacecraft coordinates, the reconstructed signal may be distorted due to the inaccurate modeling of the absorption in the back of the BGO modules (photomultipliers for instance), especially at low energy. Our analysis revealed that although the signals reconstructed from the various NaI detectors match perfectly, discrepancies between the NaI and BGO modules are observed in the overlapping energy region between 200 and $\sim$750 keV for b0 and between 200 and $\sim$600 keV for b1 for GRB~090926A. Beyond $\sim$750 keV and $\sim$600 keV for b0 and b1, respectively, NaI and BGO data are in agreement.

We used in our analysis time tagged event (TTE) data, which have the finest time (i.e., 2 $\mu$s) and energy resolution that can be achieved with the GBM. We used the NaI data from 8 keV to $\sim$900 keV cutting out the overflow high-energy channels as well as the K-edge from $\sim$30 to $\sim$40 keV. For GRB~080916C, we used BGO data from 200 keV to $\sim$39 MeV cutting out the overflow energy channels. For GRB~090926A, we also excluded the low-energy channels $<$750 keV and $<$600 keV for b0 and b1, respectively, since they may be affected by calibration problems. We then generated the response files for each GBM detector based on the best source location.

To optimize the analysis according to the instrument performance, we study the photons converted in the front and in the back of the LAT tracker as two separate data sets (hereafter FRONT and BACK, respectively) using the appropriate instrument response functions. We analyzed the FRONT and BACK data passing the Pass 7 transient cuts \citep{Ackermann:2012b} within a 10$^\circ$ region of interest (ROI) centered on the best source location and with energy above 100 MeV. To perform spectral analysis of the LAT data between 20 and 100 MeV, we used the LAT low-energy (LLE) data, which are designed to increase the LAT sensitivity below 100 MeV for transient sources as well as to make possible spectral analysis below 100 MeV \citep[for instance, see Appendix of][]{Ajello:2014}.

The background in each of the GBM detectors as well as in the LLE data was estimated by fitting polynomial functions to the light curves in various energy ranges before and after the source active time period. The background was then measured by interpolating the functions to the source active time period. For GBM data, the background was fitted to the CSPEC data files, which have the same energy resolution as the TTE data but with a coarser time resolution. CSPEC data cover a much longer time range, making the estimation of the background more reliable especially for long GRBs. We then exported the background estimated from the CSPEC data to our analysis of the TTE data. The background for the LAT FRONT and BACK data was estimated by averaging the data over several orbits (during which the GRB was not present) when \fermi~ was located at the same position in the orbit as at the trigger time and when the LAT was pointing in the same direction as at the trigger time \citep{Vasileiou:2013}.

\begin{figure*}[ht!]
\begin{center}
\includegraphics[totalheight=0.45\textheight, clip]{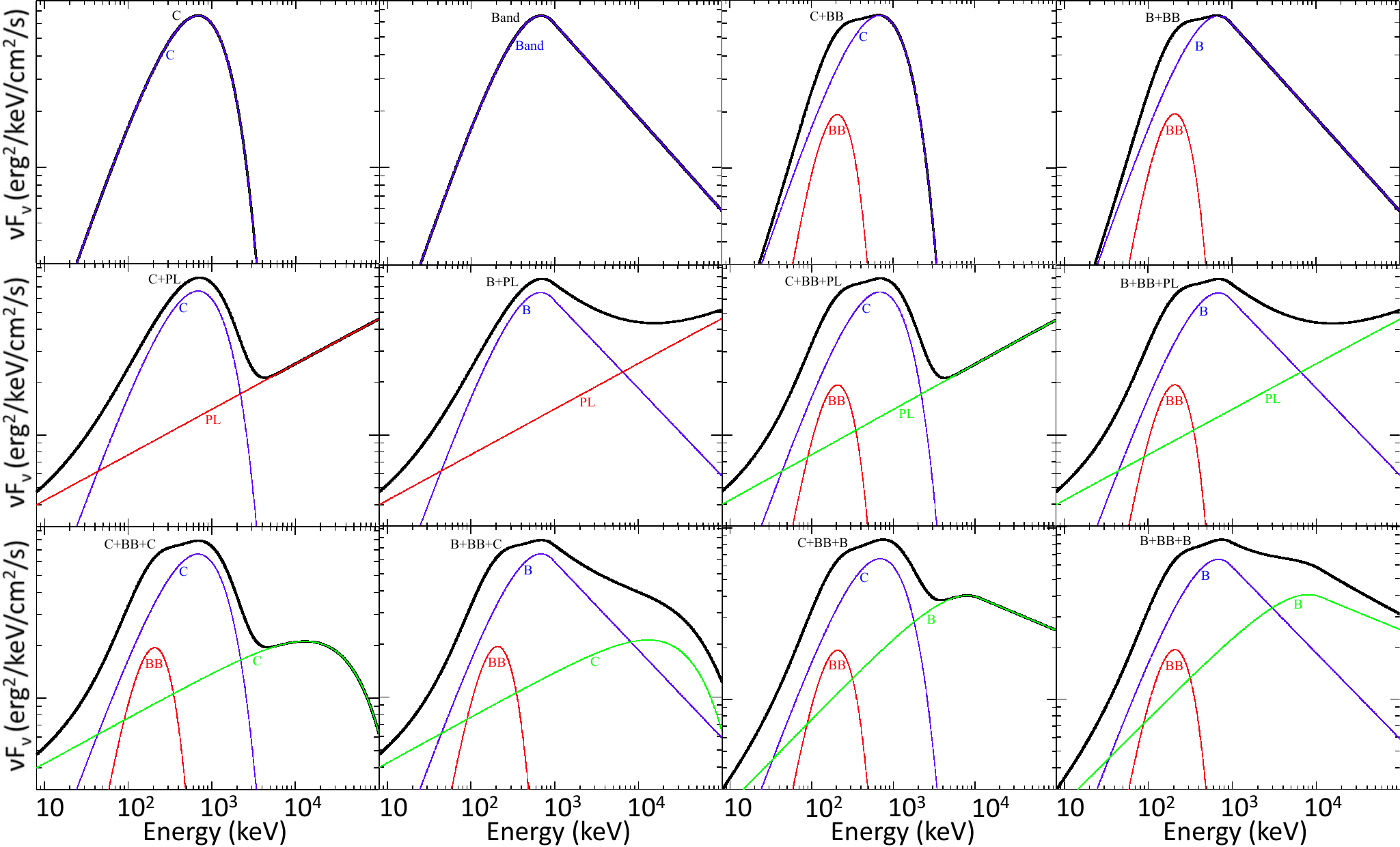}
\caption{\label{fig03}Sketches of the various tested models.}
\end{center}
\end{figure*}

\vspace{0.3cm}
\section{Spectral analysis methodology}
\label{sec:technique}

To perform the spectral analysis, we used the XSpec and Rmfit tool kits. Starting from a cutoff power law (C) or a Band function (Band), we built twelve increasingly complex spectral models with two or three components adding a power law (PL), a black body (BB), C or Band.
Sketches of the various models are presented in Figure~\ref{fig03}. The best spectral parameter values and their 1--$\sigma$ uncertainties were estimated by optimizing the Castor C-statistic (hereafter Cstat), which is a likelihood technique that converges to $\chi^2$ for a specific data set when there are enough counts. We performed the spectral analysis over three time scales---we refer to these later in the text as: time-integrated, coarse-time bins and fine-time bins.

In order to check the consistency of the NaI and BGO detectors through all tested models, we fitted both detector types simultaneously and we applied a free effective area correction (EAC) factor between b0 and each of the NaI detectors selected for the analysis. For each model fitted we then fixed these EAC values. To evaluate the consistency of the GBM data (NaI and b0) with the LAT-LLE, LAT-FRONT and LAT-BACK data, we also fitted them simultaneously adding a new, free EAC factor between b0 and each of the LAT data types. When EAC values were compatible with no correction needed, they were set to unity. We only applied the EAC corrections in the time-integrated and coarse-time bin analysis, because in the fine-time bin analysis the number of counts is much smaller, leading to unconstrained EAC factors. When significant EAC factors (i.e., significant deviation from unity) between the GBM and LAT data sets are required with the simplest models, those data sets are in much better agreement with the more complex ones. The strong corrections required with the simplest models seem to be unreasonable based on our current knowledge of the instruments; therefore, this reinforces the scenarios involving the more complex spectral shapes. The results of cross-calibration analysis are reported in greater detail in Appendix~\ref{section:crosscalibration}.

Due to the very limited number of counts at high energies, we only used the LAT-LLE data in combination with GBM in the coarse-time analysis, and we did not use any LAT data in the fine-time analysis. We then estimated the probability that the Cstat improvements obtained with the most complex models compared to the simplest ones are not merely due to signal and/or background statistical fluctuations, by performing multiple sets of Monte Carlo (MC) simulations following the same procedure as discussed in~\citet{Guiriec:2011a,Guiriec:2013a} (see Appendix~\ref{section:modelComparison}.) In Sections \ref{sec:tisa} and \ref{sec:ctr} we compare the results of joint GBM and LAT data fits with those of the GBM-only ones for the time-integrated and the coarse-time analysis, respectively; the consistency of the joint GBM and LAT analysis with the GBM-only one is important to support the results of the fine-time analysis presented in Section \ref{sec:ftr}, for which we only use GBM data.

\begin{figure*}[ht!]
\begin{center}
\includegraphics[totalheight=0.33\textheight, clip]{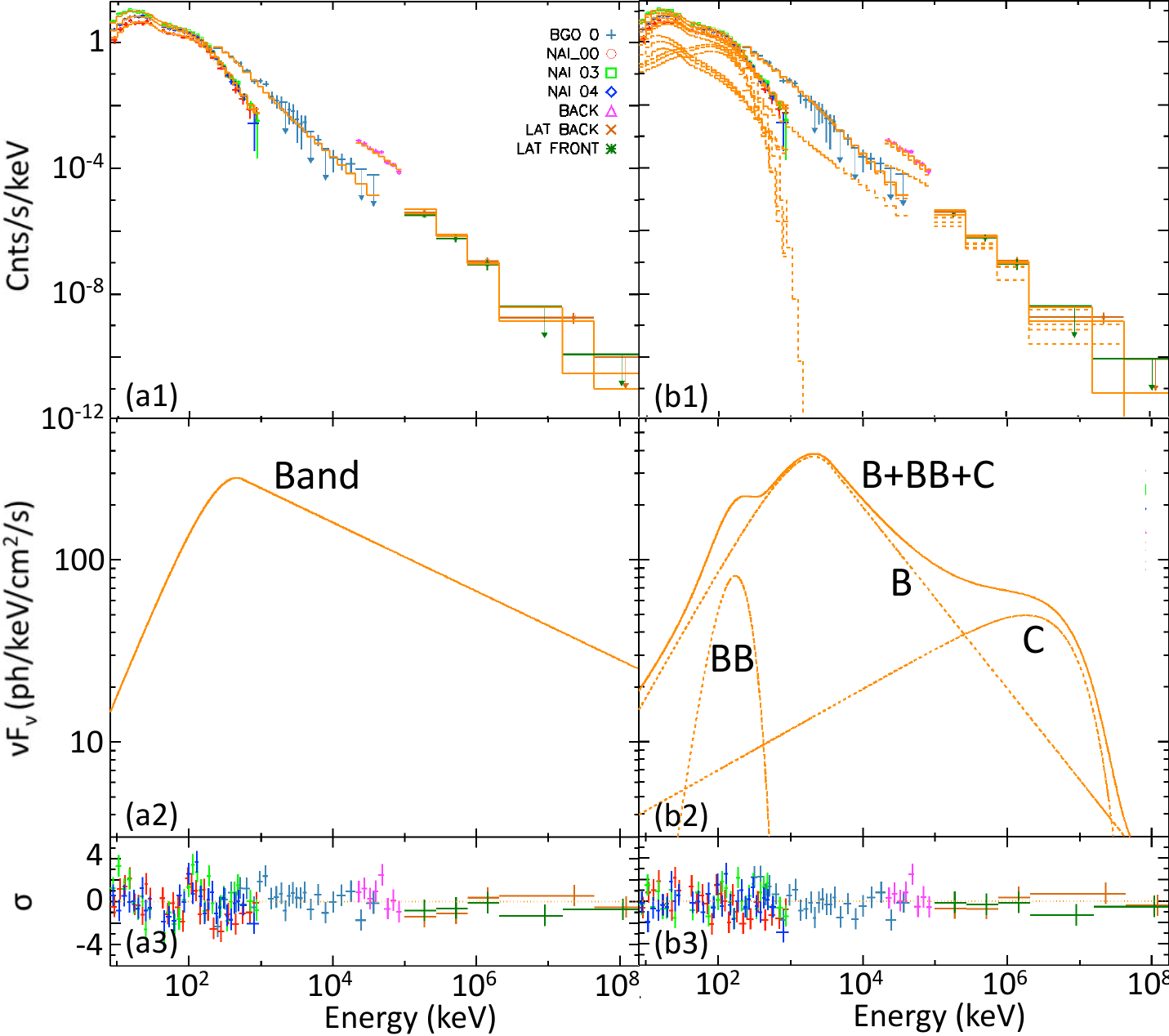}
\caption{\label{fig04}Time-integrated GBM and LAT spectra of GRB~080916C (from T$_0$-0.1 s to T$_0$+71 s) when fitted (a) with a Band function alone and (b) with B+BB+C. The count rate spectra are presented in panels (a1) and (b1)---the solid lines corresponding to the fitted model---and the deconvolved $\nu$F$_\nu$ spectra---the dashed lines correspond to the individual components of the fitted model and the solid ones to the total emission---in panels (a2) and (b2). Panels (a3) and (b3) correspond to the residuals of the fits.}
\end{center}
\end{figure*}

\begin{figure*}[ht!]
\begin{center}
\includegraphics[totalheight=0.33\textheight, clip]{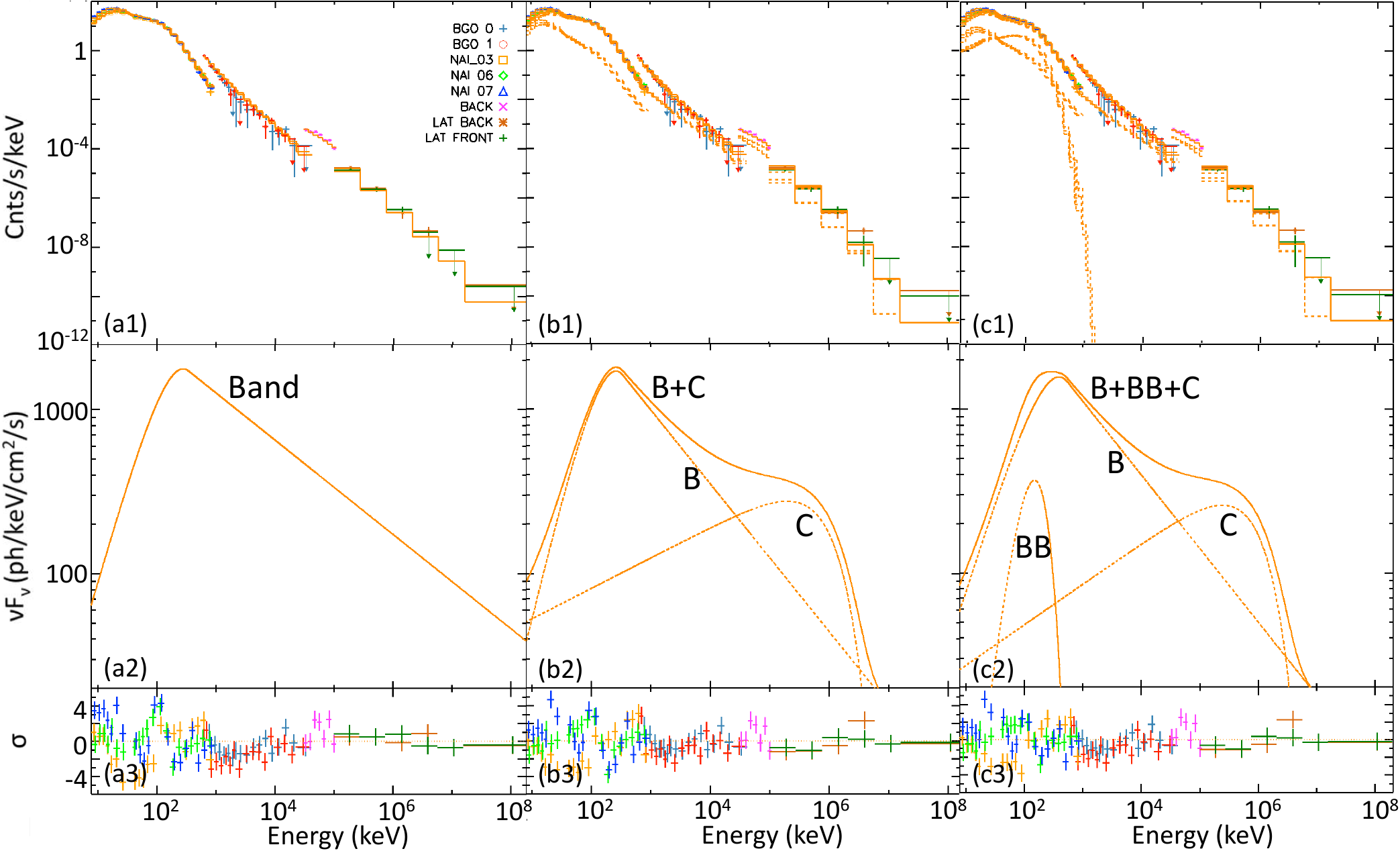}
\caption{\label{fig05}Time-integrated GBM and LAT spectra of GRB~090926A (from T$_0$ to T$_0$+20 s) when fitted (a) with a Band function alone, (b) with B+C and (c) with B+BB+C. The count rate spectra are presented in panels (a1), (b1) and (c1)---the solid lines corresponding to the fitted model---and the deconvolved $\nu$F$_\nu$ spectra---the dashed lines correspond to the individual components of the fitted model and the solid ones to the total emission---in panels (a2), (b2) and (c2). Panels (a3), (b3) and (c3) correspond to the residuals of the fits.}
\end{center}
\end{figure*}

\newpage

\section{Time-Integrated Spectral Analysis}
\label{sec:tisa}

We performed a time-integrated spectral analysis over a period corresponding to the most intense part of the prompt emission of the two GRBs in the keV--MeV energy range (i.e., from T$_\mathrm{0}$-0.10 s to T$_\mathrm{0}$+71.00 s and from T$_\mathrm{0}$ to T$_\mathrm{0}$+20.00 s for GRBs 080916C and 090926A, respectively.) The fit results using the various models discussed above are reported in Tables~\ref{tab01} \&~\ref{tab02} of Appendix~\ref{section:Spectral-analysis-results}. 
We discuss below the fits and present the most relevant ones in Figures~\ref{fig04} \&~\ref{fig05}.

\subsection{GBM-only}
\label{sec:tisa_GBM-only}

For \grbnosa, we find that the B+BB and B+PL fits improve the Band-only fit by 74 and 54 units of Cstat for 2 additional degrees of freedom (dof), respectively, while for \grbb the improvement is 139 and 81 units of Cstat, respectively\footnote{In the case of nested models and in the gaussian regime an improvement of $\sim$10 units of Cstat per additional degree of freedom corresponds to a $\sim$5 $\sigma$ level improvement.}.

As already reported in~\citet{Guiriec:2011a, Guiriec:2013a}, the addition of the BB component to the Band function results in shifting systematically the E$_\mathrm{peak}$ values toward higher energies and making both the low- and high-energy Band PL indices steeper. The values of $\alpha$ obtained with the B+BB fits are, therefore, more compatible with the pure slow-cooling synchrotron emission scenario for the two GRBs and even with the pure fast-cooling synchrotron process for \grbnosa. The impact of the parameter changes on the F$_\mathrm{\lowercase{i}}^\mathrm{NT}$--E$_\mathrm{peak,\lowercase{i}}^\mathrm{NT}$ and L$_\mathrm{\lowercase{i}}^\mathrm{NT}$--E$^\mathrm{rest,NT}_\mathrm{peak,\lowercase{i}}$ relations as well as on the interpretation of the prompt emission mechanisms  will be discussed in great detail in Sections~\ref{sec:flux-epeak} and~\ref{section:interpretation}. 
The temperatures of the BB are $\sim$40 keV for \grba and \grbb corresponding to the Planck function peaking at $\sim$120 keV (right within the sensitivity range of the NaI detectors.)

It must be noted that when the high-energy PL of the Band function has an index $<$-3, the Band function can be replaced with C without affecting much the Cstat value of the fit (see for instance Table~\ref{tab01} of Appendix~\ref{section:Spectral-analysis-results} 
or time interval from T$_\mathrm{0}$-0.1 s to T$_\mathrm{0}$+4.3 s in Table~\ref{tab03} of Appendix~\ref{section:Spectral-analysis-results}.
) This does not mean that the high-energy slope is as steep as an exponential cutoff, but that the slope is at least as steep as $\sim$-3 and that our data may not allow us to better measure $\beta$. In the time-resolved analysis and in the context of the multi-component models, we will not make any distinction between Band and C for the main keV--MeV spectral contribution.

Adding a PL to the Band function (i.e., B+PL) has the opposite effect on the Band parameters, as reported in~\citet{Guiriec:2010}: E$_\mathrm{peak}$ is shifted toward lower energies and both low- and high-energy Band PLs become less steep.
The index of the additional PL is $\sim$-2.00 for both GRBs.

Although we performed simulations to compare all models, here we only describe the results obtained when comparing Band-only fits with C+BB+PL because they are the most relevant ones for the rest of the analysis. Indeed, Band-only is traditionally considered as a good fit to the data and can, therefore, be considered as a reference---for instance, it has been the case for \grba\citep{Abdo:2009:GRB080916C}---but we show throughout this article that C+BB+PL is a globally better description of the data for the two GRBs by following the statistical test procedure presented in Appendix~\ref{section:modelComparison}. For \grbnosa,  C+BB+PL improves the Band-only fit by 77 units of Cstat for 3 additional dof, while for \grbnosb, C+BB+PL leads to an improvement of 162 units of Cstat for 3 additional dof.
For both GRBs, none of the 10$^\mathrm{5}$ synthetic spectra using the Band-only fits as null hypothesis give a $\Delta$Cstat value as high as the observed ones, thus supporting the statement that the probability that C+BB+PL better fits the data than Band is due to statistical fluctuation of signal and background around the Band function is $<$10$^\mathrm{-5}$. Moreover, the Cstat value resulting from the fit of the real data with Band-only is much higher than the Cstat values obtained when fitting the synthetic spectra, which would not be expected if the true spectrum were a Band function. We performed the same exercise, this time using the fit of C+BB+PL to the real data as the null hypothesis and recovered Cstat values for the synthetic spectra that were compatible with the ones obtained when fitting the real data with either Band or C+BB+PL. All the above results reinforce the hypothesis that the  Band function is not the best description of the data and that C+BB+PL is a significantly better model.

\subsection{GBM+LAT}

Panels (a3)
of Figures~\ref{fig04} \& \ref{fig05} show strong wavy patterns in the residuals of the fits when a Band function alone is fitted to the GBM+LAT time-integrated spectra of GRBs 080916C and 090926A.The spectral parameter values resulting from the simultaneous fit of GBM and LAT data with the various models---and especially the most complex ones---are very well compatible with those reported from the GBM-only fits (see Tables~\ref{tab01} \&~\ref{tab02} of Appendix~\ref{section:Spectral-analysis-results}.
). 
While the high-energy power law of the Band function is statistically compatible with an exponential cutoff when fitting B+BB+PL and C+BB+PL to the time-integrated GBM data alone, B+BB+PL is preferred over C+BB+PL in the simultaneous fit of the time-integrated GBM and LAT data.

We investigated the effect of a possible cutoff in the additional PL of the B+BB+PL model by replacing the PL with a PL with an exponential cutoff (i.e., B+BB+C) or with a second Band function (i.e., B+BB+B). We find that for \grbnosa, B+BB+C does not reduce the Cstat value obtained with B+BB+PL.  For \grbnosb, both B+BB+C and B+BB+B are significantly better than B+BB+PL.  

%

\section{Coarse-Time Spectroscopy}
\label{sec:ctr}

We concluded above that both \grba and \grbb were better fitted with a three-component model. However, it is impossible to conclude from the time-integrated spectra whether these separate components are artifacts due to a strong spectral evolution during the event duration, and whether they can be associated directly to physical processes. To address these questions, we performed a coarse-time spectral analysis selecting time intervals including the main structures observed in each burst light curve (see Table~\ref{tab03} and~\ref{tab04} of Appendix~\ref{section:Spectral-analysis-results} for \grba and \grbnosb, respectively.) This selection resulted in 3 and 5 time intervals for \grba and \grbnosb, respectively.

We fitted either the GBM data alone or in combination with the LLE data. We did not include the LAT transient data in this analysis to avoid complications and possible biases due to the low number of photons and possible calibration inconsistencies. We also tested the need of EAC either among the GBM detectors or between the GBM and the LLE data; the results are reported in Appendix~\ref{section:crosscalibration}. 

\subsection{\grbnosa}

\subsubsection{GBM-only}

We find that C+BB is a significantly better fit than Band only in the first and third time intervals, improving the Cstat values by 28 and 45 units, respectively, for only one additional dof. As reported in Section~\ref{sec:tisa_GBM-only}, the addition of a BB component to the Band function systematically shifts E$_\mathrm{peak}$ to higher energy and decreases both values of $\alpha$ and $\beta$. In fact, the slope of the high-energy PL of the Band function becomes so steep that it can be replaced with C. That is why C+BB and B+BB have similar Cstat values in these two time intervals. In the second time interval, the addition of the BB component to a Band function does not improve the fit significantly (i.e., 11 units of Cstat for two additional dof), and it impacts much less the Band parameters. In particular, the value of $\beta$ remains pretty high and inconsistent with an exponential cutoff. In all intervals, the temperature of the BB is similar to the temperature measured in the burst time-integrated spectrum.

Adding a PL component to the Band function does not work for the first time interval (i.e., no convergence is obtained with a PL compatible with a positive flux). Although no improvement of the Cstat value is observed in the second time interval either, it is possible to fit B+PL to this time interval, and the fit results in PL with a positive flux (i.e., the flux of the additional PL at 100 keV is 3.7 $\sigma$ above 0.) In the latter case, the fitting engine prefers a solution for which most of the high-energy emission is associated with the PL and not with the Band function. To enable  this solution, the high-energy slope of the Band function becomes very steep and compatible with an exponential cutoff, becoming identical to slopes of the two other intervals when adding a BB to the Band function (i.e., B+BB is equivalent to C+BB in term of Cstat values.) Therefore, C+PL is equivalent to B+PL (i.e., similar Cstat values) in the second time interval with one less dof. In the third time interval, a B+PL fit significantly improves the Band only fit (i.e., 39 units of Cstat for 2 additional dof); however, C+BB remains the best two-component model. In the second and third time intervals, the indices of the additional PL are compatible with the values resulting from the time-integrated analysis.

Since for the three time intervals, C+BB (or B+BB) is significantly better than B+PL (or C+PL), we investigated the effect of an additional PL to the C+BB model (i.e., C+BB+PL). In the first time interval, we could not obtain an additional PL with a positive flux. In the second time interval, adding a PL to the B+BB model reduces the Cstat value by 13 more units and impacts the parameters of the Band function. With the additional PL, E$_\mathrm{peak}$ is shifted toward lower energy and while the value of the low-energy PL index of the Band function increases, its high-energy PL becomes compatible with an exponential cutoff. The changes in both slopes mark the contribution of the additional PL not only at high-energy but also at low-energy. C+BB+PL is equivalent to B+BB+PL (i.e., similar Cstat values) and an improvement of 13 units of Cstat is obtained compared to B+BB for one additional dof. In the third time interval, the Cstat values obtained with C+BB and C+BB+PL are very similar; however, the addition of the PL affects the spectral parameters of the other components in a way that the PL has a positive contribution to the total flux (i.e., the flux of the additional PL at 100 keV is $\sim$13 $\sigma$ above 0 for the 2$^\mathrm{nd}$ and the 3$^\mathrm{rd}$ time intervals, respectively.) Finally, we investigate the possibility of a break in the additional PL of the C+PL scenario in the third time interval by replacing the PL with a second cutoff PL (i.e., C+C2). With this additional dof, the fitting engine prefers to have C2 mimicking a thermal shape instead of the non-thermal shape of the additional PL. Indeed, the index of the C2 is +0.95$\pm0.22$ and E$_\mathrm{0}$ is 64$\pm7$ keV which matches very well with the BB component measured in the B+BB and C+BB scenarios.

\subsubsection{GBM+LLE}

The spectral results obtained when fitting simultaneously GBM and LLE data remain globally similar to those derived from the fits of GBM data alone, both for the parameters and for the statistics. However, the LLE data allow for a better constraint of the high-energy PL of the Band function in the most complex models such as B+BB, B+PL and B+BB+PL, and confirm steep slopes for the high-energy PL of the Band function with indices below -2.5. In addition, possible breaks have been identified in the additional PL when using LLE data with E$_\mathrm{0}$ between $\sim$100 and $\sim$200 MeV in the second time interval, and between $\sim$250 and $\sim$700 MeV in the third time interval---because E$_\mathrm{0}$=E$_\mathrm{peak}$/(2-$\alpha_\mathrm{PL}$) with E$_\mathrm{peak}$ being the energy at the break in the photon PL spectrum and $\alpha_\mathrm{PL}$ being the PL index, the actual break energy E$_\mathrm{peak}$ in the photon PL spectrum is between $\sim$50 and $\sim$100 MeV in the second time interval and between $\sim$75 and $\sim$210 MeV in the third one.

\subsection{\grbnosb}

\subsubsection{GBM-only}

A B+BB fit improves significantly compared to the Band-alone fit in all the time intervals but the last, in which the intensity of the signal is weaker. A B+PL leads to similar improvements although the B+BB remains slightly better. Before T$_\mathrm{0}$+5s, the Cstat value decreases by 67 units for two dof when adding a BB to the Band function; a B+PL model, however, cannot be constrained. An additional PL is identified only after T$_\mathrm{0}$+5s in the coarse-time analysis. In that case, both E$_\mathrm{peak}$ and $\beta$ values remain similar to those measured in the Band-only scenario; however, the value of $\alpha$ increases. Both the temperature of the BB and the index of the additional PL are similar to those measured in the time-integrated analysis. This reinforces the idea that the three components identified in the time-integrated spectrum are not artifacts due to strong spectral evolution as they are recovered in the time-resolved analysis and from one time-interval to the other.

From 5s to 9s after the trigger time, B+BB+PL improves B+BB by 13 units of Cstat. The resulting high-energy PL of the Band function is very steep and compatible with an exponential cutoff. Therefore, C+BB+PL is as good as B+BB+PL and an improvement of 9 units of Cstat compared to B+BB is measured for only one additional dof. After T$_\mathrm{0}$+9s, a limited improvement is obtained when adding a PL to B+BB; however, the fit results indicate that in the C+BB+PL model, the fluxes of the three components are positive (for instance the fluxes of the additional PL at 100 keV are 9.0, 4.4 and 1.2 $\sigma$ above 0 for the 3$^\mathrm{rd}$, the 4$^\mathrm{th}$ and the 5$^\mathrm{th}$ time intervals, respectively.) The reason the Cstat values do not change much is that the global shape of the whole spectrum changes, especially the low- and high-energy PLs of the Band function.

\subsubsection{GBM+LLE}

The simultaneous fits of GBM and LLE data give similar spectral results to those obtained when fitting GBM data alone with the various tested models; however, LLE data enhance the significance of the additional BB and PL components. Indeed, before T$_\mathrm{0}$+5s, B+BB is better than Band alone by 97 units of Cstat while an improvement of 67 units of Cstat was obtained from GBM only data. In the second and third time intervals, B+PL is clearly a better improvement than B+BB, while the two models were competitive when using only GBM data. This is expected since only the additional PL contributes to the emission detected at high-energy with the LAT. Finally, we included an exponential cutoff in the additional PL to look for a high-energy break, but the significance of this result remains weak due to the low data counts in these energies.

\subsection{Partial Conclusion}

The coarse-time spectral analysis supports the complex spectral shape identified in the time-integrated spectra of GRBs 080916C and 090926A as well as the possible presence of multiple spectral components. The various components identified in the time-integrated spectra of these two GRBs cannot be explained as artifacts due to strong spectral evolution during the event durations. Indeed, we recovered the multiple components in the smaller time intervals with values of their spectral parameters consistent with those obtained in the time-integrated analysis. Furthermore, this refined analysis suggests that the three components do not have the same relative intensities during the burst durations. The BB component and the Band function seem to be the most intense at early times, while the contribution of the additional PL is stronger at later times.

\section{Fine-Time Spectroscopy}
\label{sec:ftr}

To better understand the behavior of the various spectral components as well as their evolution with time, we performed a very fine-time spectral analysis.
At very fine time scales, GBM and LAT data are in very different count regimes; therefore, to avoid any complication related to this difference as well as to possible cross-calibration issue we only consider GBM data throughout this Section.

\begin{figure}[ht!]
\begin{center}
\includegraphics[totalheight=0.28\textheight, clip]{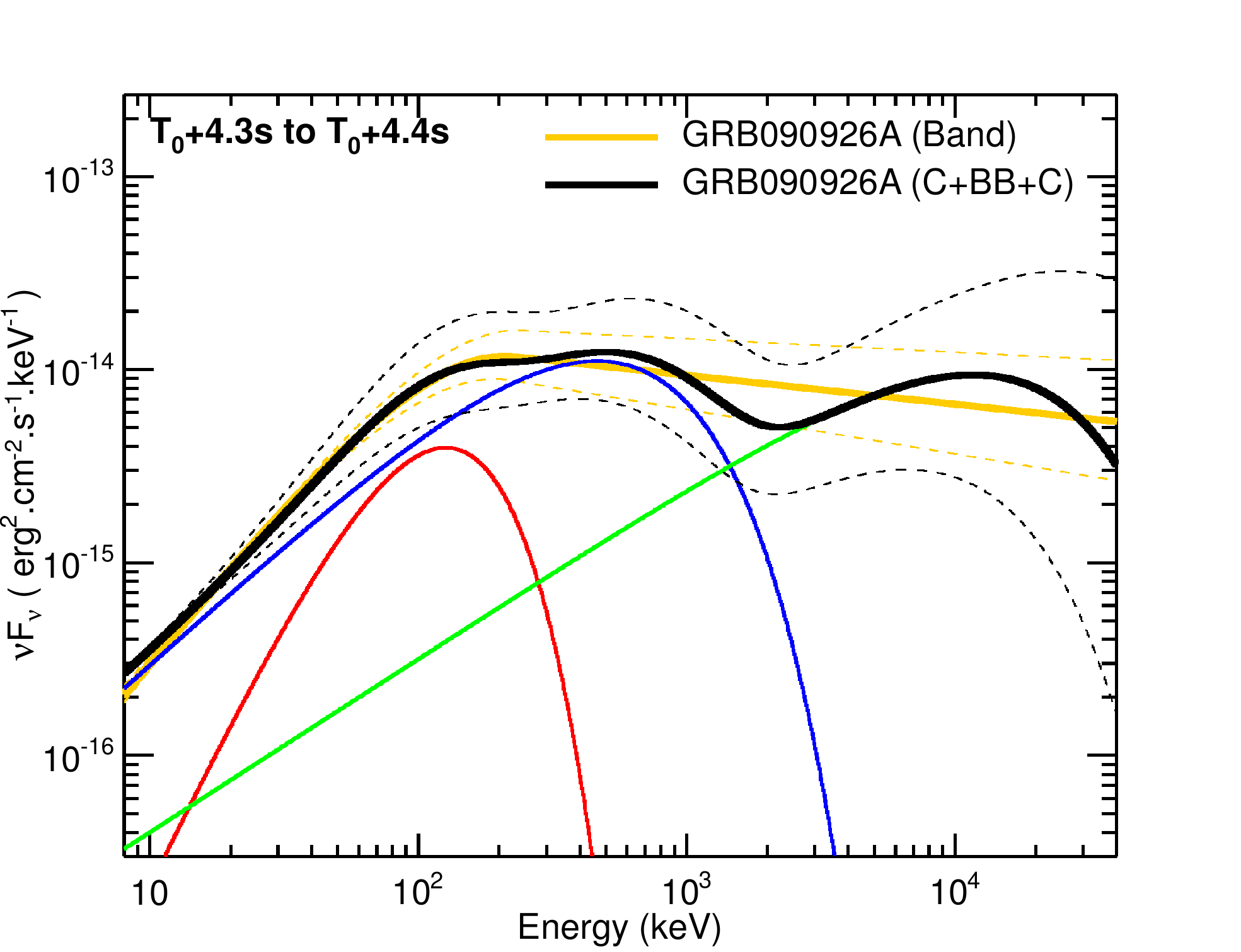}
\caption{\label{fig06}Snapshot of a $\nu$F$_\nu$ spectrum resulting from the fine-time analysis of GRB~090926A. This shows the new three-component model (i.e., (1) a non-thermal Band function with a high-energy power law index $<$-3 and which is statistically equivalent to a cutoff power law, C, (2) a thermal-like component adequately approximated with a black body component, BB, and (3) a non-thermal additional cutoff power law, C) overplotted with the Band-only fit for GRB~090926A. The solid yellow and black lines correspond to the best Band-only and C+BB+C fits, respectively. The dashed yellow and black lines correspond to the 1--$\sigma$ confidence regions of the Band-only and C+BB+C fits, respectively. The solid blue, red and green lines correspond to the cutoff power law, to the BB component and to the additional cutoff power-law resulting from the best fit with the C+BB+C model (i.e., solid black line) to the data, respectively. The C+BB+C model exhibits a break in the second C component below 100 MeV.}
\end{center}
\end{figure}

\begin{figure*}[ht!]
\begin{center}
\includegraphics[totalheight=0.7\textheight, clip]{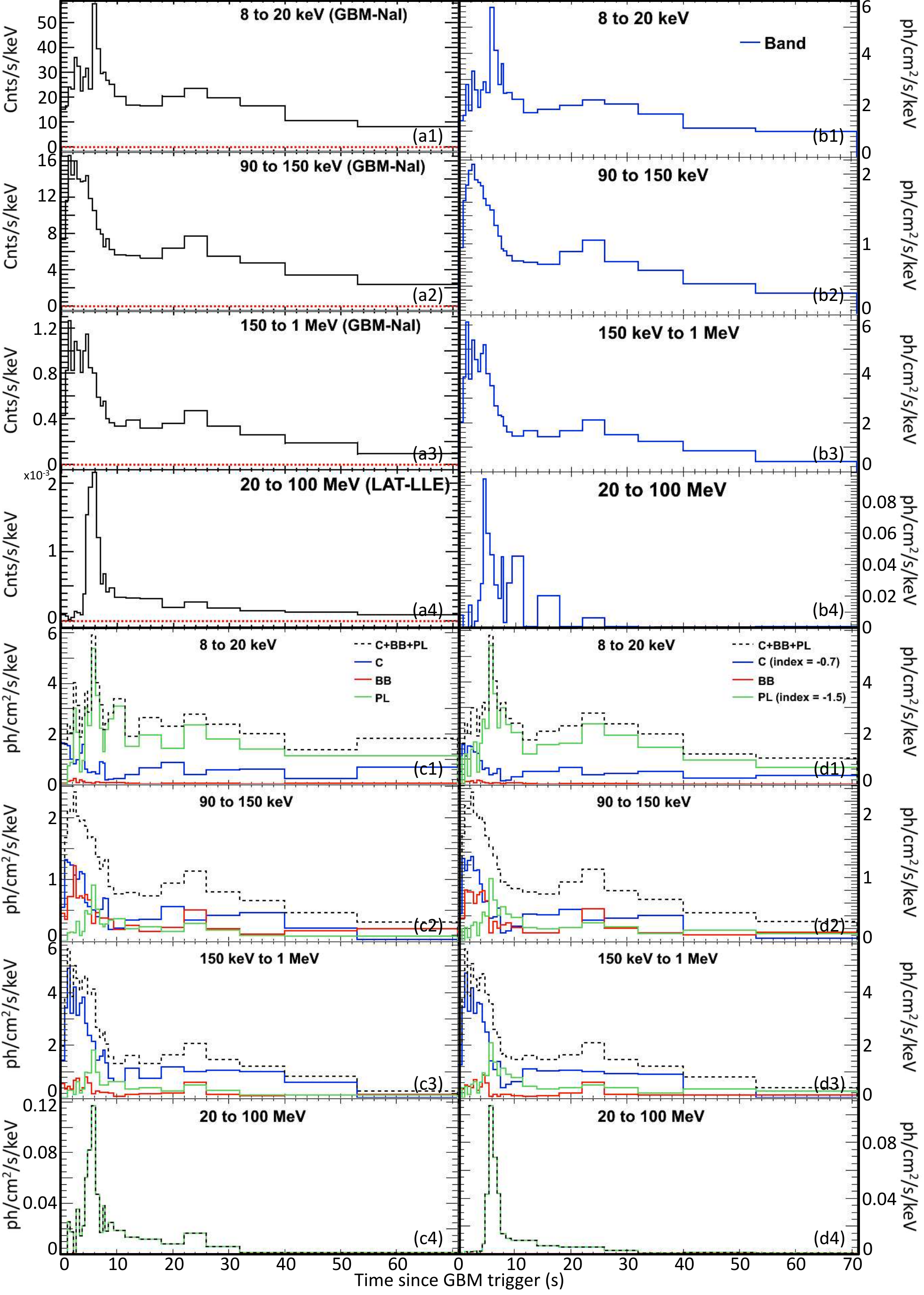}
\caption{\label{fig07}GRB 080916C : (a1--4) Count light curves as observed with GBM between 8 keV and 1 MeV, and with LAT/LLE between 20 and 100 MeV. The time bins correspond to those used for the fine-time analysis; (b1--4) Reconstructed photon light curves resulting from the fine-time analysis with a Band function alone. The energy bands and time intervals are the same as for the count light curves (i.e., panels (a)) for comparison. The 20--100 MeV light curve is an extrapolation of the model (i.e., Band) fitted to the GBM energy range into the LAT/LLE energy domain. (c1--4) Reconstructed photon light curves resulting from the fine-time analysis with the C+BB+PL model. The blue, green and red lines correspond to the cutoff power law, to the black-body and to the additional power law, respectively. The black dashed line corresponds to the sum of the 3 components. The energy bands and time intervals are the same as for the count light curves (i.e., panels (a)) for comparison. The 20--100 MeV light curve is an extrapolation of the model (i.e., Band) fitted to the GBM energy range into the LAT/LLE energy domain; (d1--4) Reconstructed photon light curves resulting from the fine-time analysis with the C+BB+PL$_\mathrm{5params}$ model. The blue, green and red lines correspond to the cutoff power law, to the black-body and to the additional power law, respectively. The black dashed line corresponds to the sum of the 3 components. The energy bands and time intervals are the same as for the count light curves (i.e., panels (a)) for comparison. The 20--100 MeV light curve is an extrapolation of the model (i.e., Band) fitted to the GBM energy range into the LAT/LLE energy domain. For clarity, the uncertainties on the count light curves and on the reconstructed photon light curves are not shown. Therefore, those figures are for qualitative purpose.}
\end{center}
\end{figure*}

\begin{figure*}[ht!]
\begin{center}
\includegraphics[totalheight=0.7\textheight, clip]{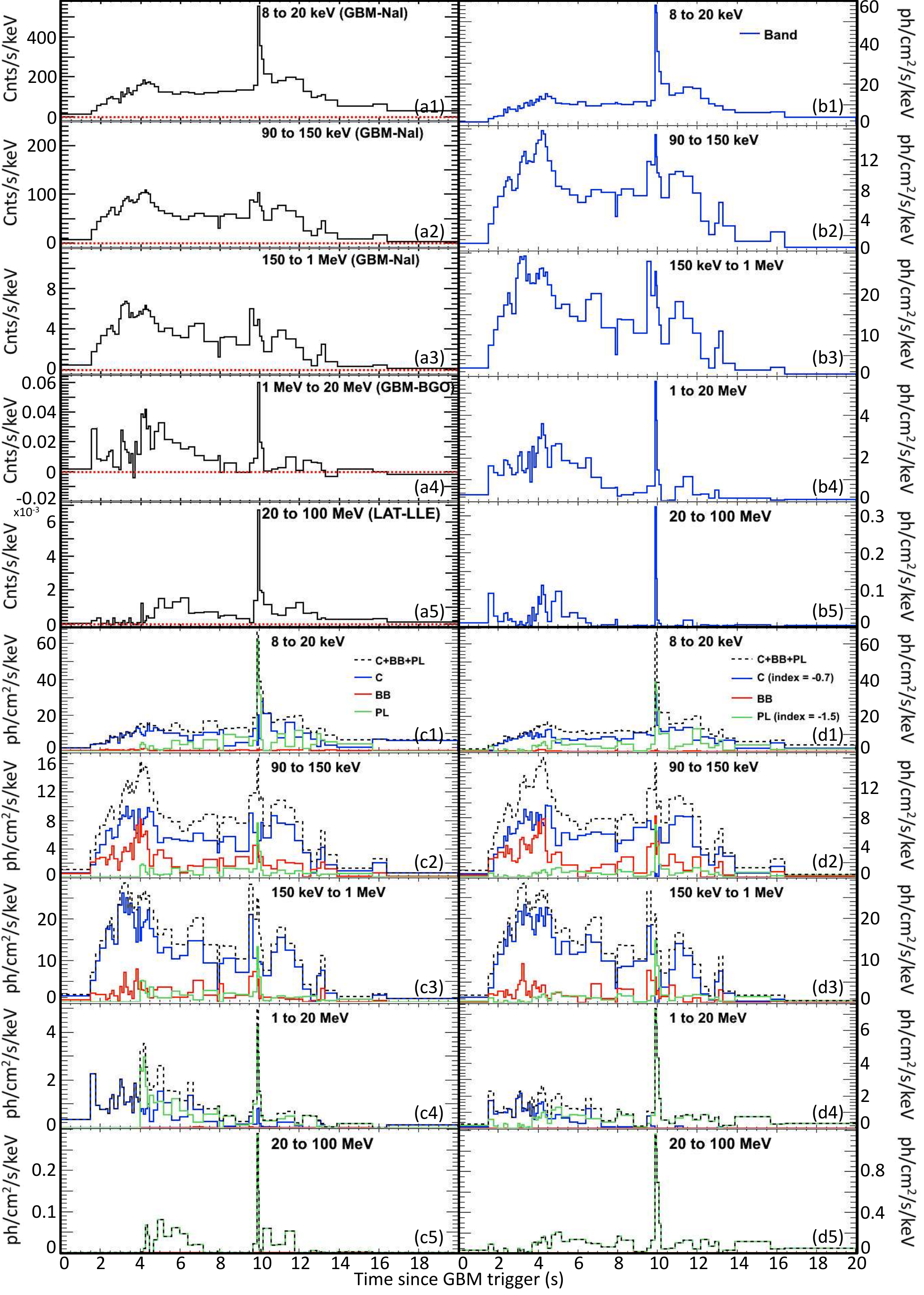}
\caption{\label{fig08}GRB 090926A : (a1--4) Count light curves as observed with GBM between 8 keV and 1 MeV, and with LAT/LLE between 20 and 100 MeV. The time bins correspond to those used for the fine-time analysis; (b1--4) Reconstructed photon light curves resulting from the fine-time analysis with a Band function alone. The energy bands and time intervals are the same as for the count light curves (i.e., panels (a)) for comparison. The 20--100 MeV light curve is an extrapolation of the model (i.e., Band) fitted to the GBM energy range into the LAT/LLE energy domain. (c1--4) Reconstructed photon light curves resulting from the fine-time analysis with the C+BB+PL model. The blue, green and red lines correspond to the cutoff power law, to the black-body and to the additional power law, respectively. The black dashed line corresponds to the sum of the 3 components. The energy bands and time intervals are the same as for the count light curves (i.e., panels (a)) for comparison. The 20--100 MeV light curve is an extrapolation of the model (i.e., Band) fitted to the GBM energy range into the LAT/LLE energy domain; (d1--4) Reconstructed photon light curves resulting from the fine-time analysis with the C+BB+PL$_\mathrm{5params}$ model. The blue, green and red lines correspond to the cutoff power law, to the black-body and to the additional power law, respectively. The black dashed line corresponds to the sum of the 3 components. The energy bands and time intervals are the same as for the count light curves (i.e., panels (a)) for comparison. The 20--100 MeV light curve is an extrapolation of the model (i.e., Band) fitted to the GBM energy range into the LAT/LLE energy domain. For clarity, the uncertainties on the count light curves and on the reconstructed photon light curves are not shown. Therefore, those figures are for qualitative purpose.}
\end{center}
\end{figure*}

With fine time intervals it is often difficult to unambiguously conclude that the most complex models are significantly better than the simpler ones based on the pure statistical considerations of the likelihood ratio test solely. Instead, our goal here is to verify that the complex shapes identified as significantly better than the Band function in the time-integrated and the coarse-time spectral analyses are also valid options at very fine time-scales. It is important to keep in mind that even if the most complex models are not statistically better than the Band function based on the likelihood ratio test results, it does not necessarily mean that the contributions of the additional components (i.e., BB and PL) to the total flux are negligible and compatible with a null flux. As presented in the previous Sections, the addition of a BB and a PL to the Band function significantly modifies the parameters of the Band function. Therefore, it is possible to have complex models that are statistically not favored compared to a Band function alone---based solely on the likelihood ratio test---but whose additional component fluxes are clearly positive. Moreover, we show in Appendix~\ref{section:modelComparison} that the results of the likelihood ratio test are not always sufficient to decide whether the simpler model is truly a better description of the data than a more complex one and that additional statistical tests are sometime necessary; following the procedure described in Appendix~\ref{section:modelComparison}, we conclude that C+BB+PL is globally a better description of the data than all the other simpler models. In this Section, we will focus mostly on the consistency and continuity of the spectral results from time interval to time interval and on how the complex models modify the values of the spectral parameters of the Band function.

\begin{figure}[ht!]
\begin{center}
\includegraphics[totalheight=0.42\textheight, clip]{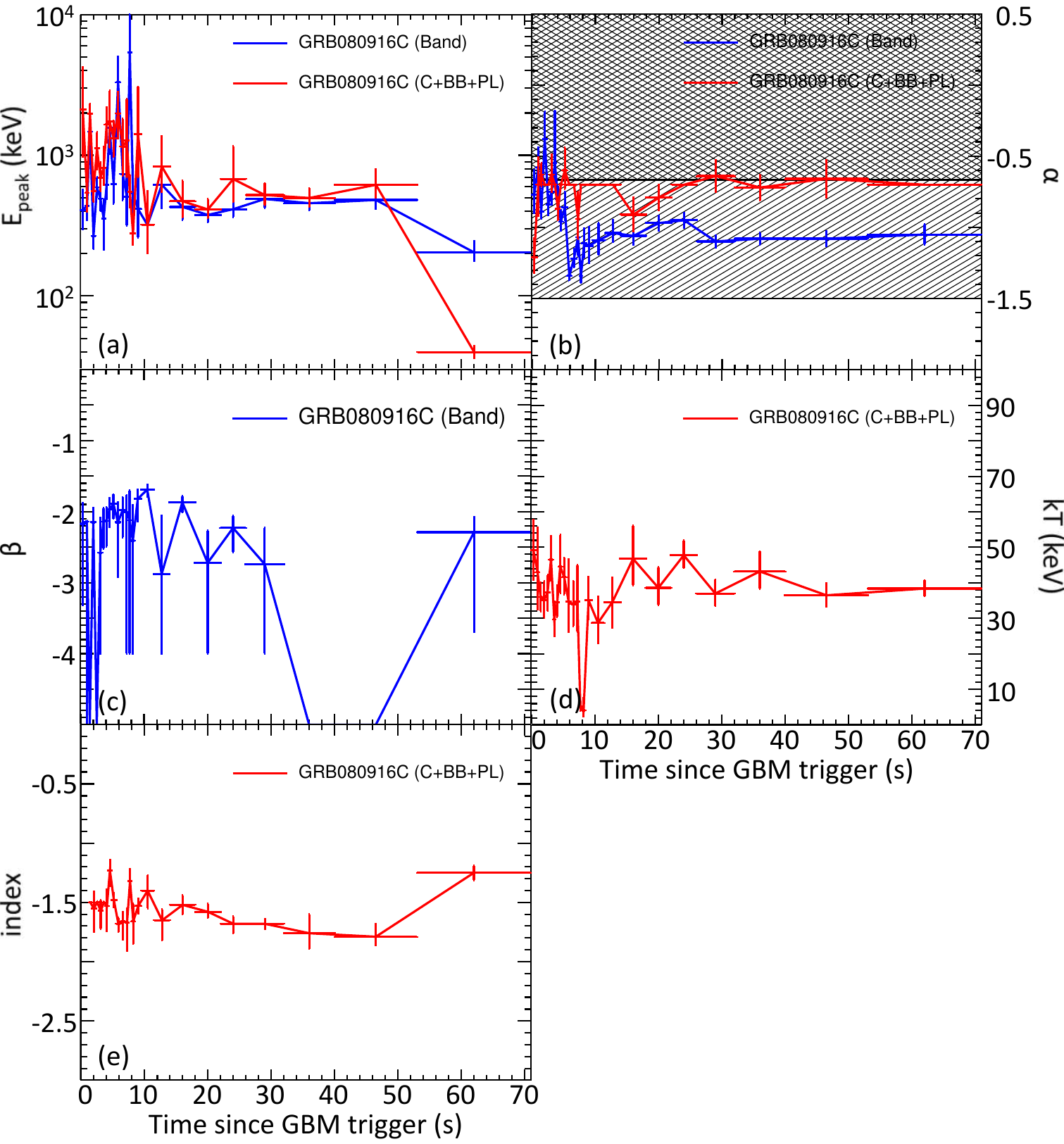}
\caption{\label{fig09}GRB 080916C : Evolution with time of the spectral parameters resulting from the fits of a Band function alone (blue) and of a C+BB+PL model (red) to the fine time intervals. The dashed regions $>$-2/3 and $>$-3/2 correspond to the domains in which the values of $\alpha$ are incompatible with pure synchrotron emission from electrons in both the slow and fast cooling regimes, and with synchrotron emission from electrons in the fast cooling regime only, respectively.}
\end{center}
\end{figure}

\begin{figure}[ht!]
\begin{center}
\includegraphics[totalheight=0.42\textheight, clip]{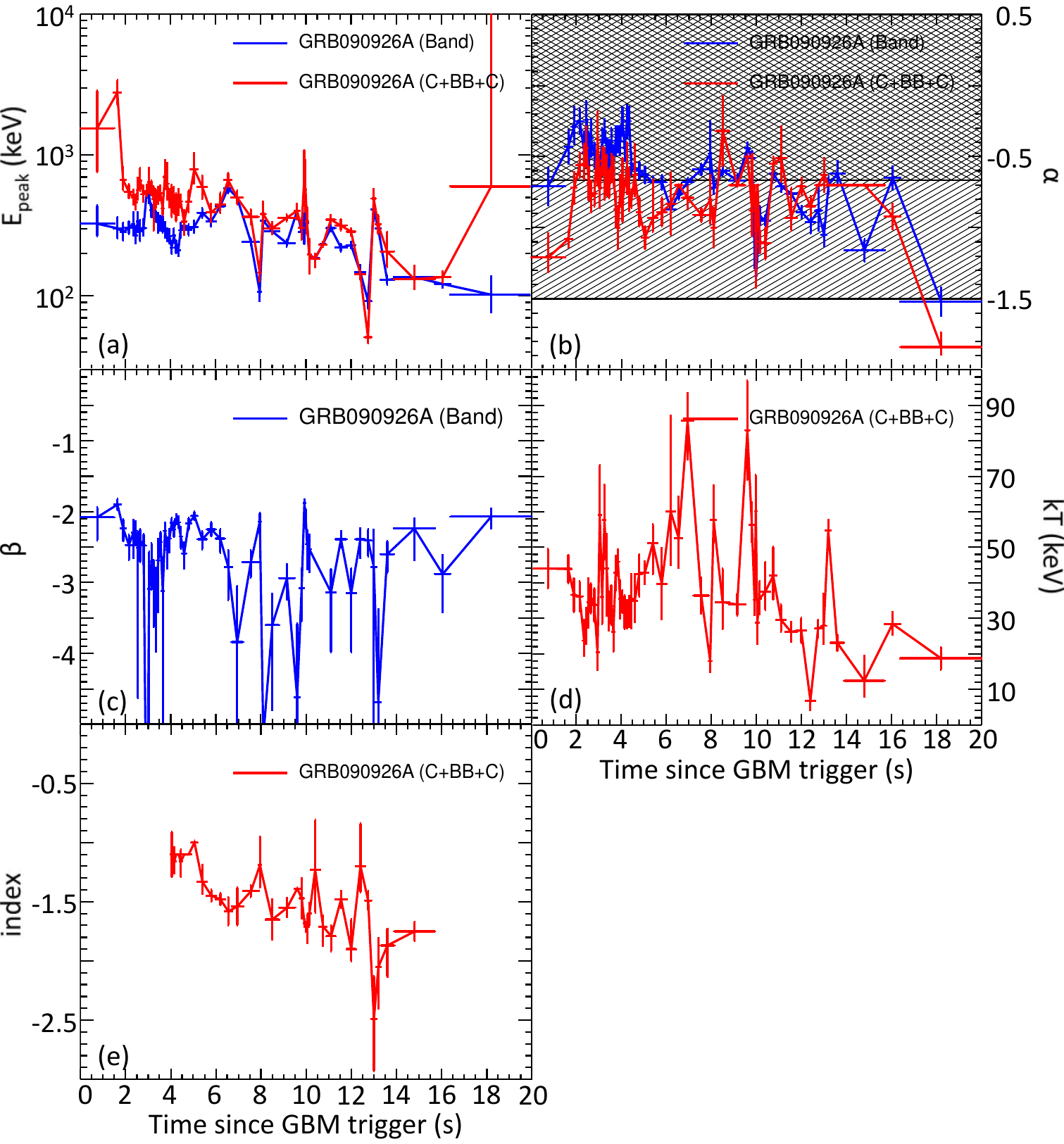}
\caption{\label{fig10}GRB 090926A : Evolution with time of the spectral parameters resulting from the fits of a Band function alone (blue) and of a C+BB+PL model (red) to the fine time intervals. The dashed regions $>$-2/3 and $>$-3/2 correspond to the domains in which the values of $\alpha$ are incompatible with pure synchrotron emission from electrons in both the slow and fast cooling regimes, and with synchrotron emission from electrons in the fast cooling regime only, respectively.}
\end{center}
\end{figure}

\begin{figure}[ht!]
\begin{center}
\includegraphics[totalheight=0.334\textheight, clip]{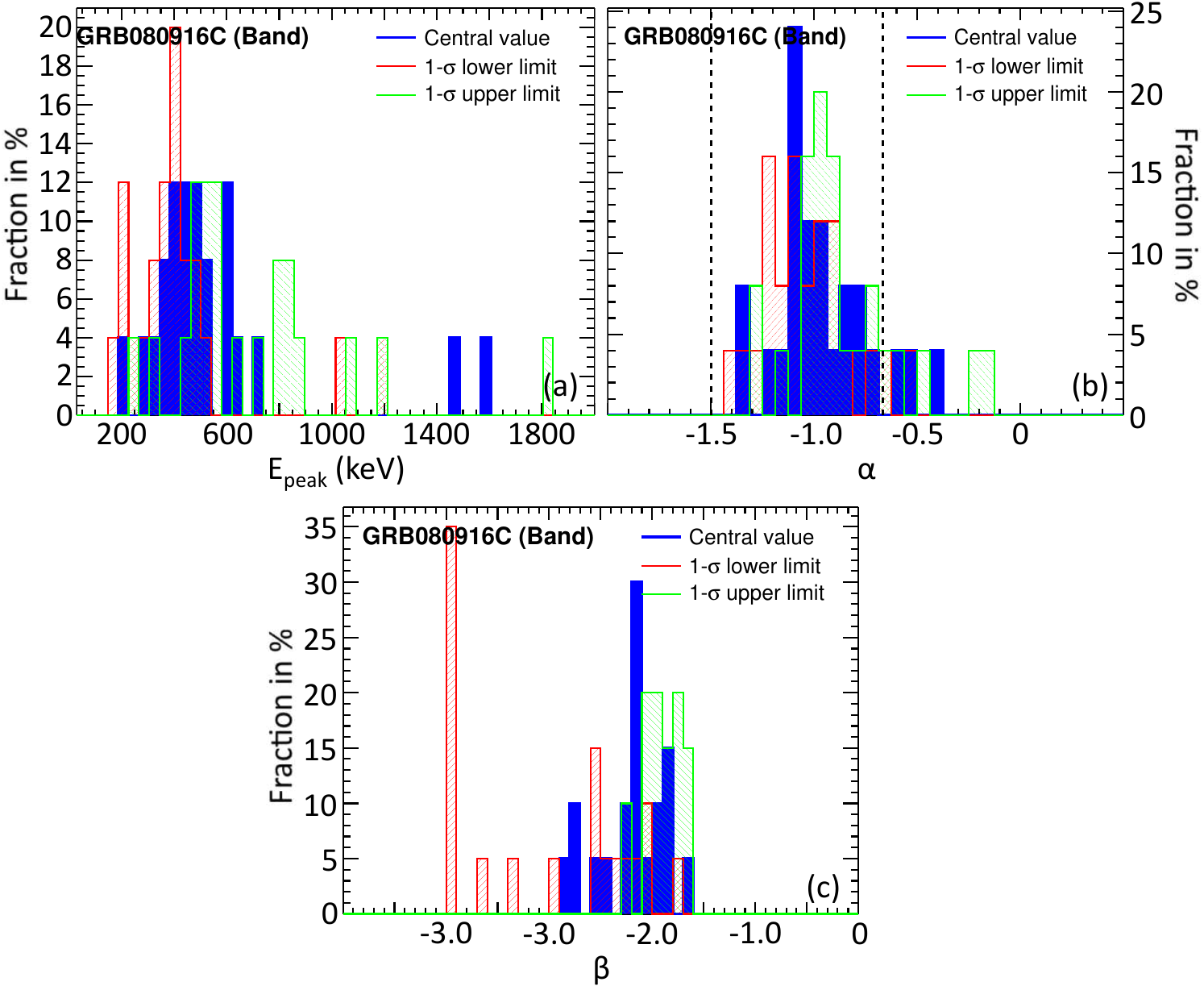}
\caption{\label{fig11}GRB 080916C : Distribution of the spectral parameters resulting from the fine-time analysis using a Band function alone. The filled blue histograms correspond to the distribution of the best parameters (i.e., central values) and the dashed red and green histograms corresponds to the distributions of the 1--$\sigma$ lower and upper limits, respectively. The vertical black dashed line at-2/3 and at -3/2 correspond to the limits above which the values of $\alpha$ are incompatible with pure synchrotron emission from electrons in both the slow and fast cooling regimes, and with synchrotron emission from electrons in the fast cooling regime only, respectively.}
\end{center}
\end{figure}

We have shown in panels (a1--6) of Figure~\ref{fig01}~and~\ref{fig02} the two burst light curves in various energy bands with a 0.1 s time resolution; we show in panels (b1--6) the same light curves binned at the intervals we used for the fine-time spectral analysis (for the bin size determination see Appendix~\ref{section:timeBinSelection}). For this analysis we started by fitting to every time interval both Band-only and the model that we identified in the previous sections as the best choice, namely C+BB+PL. When all three components could not be fitted simultaneously, we kept as many of them as possible (i.e., if C+BB+PL could be fitted but only C+BB, we report the results from C+BB.) An example of the three-component fit using C+BB+C for one time interval of GRB~090926A is presented in Figure~\ref{fig06}.  The complete fit results are reported in Appendix~\ref{section:Spectral-analysis-results} in Table~\ref{tab05} for \grbnosa, and in Table~\ref{tab06} for \grbnosb (See also Appendix~\ref{section:Time-resolved-analysis}. 
 Figures~\ref{fig07} and \ref{fig08} show the reconstructed photon light curves in multiple energy bands using the Band-only and the C+BB+PL analyses of GRBs 080916C and 090926A, respectively. These figures show the time evolution of the individual spectral components as well as the time evolution of the total emission, to be compared to the observed count light curves in the same energy bands. The evolution of the spectral parameters of the Band-only and C+BB+PL models are presented in Figure~\ref{fig09} and \ref{fig10} for GRBs 080916C and 090926A, respectively, and the parameter distributions of the two fits are presented in Figure~\ref{fig11} to \ref{fig14}.

Note that in the following we sometimes use the terminology Band when we talk about the C component of the C+BB+PL scenario; indeed, the C component of C+BB+PL corresponds globally to the traditional Band function and it is sometimes easier to refer to it as Band. The C and Band component are usually statistically equivalent as long as Band $\beta$ is $<$-3.

\clearpage

\subsection{Evolution of the Various Spectral Components in the C+BB+PL Scenario}

\begin{center}
The Additional PL Component
\end{center}

At early times we were able to fit only C+BB for both GRBs, i.e., until T$_\mathrm{0}$+1.2 s and T$_\mathrm{0}$+4.0 s for GRBs 080916C and 090926A, respectively, while keeping all parameters of the C+BB+PL model free;  no additional PL was required. This is consistent with the results reported in Section~\ref{sec:ctr}. Interestingly, while here we only fitted GBM data (i.e., $<$40 MeV), the additional PLs appear contemporaneously with the rise of the LAT-LLE emission above 20 MeV for both GRBs as shown in Figure~\ref{fig07} and \ref{fig08}. This is consistent with the fact that the mechanisms responsible for the additional PL extends from low energies in GBM (i.e., $<$1 MeV) to more than several tens of MeV in the LAT. 

After it kicks off, the additional PL remains present until the end of each burst. At first, its intensity rises until it overpowers the other components and reaches its maximum at $\sim$5 s and $\sim$10 s for GRBs 080916C and 090926A, respectively. The peak intensity of the PL is perfectly correlated in time with a very intense and sharp structure observed in both light curves (see Panels (a) of Figures~\ref{fig07} and \ref{fig08}). For \grbnosb, the PL is so intense at its maximum that it overpowers the two other components at all energies: this  results in a sharp spike lasting $\sim$0.1\,s visible at all energies in the count light curves. Since the relative intensity of the PL to the other components---between 100 keV and 1 MeV---is lower at the peak for \grbnosa, the sharp structure---also lasting for $\sim$0.1\,s---is only clearly visible $<$20 keV and $>$20 MeV in the count light curves. In \grba the PL remains after the peak the most intense component at late time especially below 20 keV.  In the case of \grbnosb, the PL seems to fade faster than the C component in the C+BB+PL model, but it remains until late times.

\begin{figure}[ht!]
\begin{center}
\includegraphics[totalheight=0.334\textheight, clip]{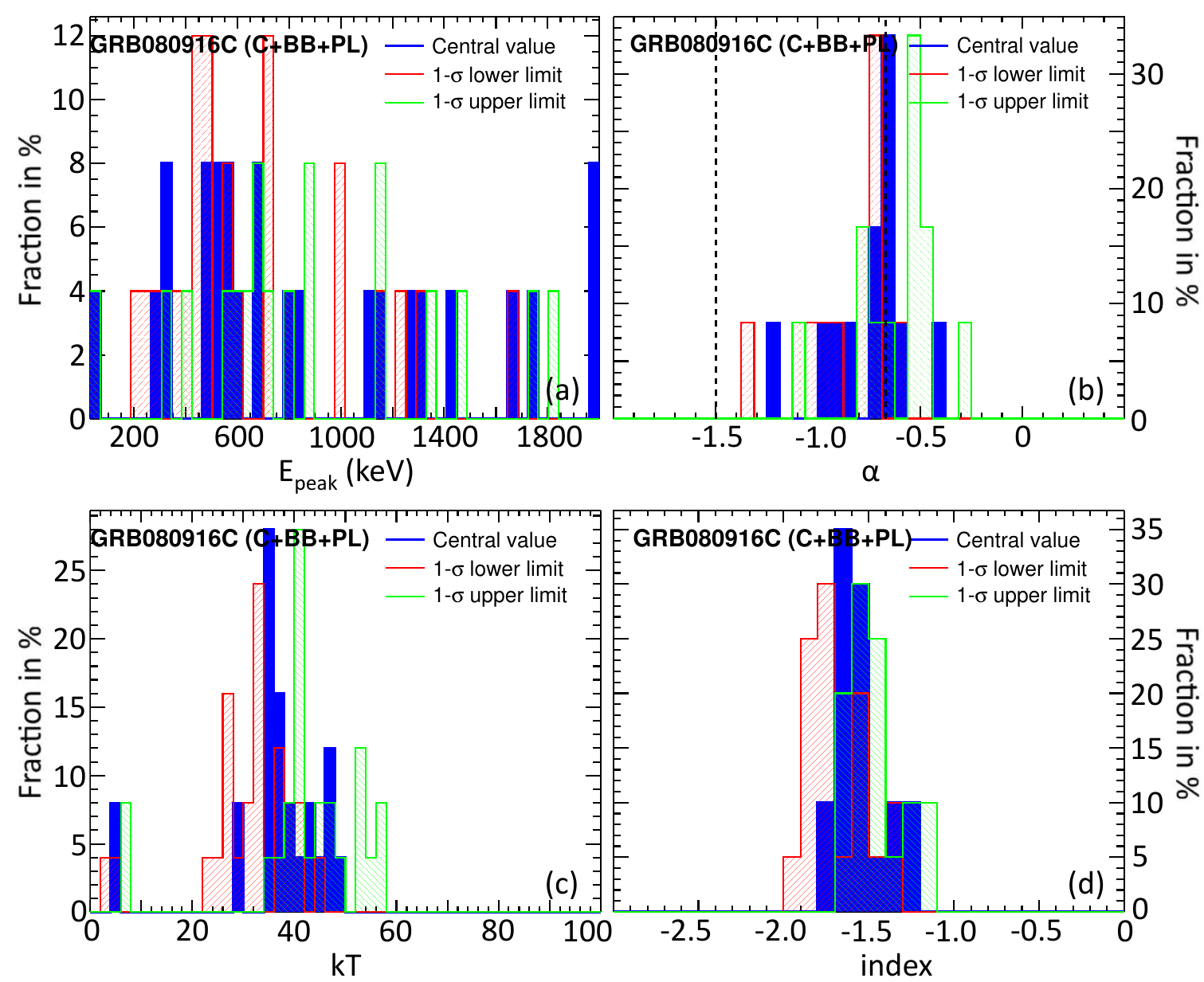}
\caption{\label{fig12}GRB 080916C : Distribution of the spectral parameters resulting from the fine-time analysis using a the C+BB+PL model. The filled blue histograms correspond to the distribution of the best parameters (i.e., central values) and the dashed red and green histograms corresponds to the distributions of the 1--$\sigma$ lower and upper limits, respectively. The vertical black dashed line at -2/3 and at -3/2 correspond to the limits above which the values of $\alpha$ are incompatible with pure synchrotron emission from electrons in both the slow and fast cooling regimes, and with synchrotron emission from electrons in the fast cooling regime only, respectively.}
\end{center}
\end{figure}

\begin{figure}[ht!]
\begin{center}
\includegraphics[totalheight=0.334\textheight, clip]{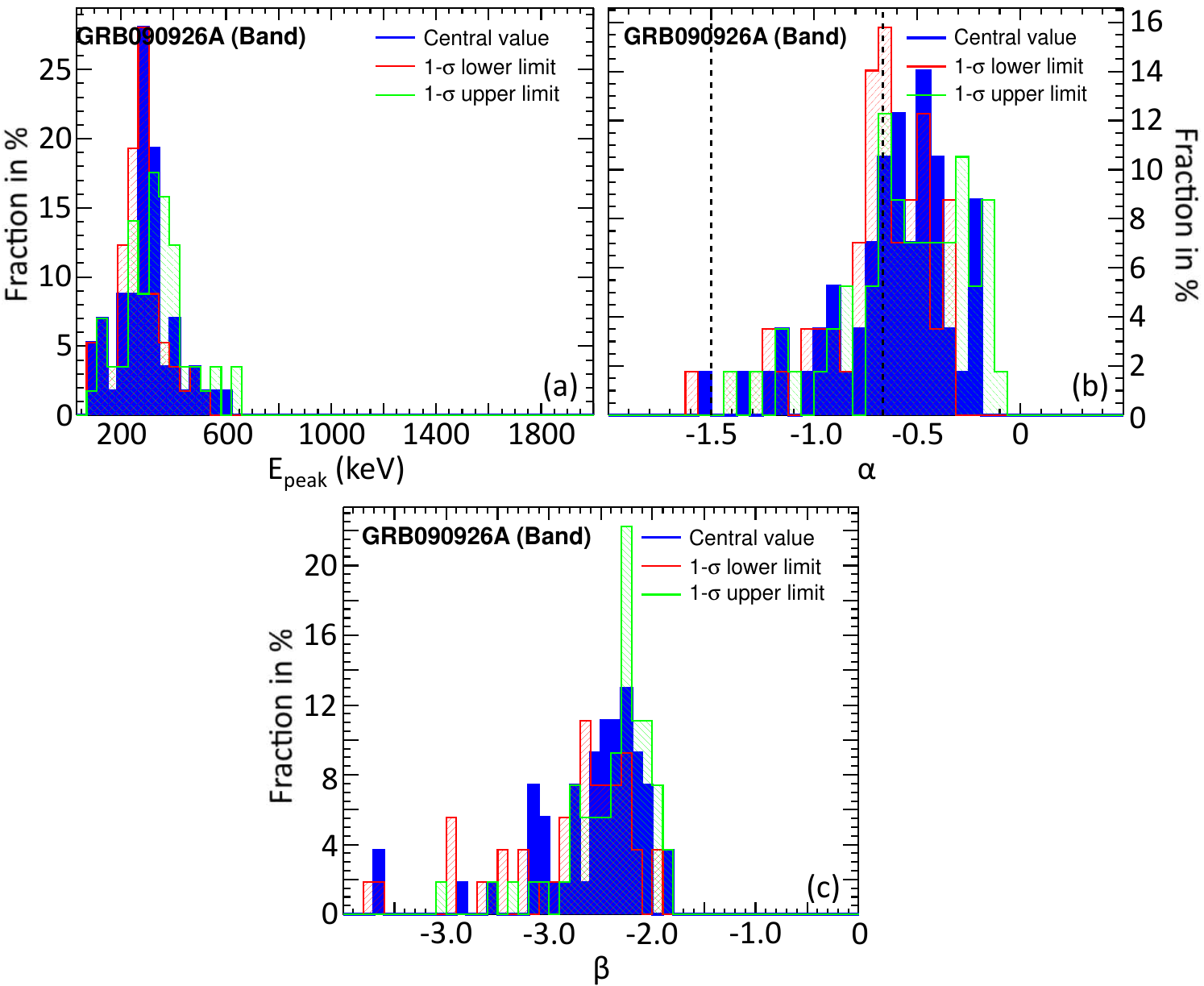}
\caption{\label{fig13}GRB 090926A : Distribution of the spectral parameters resulting from the fine-time analysis using a Band function alone. The filled blue histograms correspond to the distribution of the best parameters (i.e., central values) and the dashed red and green histograms corresponds to the distributions of the 1--$\sigma$ lower and upper limits, respectively. The vertical black dashed line at -2/3 and at -3/2 correspond to the limits above which the values of $\alpha$ are incompatible with pure synchrotron emission from electrons in both the slow and fast cooling regimes, and with synchrotron emission from electrons in the fast cooling regime only, respectively.}
\end{center}
\end{figure}

\begin{figure}[ht!]
\begin{center}
\includegraphics[totalheight=0.334\textheight, clip]{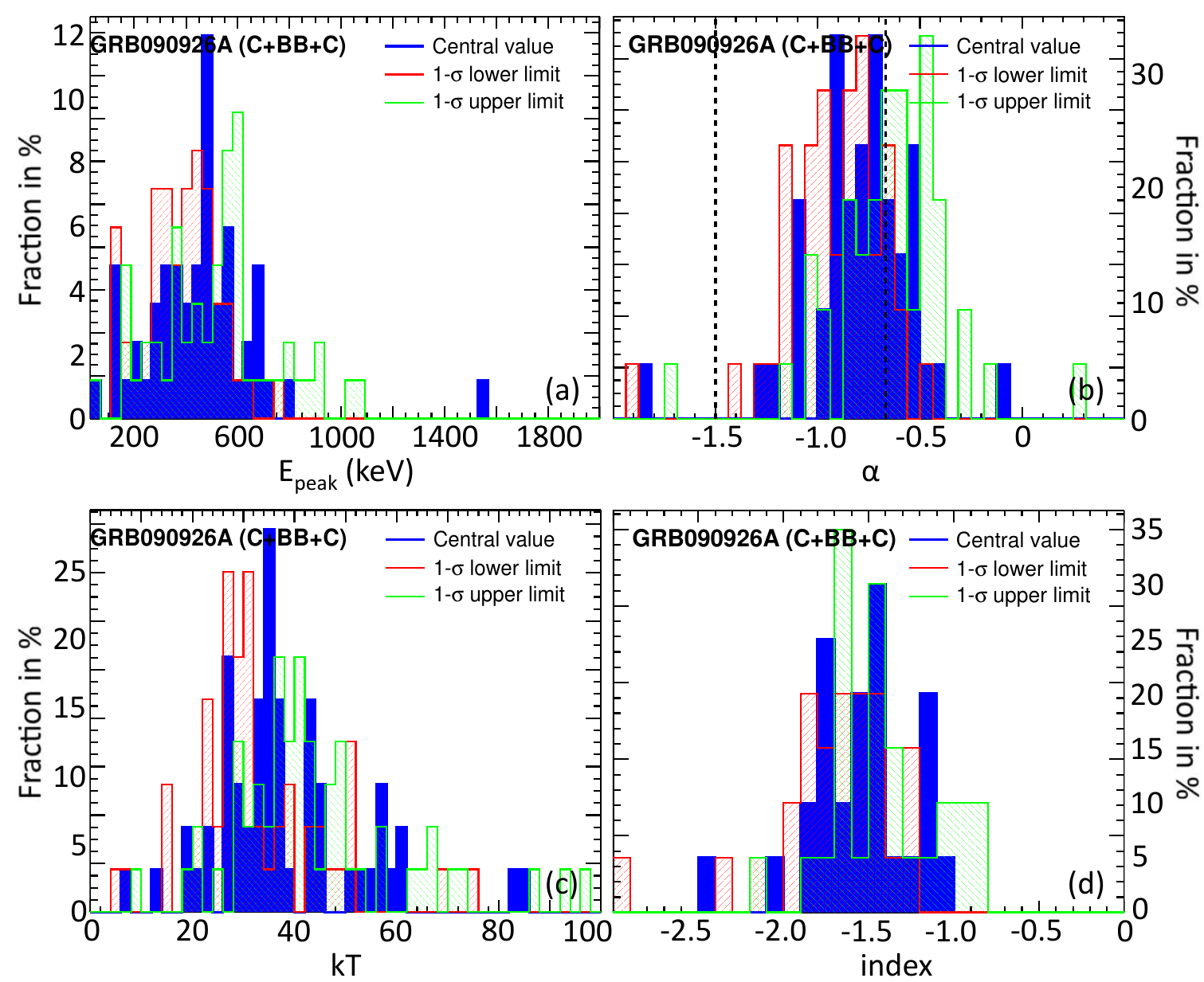}
\caption{\label{fig14}GRB 090926A : Distribution of the spectral parameters resulting from the fine-time analysis using a the C+BB+PL model. The filled blue histograms correspond to the distribution of the best parameters (i.e., central values) and the dashed red and green histograms corresponds to the distributions of the 1--$\sigma$ lower and upper limits, respectively. The vertical black dashed line at -2/3 and at -3/2 correspond to the limits above which the values of $\alpha$ are incompatible with pure synchrotron emission from electrons in both the slow and fast cooling regimes, and with synchrotron emission from electrons in the fast cooling regime only, respectively.}
\end{center}
\end{figure}

While the intensities of the additional PLs evolve with time, their indices remain between $-1.0$ and $-2.0$ across the burst durations (see Figure~\ref{fig09}e \& \ref{fig10}e and Figure~\ref{fig12}d \& \ref{fig14}d). For \grba the values of the PL index are $\sim$\mbox{-1.5} during the whole burst (see Figure~\ref{fig09}e \&~\ref{fig12}d.) For \grbnosb, the PL index values decline steadily and also cluster around -1.5 (see Figure~\ref{fig10}e \& \ref{fig14}d).

We note that the index of the additional PL during the prompt phase is similar with the PL indices reported for the extended LAT emission seen up to thousands of seconds after the prompt phase and contemporaneous with the afterglow emission~\citep[e.g.,][]{Ackermann:2014}. This may suggest that the latter is the result of the same physical process that produced the PL in the prompt emission.

\clearpage

\begin{center}
The BB Component
\end{center}

For both GRBs, the BB is more intense at early time (i.e., during the first 4 to 5 s of each burst) than at late time (see red curves in Figure~\ref{fig07}c \& \ref{fig08}c). Although its intensity decreases with time, it is possible to identify this component during the entire prompt phase. For both GRBs, the BB temperature peaks around 30--40 keV with a larger spread of values for \grbb (see Figure~\ref{fig12}c \& \ref{fig14}c). While a global cooling trend may be observed for \grbnosb, the temperature seems to remain constant for \grba (see Figure~\ref{fig09}d \& \ref{fig10}d). Although the BB is a subdominant component over the entire energy range, it approaches and even sometimes overpowers the C component at the peak energy of the Planck spectrum around 100 keV (see Figure~\ref{fig07}c \& \ref{fig08}c).

\begin{center}
The C Component
\end{center}
\label{sec:CPL-component}

The C component carries most of the prompt emission energy and it clearly overpowers the other components between 100 keV and 1 MeV. After a fast rise at early times, the intensity of the C component decays steadily with time (see Figure~\ref{fig07}c \& \ref{fig08}c). This simple shape structure is much less evident when a Band function alone is fitted to the data. For both GRBs, the $\alpha$ index of C, remains more or less constant with time, especially for \grba (see Figure~\ref{fig09}b \& \ref{fig10}b). Although the values are more spread for \grbnosb, both distributions are centered around $-0.7$ (see Figure~\ref{fig12}b \& \ref{fig14}b). In addition, the 1--$\sigma$ lower limits of $\alpha$ are between $\sim$$-1.5$ and $\sim$$-0.7$---beside a few outliers---, which are the limits for synchrotron fast and slow cooling, respectively.

It is very important to note that the values of $\alpha$ are much lower (i.e., $<$-1) in some time intervals where only C+BB can be fitted to the data (see first time interval of GRB 080916C as well as the two first and last time intervals of GRB 090926A in Figure~\ref{fig09}b \& \ref{fig10}b, respectively.) In these intervals, the values of $\alpha$  are closer to the mean value of the additional PL index than in the other time intervals. As we will see in Section~\ref{sec:newmodel}, this most likely indicates that what we think is the contribution of the C component at early times might actually be due to the additional PL contribution, which is very intense at early and late times. Therefore, {\it the additional PL may be the first visible component in some bursts, and this component may last much longer and even extend into the afterglow phase}. 

Finally, the E$_\mathrm{peak}$ globally decreases with time from $>$several MeV to hundreds and even tens of keV at late times (see Figure~\ref{fig09}a \& \ref{fig10}a). The E$_\mathrm{peak}$ also tracks the Band function energy flux as we will see in Section~\ref{sec:flux-epeak}.

\subsection{Band vs C+BB+PL}
\label{sec:Band-vs-C+BB+PL}

If the prompt emission is indeed composed of separate spectral components as proposed in the C+BB+PL scenario, the fit of any single component (i.e., a Band function) to the data should correspond to an average of the complex model shape. It is indeed what we observe for \grba and \grbb (see Figure~\ref{fig19} \& \ref{fig20} of Appendix~\ref{section:Time-resolved-analysis}):  the yellow lines corresponding to the Band function fits are averages of the black ones, which correspond to the sum of the three components of the C+BB+PL model. Therefore, in the context of the C+BB+PL scenario, the systematic wavy pattern observed in the fit residuals when a Band function alone is fitted to the data (see panel (a3) of Figure~\ref{fig04} \& \ref{fig05}) is naturally expected: taking into account the various components not only reduces the Cstat value, but it globally flattens the systematic wavy pattern of the fit residuals.

Generally, the fit of a Band function alone to a spectrum composed of a Band function and a subdominant BB component results in spectral parameters that are systematically biased: both $\alpha$ and $\beta$ are greater than their `real' values, and the value of E$_\mathrm{peak}$ is underestimated. On the other hand, a Band-only fit on a spectrum which is truly Band+PL overestimates $\beta$ and underestimates E$_\mathrm{peak}$ and $\alpha$. This is in perfect agreement with the results presented in Sections~\ref{sec:tisa} and \ref{sec:ctr}, as well as with those already published in \citet{Guiriec:2010,Guiriec:2011a,Guiriec:2013a}.

We discuss below two cases, one where the additional PL is intense compared to C but  the BB contribution is limited (e.g., time interval 26.0--32.0 s of \grbnosa---see Figure~\ref{fig19} of Appendix~\ref{section:Time-resolved-analysis}), and one where the BB component is intense relative to C but the additional PL is limited (e.g., time interval 1.5--1.8 s of \grbnosb---see Figure~\ref{fig20} of Appendix~\ref{section:Time-resolved-analysis})

(i) time interval ``26.0--32.0 s'' of \grbnosa:  The E$_\mathrm{peak}$ of the Band-only fit is located between the two humps of the C+BB+PL model; the low-energy hump of the latter corresponds to the energy at the maximum of the BB component and the high-energy one to the E$_\mathrm{peak}$ of C. At the same time, the high-energy slope of the Band-only fit (i.e., $\beta$) is much harder than the high-energy slope of C (i.e., exponential cutoff) of the C+BB+PL fit, because the high-energy power law of the Band-only fit is an average between the high-energy emission of C and of the additional PL component. Finally, and more importantly, the value of the Band-only $\alpha$ index is mostly an average of C and PL, since these are the most intense components in this time interval. Indeed, $\alpha$ of the Band-only fit is $\sim-1.1$, which is about the average of the C and the PL component slopes, whose indices are $\sim-0.7$ and $\sim-1.5$, respectively. The results derived here can be extrapolated to all intervals for \grbnosa. Indeed, if we closely examine the evolution of $\alpha$ over time for \grba (see Figure~\ref{fig09}b) and compare it to the intensity evolution of the  various spectral components in the C+BB+PL scenario (see panels (c) of Figure~\ref{fig07}), we can easily explain the evolution of the parameters of the Band function fits. Figure~\ref{fig09}b shows that the Band $\alpha$ values (blue line) exhibit a dramatic change around 4--5\,s: $\alpha$ drops from $\sim-0.7$ before the discontinuity down to $<-1.0$ after it. This discontinuity in the values of $\alpha$ matches perfectly with the time at which the additional PL starts to overpower the other components, especially at low energies that have the strongest impact on the Band $\alpha$ (see panels (c) of Figure~\ref{fig07}). The values of Band-only $\alpha$ are the lowest (i.e., $<$-1.3) around 5--6\,s which corresponds to the intensity peak of the additional PL in the C+BB+PL model. However, the $\alpha$ of the C+BB+PL fit remains constant around $-0.7$ during the entire burst, except for the first time interval where it is $<-1.1$ (red line). We will see in Section~\ref{sec:newmodel} that this low value of $\alpha$ at early times in the C+BB+PL model may be an indication that the additional PL is already present at very early times in some GRBs. Finally we note the strong changes in the values of $\beta$, which vary from $\sim-1.7$ to $<-5.0$ (see Figure~\ref{fig09}c). The values of $\beta$ are the highest from $\sim$5 to $\sim$10\,s, which is when the additional PL is the most intense in the C+BB+PL scenario (see panels (c) of Figure~\ref{fig07}.) Also, values of $\beta$ above $-2.0$ are clearly not physical as these imply an infinite amount of energy, so if the Band function were indeed the correct fit, then a cutoff in its high-energy PL would be required to adequately explain the data when $\beta$$>$-2. Therefore, it is perfectly natural to account for such strong variations of the Band $\beta$ index with a model including an additional PL at high energies as explained earlier in this Section.

(ii) time interval ``1.5--1.8 s'' of \grbnosb: The E$_\mathrm{peak}$ of the Band-only fit is also located in between the two humps of the C+BB+PL model. The high value of $\beta$ (i.e., $>-2$) in the Band-only scenario is too hard to be physical and a spectral cutoff at high-energy should be present. On the other hand, such a high $\beta$ value is naturally accommodated in the C+BB+PL model, where $\beta$ would be an average of the decaying parts of BB and C. The value of Band $\alpha$ would also be an average of the BB and C components. Since a pure Planck function is adequately approximated with a C with an index of $+1.0$, and since the C component of the C+BB+PL model has an index of $\sim-1.1$, we expect to recover a value of $\alpha$ greater than $\sim-1.1$ when fitting Band-alone to the data if the true model were a combination of a C and BB. Indeed, when fitting Band-alone to the data, we measure a value of $\alpha$ $>-0.6$. We will see in Section~\ref{sec:newmodel} that the low value of $\alpha$ (i.e., $\sim$-1.1) resulting from the C+BB+PL fit may indicate the presence of an underlying additional PL component at early times in \grbnosb. With the same considerations as for \grba in the previous paragraph, we can extend the results obtained in this time interval to the entire burst by comparing the evolution of $\alpha$ between the Band-only and the C+BB+PL models (see Figure~\ref{fig10}b) with the intensity evolution of the various components with time (see panels (c) of Figure~\ref{fig08}). The BB component is the most intense during the first 4--5 s of the burst, which also corresponds to the time intervals where the highest values of $\alpha$ are measured when fitting Band-alone to the data. The BB component biases Band-only $\alpha$ towards higher values in this time period. From $\sim5$ to $\sim10$\,s, C clearly overpowers the other components, which explains the values of $\alpha$ around $-0.7$ obtained during this time. Finally, after 10\,s, the additional PL becomes intense and biases the values of $\alpha$ down to values as low as $\sim-1.5$.

While $\alpha$ values for \grba in the Band-only scenario are mostly between $-1.5$ and $-0.7$---namely the fast and slow synchrotron cooling regimes, respectively---most of them are greater than $-0.7$ for \grbb (see Figure~\ref{fig11}b \& \ref{fig13}b).

In the C+BB+PL scenario, the distributions of $\alpha$ values peak around $-0.7$ for both GRBs. For \grba the $\alpha$ distribution is very sharp, but it is wider for \grbb (see Figure~\ref{fig12}b \& \ref{fig14}b); however, the 1--$\sigma$ lower limits are nearly all below $-0.7$. Conversely, the values of E$_\mathrm{peak}$ are spread over hundreds of keV with C+BB+PL, while they are all clustered around 300--400 keV with Band-alone fits.

An interesting result appears when we extrapolate the reconstructed photon light curves obtained by fitting either Band or C+BB+PL to the GBM data (i.e., between 8 keV and 40 MeV) into the LLE energy band (i.e., 20 to 100 MeV). While both Band and C+BB+PL well reproduce the count light curves in the energy bands ranging from 8 keV up to 1 MeV, only C+BB+PL adequately mimics the LLE light curve between 20 and 100 MeV (see panels (c) of Figure~\ref{fig07} \& \ref{fig08}). This is particularly true for \grbnosa. Indeed, while C+BB+PL adequately reproduces the observed position of the peak intensity of the 20--100 MeV light curve as well as its relative intensity after 8\,s, the Band-only fits result in a peak position which is shifted to earlier times and in an extended high-energy emission which is too intense before 20\,s and too weak afterwards.

Finally, we performed MC simulations to investigate the validity of the C+BB+PL model based on the method described in Appendix~\ref{section:modelComparison}. For each time interval, we generated 10$^\mathrm{5}$ synthetic spectra choosing the Band-only fit as the null hypothesis; these were then fit either with Band or C+BB+PL. When fitting a Band function alone to the synthetic spectra, we adequately recovered the input parameters. However, the Cstat values were much lower than those resulting from the fit to the real data. It was often not possible to fit C+BB+PL to synthetic spectra corresponding to time intervals where the input $\beta$ values were high (i.e., $\sim-2$) because the fitting engine did not converge. In the cases it did converge and the resulting parameter values were ``compatible'' with the observed ones, the resulting Cstat values were much higher than those obtained when fitting C+BB+PL to the real data.  Therefore, Band was always a better description of the synthetic spectra. C+BB+PL was usually an adequate model to fit the synthetic spectra corresponding to time intervals with steep Band-function $\beta$ ($<-5$); in these intervals C+BB+PL and Band-alone fits led to similar Cstat values. However, C+BB+PL fits resulted in null fluxes for BB and PL, leaving only the C component. In addition, the C parameters were corresponding to the input parameters of the Band function and not to those observed when fitting C+BB+PL to the real data. We then performed the same exercise but choosing C+BB+PL as the null hypothesis. When fitting the synthetic spectra with C+BB+PL, the fitting engine converged $\geq$80\% of the time, allowing us to recover the input parameters. The resulting Cstat values were also similar to those obtained when fitting the real data though a little bit lower. When fitting a Band function alone to these synthetic spectra, we recovered the parameters obtained when fitting the same function to the real data. In addition, the Cstat values were also similar to those resulting from the real data fits. This reinforces the idea that C+BB+PL is an overall better description of the time-resolved spectra for both GRBs. We note that this exercise could not be performed in all time intervals while leaving all parameters from the C+BB+PL model free; it was, however, possible to simulate all time intervals where the additional PL clearly overpowers the two other components. 

\begin{figure*}[ht!]
\begin{center}
\includegraphics[totalheight=0.55\textheight, clip]{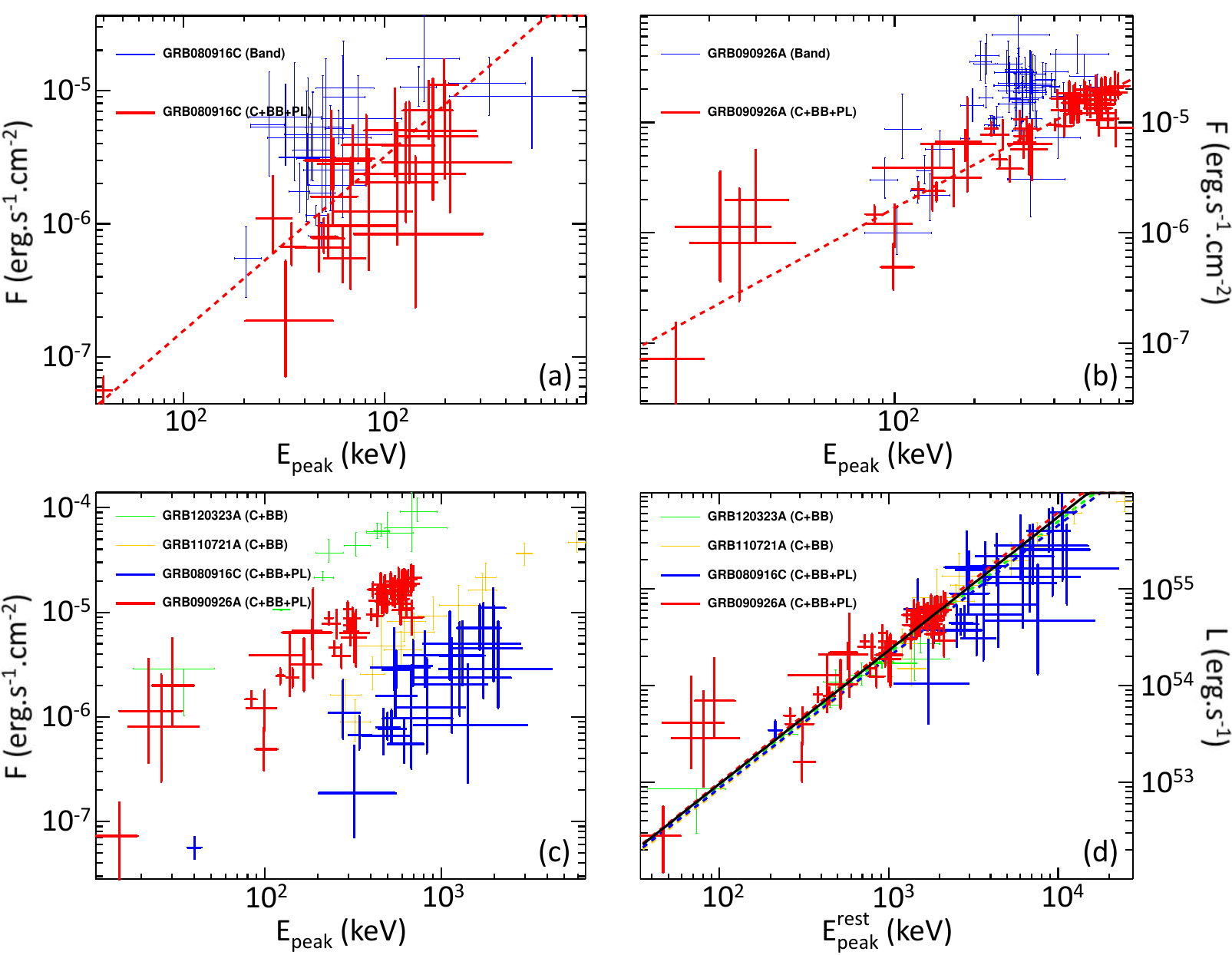}
\caption{\label{fig15}(a) Flux of the Band function (or of the cutoff power law) versus its E$_\mathrm{peak}$ when fitting Band-only (blue) or C+BB+PL (red) to the fine time intervals of GRB~080916C. The dashed red line corresponds to the best fit of the red data points with a power law; (b) Flux of the Band function (or of the cutoff power law) versus its E$_\mathrm{peak}$ when fitting Band-only (blue) or C+BB+PL (red) to the fine time intervals of GRB~090926A. The dashed red line corresponds to the best fit of the red data points with a power law; (c) Flux of the cutoff power law (i.e., non-thermal component, NT) versus its E$_\mathrm{peak}$ when C+BB or C+BB+PL are fitted to the time resolved data of four GRBs including short (green) and long ones (red, yellow and green). This is a refined version of the figure published in~\citet{Guiriec:2013a} showing the similarity of the  F$_\mathrm{i}^\mathrm{NT}$--E$_\mathrm{peak,i}^\mathrm{NT}$ relation for both short and long GRBs; (d) Luminosity of the cutoff power law versus its E$_\mathrm{peak}$ when corrected for both the redshift and the k-correction (i.e., L$_\mathrm{i}^\mathrm{NT}$--E$_\mathrm{peak,i}^\mathrm{rest,NT}$ relation) for the same sample of GRBs as in panel (c). The color dashed lines correspond to the best power law fit to the data of each individual GRB. The solid black line corresponds to the best power law fit to the whole sample simultaneously.}
\end{center}
\end{figure*}

\vspace{2cm}
\section{F$_\mathrm{\lowercase{i}}^\mathrm{NT}$--E$_\mathrm{peak,\lowercase{i}}^\mathrm{NT}$ and L$_\mathrm{\lowercase{i}}^\mathrm{NT}$--E$^\mathrm{rest,NT}_\mathrm{peak,\lowercase{i}}$ relations}
\label{sec:flux-epeak}

Figure~\ref{fig15}a and b show the energy fluxes of the Band function (or C in the case of C+BB+PL) as a function of E$_\mathrm{peak}$ when fitting either Band-alone (blue) or C+BB+PL (red) to the time-resolved spectra of \grba (Figure~\ref{fig15}a) and \grbb (Figure~\ref{fig15}b). We note that the correlation between the two quantities is much stronger with C+BB+PL. As reported for the first time in \citet{Guiriec:2013a}, if this new F$_\mathrm{i}^\mathrm{NT}$--E$_\mathrm{peak,i}^\mathrm{NT}$ relation is a specific property of the non-thermal component only (i.e., Band or C in the new multi-component model), it is expected that the fits of a Band function alone to GRBs that have additional spectral components will lead to large scatter in the data and in a weaker and biased relation or even in no correlation at all if the additional components are very intense.

In Figure~\ref{fig15}c, the F$_\mathrm{i}^\mathrm{NT}$--E$_\mathrm{peak,i}^\mathrm{NT}$ relations of several GRBs are overplotted. The results on short GRB 120323A and long GRB 110721A were already reported in \citet{Guiriec:2013a} and preliminary results of GRBs 080916C and 090926A using C+BB+PL were also shown in the same article. Here, the detailed analysis of GRBs 080916C and 090926A confirms the results reported in \citet{Guiriec:2013a}: in the observer frame, the F$_\mathrm{i}^\mathrm{NT}$--E$_\mathrm{peak,i}^\mathrm{NT}$ relations have similar slope for all GRBs whatever is the duration of the burst. In addition, when corrected for redshift, all GRBs lie along the same L$_\mathrm{i}^\mathrm{NT}$--E$^\mathrm{rest,NT}_\mathrm{peak,i}$ relation---with a very limited scatter---suggesting a universality for the phenomenon (see Figure~\ref{fig15}d). This new detailed analysis tightened even more the relation presented in \citet{Guiriec:2013a}.

The PL fit of the L$_\mathrm{i}^\mathrm{NT}$--E$_\mathrm{peak,i}^\mathrm{rest,NT}$ relation for each individual GRB result is shown in colored dashed lines:
\begin{figure}[ht!]
\begin{center}
\includegraphics[totalheight=0.334\textheight, clip]{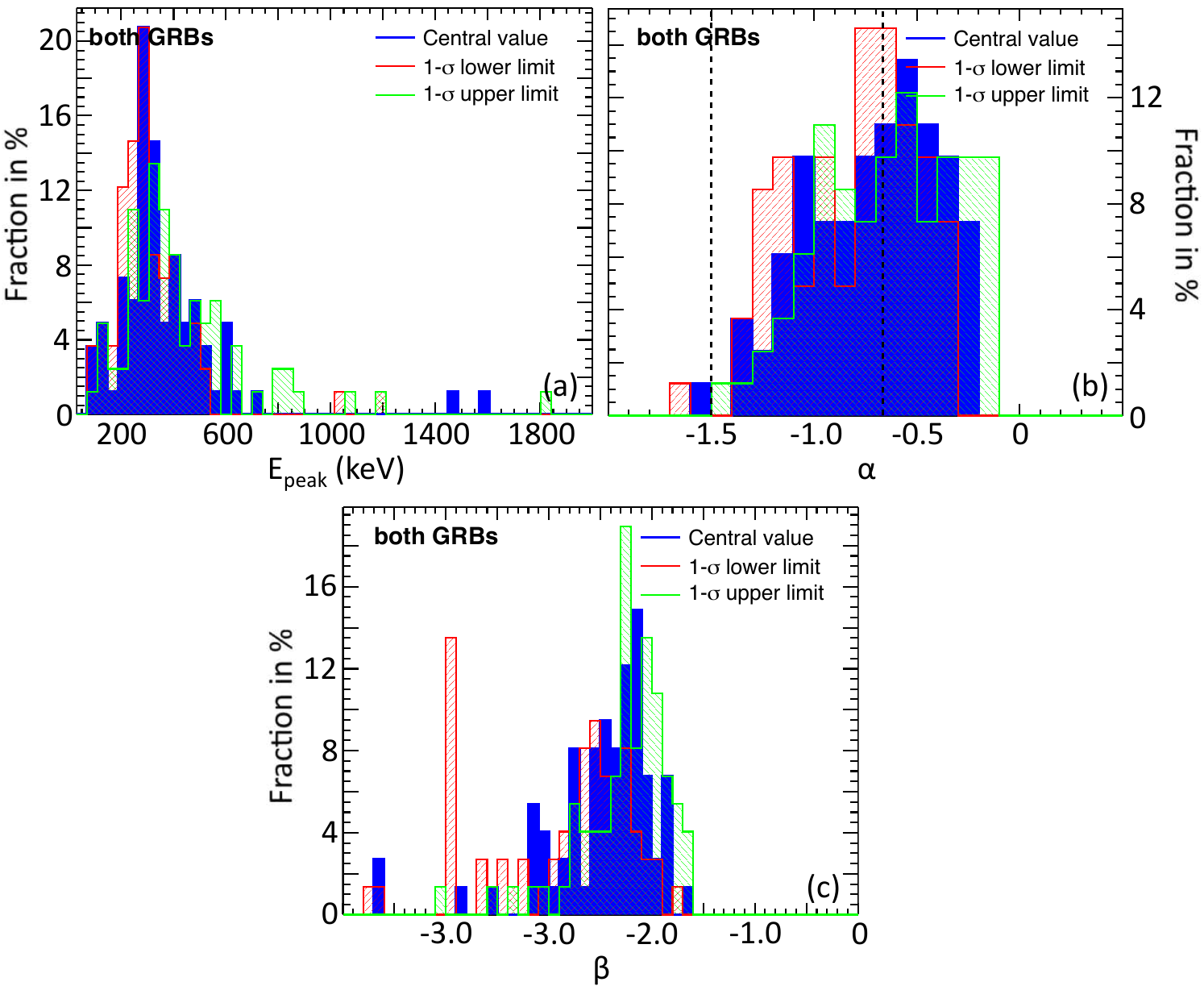}
\caption{\label{fig16}Distribution of the spectral parameters resulting from the fine-time analysis of both GRBs~080916C and~090926A using a Band function alone. The filled blue histograms correspond to the distribution of the best parameters (i.e., central values) and the dashed red and green histograms corresponds to the distributions of the 1--$\sigma$ lower and upper limits, respectively. The vertical black dashed line at -2/3 and at -3/2 correspond to the limits above which the values of $\alpha$ are incompatible with pure synchrotron emission from electrons in both the slow and fast cooling regimes, and with synchrotron emission from electrons in the fast cooling regime only, respectively.}
\end{center}
\end{figure}
\[
\mathrm{L_{080916C,i}^{NT}}=\mathrm{(8.7\pm3.4)10^{52}~\left(\frac{E_{peak,i}^{rest,NT}}{100~keV}\right)^{1.36\pm0.12}~erg.s^{-1}}
\]
\[
\mathrm{L_{090926A,i}^{NT}}=\mathrm{(10.0\pm1.8)10^{52}~\left(\frac{E_{peak,i}^{rest,NT}}{100~keV}\right)^{1.40\pm0.07}~erg.s^{-1}}
\]
\[
\mathrm{L_{120323A,i}^{NT}}=\mathrm{(9.3\pm3.0)10^{52}~\left(\frac{E_{peak,i}^{rest,NT}}{100~keV}\right)^{1.36\pm0.15}~erg.s^{-1}}
\]
\[
\mathrm{L_{110721A,i}^{NT}}=\mathrm{(8.4\pm2.0)10^{52}~\left(\frac{E_{peak,i}^{rest,NT}}{100~keV}\right)^{1.36\pm0.12}~erg.s^{-1}}
\]

The simultaneous fit of all the GRBs with a PL results in the solid black line:

\begin{equation}
\mathrm{L_\mathrm{i}^\mathrm{NT}}=\mathrm{(9.6\pm1.1)10^{52}~\left(\frac{E_{peak,i}^{rest,NT}}{100~keV}\right)^{1.38\pm0.04}~erg.s^{-1}}
\label{eq:L-Epeak_both}
\end{equation}

The strong compatibility of the results for all the tested GRBs reinforces the possibility to eventually use this relation as a redshift estimator.

We must note that the initial rising part of the first pulse of a burst usually does not follow the typical F$_\mathrm{i}^\mathrm{NT}$--E$_\mathrm{peak,i}^\mathrm{NT}$ relation as already reported in \citet{Guiriec:2013a}, but an anti-correlation is observed between F$_\mathrm{i}^\mathrm{NT}$ and E$_\mathrm{peak,i}^\mathrm{NT}$ instead.

\begin{figure}[ht!]
\begin{center}
\includegraphics[totalheight=0.334\textheight, clip]{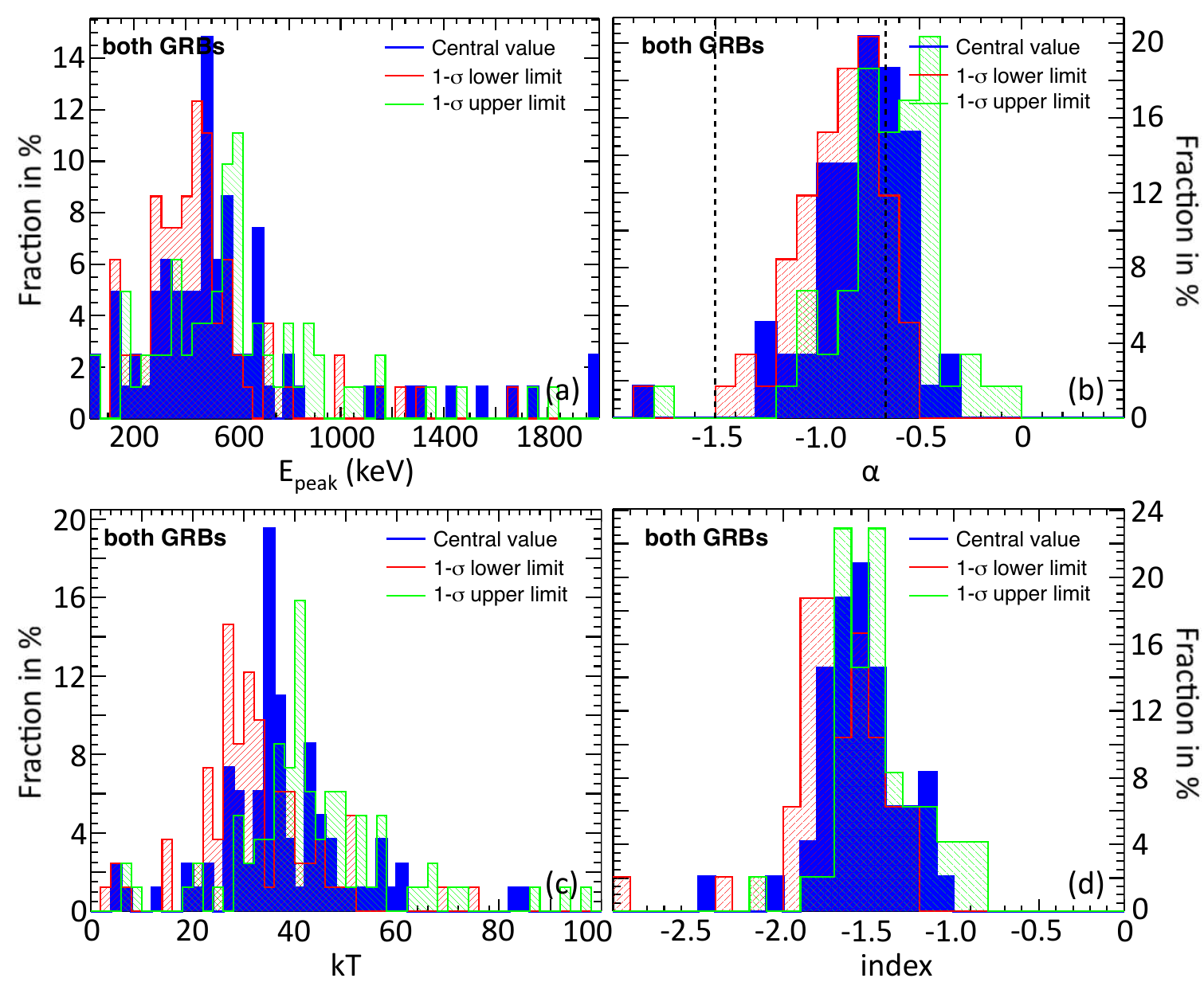}
\caption{\label{fig17}Distribution of the spectral parameters resulting from the fine-time analysis of both GRBs~080916C and~090926A using a the C+BB+PL model. The filled blue histograms correspond to the distribution of the best parameters (i.e., central values) and the dashed red and green histograms corresponds to the distributions of the 1--$\sigma$ lower and upper limits, respectively. The vertical black dashed line at -2/3 and at -3/2 correspond to the limits above which the values of $\alpha$ are incompatible with pure synchrotron emission from electrons in both the slow and fast cooling regimes, and with synchrotron emission from electrons in the fast cooling regime only, respectively.}
\end{center}
\end{figure}

\section{A new model for GRB prompt emission}
\label{sec:newmodel}

In Section~\ref{sec:ftr}, we reported the constancy in the values of the $\alpha$ and the indices of the additional PL when fitting C+BB+PL to the time-resolved data of both GRBs. Not only did these values remain mostly constant during each burst (see panels (b) and (e) of Figure~\ref{fig09} and \ref{fig10}), but they are also nearly the same for both events (see panels (b) and (d) of Figure~\ref{fig12} and \ref{fig14}). Figures~\ref{fig16} and \ref{fig17} show the Band-only and C+BB+PL combined parameter distributions, respectively, for GRBs 080916C and 090926A. This reinforces the conclusions reported in Section~\ref{sec:Band-vs-C+BB+PL}: the Band-only fits result in a narrow distribution of the E$_\mathrm{peak}$ values around 300 keV and in a broader distribution for $\alpha$, with a large fraction of the latter values $>-0.7$. In contrast, the E$_\mathrm{peak}$ values for C+BB+PL are much more spread with a maximum of the distribution around 500 keV, the values of $\alpha$ are narrowly peaked around $-0.7$ and the $\alpha$ 1--$\sigma$ lower limits are all bellow $-0.7$ besides a few cases, and the values of the additional PL index are clustered between $-1$ and $-2$ with a distribution peak around $-1.5$.

Since the slopes of the low-energy PL of C and of the additional PL in the C+BB+PL model do not change much with time and remain similar in the two bursts, those two parameters can be frozen to their typical values in the fits, reducing the complexity of the C+BB+PL model by two dof. This new model, therefore, which we will call C+BB+PL$_\mathrm{5params}$ hereafter, has only 5 parameters 
. C+BB+PL$_\mathrm{5params}$ has only one parameter more than Band-alone, which makes the two models very competitive from a statistical point of view.

\begin{figure*}[ht!]
\begin{center}
\includegraphics[totalheight=0.50\textheight, clip]{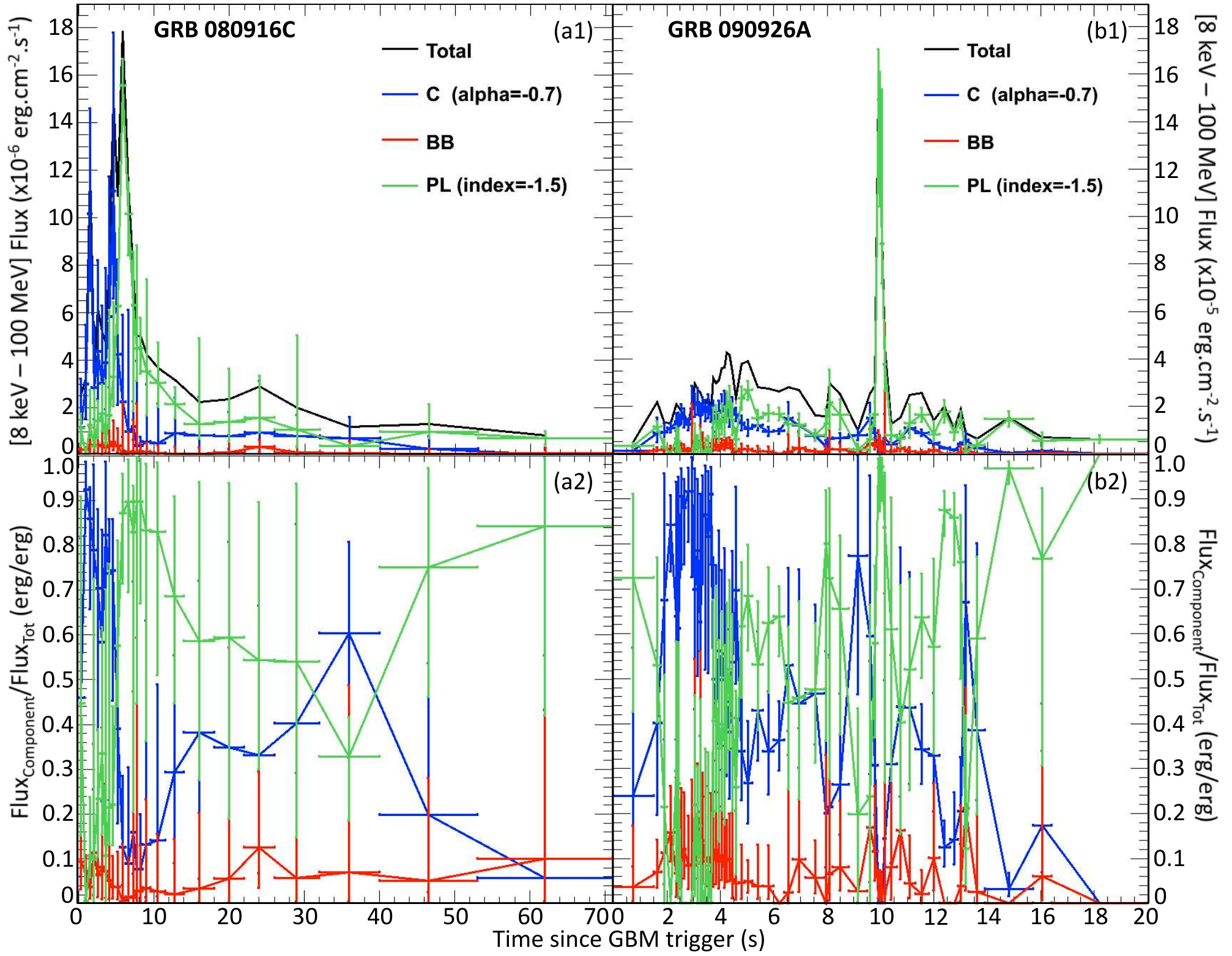}
\caption{\label{fig18}Energy flux evolution between 8 keV and 100 MeV for (a1) GRB~080916C and (b1) GRB~090926A in the context of the C+BB+PL$_\mathrm{5params}$ model. Panels (a2) and (b2) of each figure show the contribution of each component to the total energy flux of GRBs~080916C and~090926A, respectively. For clarity, no uncertainty is displayed for the total energy flux (black line). Because of its narrow shape, the contribution of the BB component to the total 8 keV--100 MeV flux is very limited; however, it has an important relative contribution to the total emission at the BB spectral peak around 100--200 keV (see Figure~\ref{fig21}~\&~\ref{fig22} of Appendix~\ref{section:Time-resolved-analysis}).}
\end{center}
\end{figure*}

We fitted C+BB+PL$_\mathrm{5params}$ to the time-resolved spectra and the results are reported in Table~\ref{tab05} of Appendix~\ref{section:Spectral-analysis-results} \& Figure~\ref{fig21} of Appendix~\ref{section:Time-resolved-analysis} and Table~\ref{tab06} of Appendix~\ref{section:Spectral-analysis-results} \& Figure~\ref{fig22} of Appendix~\ref{section:Time-resolved-analysis} for GRBs 080916C and 090926A, respectively. As expected, C+BB+PL$_\mathrm{5params}$ results in similar Cstat values as C+BB+PL except for a few cases where the additional PL slope appears to be a bit steeper than $-1.5$. However, this can be easily explained by the presence of a cutoff in the additional PL. Indeed, the C+BB+C$_\mathrm{6params}$ fits (i.e., C+BB+PL$_\mathrm{5params}$ with a exponential cutoff in the additional PL) seem to indicate the presence of a cutoff whose energy evolves smoothly with time. The cutoff could be hidden by the C component at times, and it could be at much higher energies at others. This result may question the use of the additional PL cutoff to estimate the Lorentz factor as proposed in~\citet{Ackermann:2011:GRB090926A}. Instead, such an evolution of the cutoff energy may be a signature of the emission mechanism(s) producing this component and not an effect of the $\gamma$--$\gamma$ opacity.

It is very important to note that the number of time intervals in which the three components cannot be simultaneously fitted is very limited for C+BB+PL$_\mathrm{5params}$ compared to C+BB+PL. Even more interestingly, C+BB+PL$_\mathrm{5params}$ results indicate that the additional PL is present from the very beginning of the prompt emission in both GRBs. Moreover, the additional PL is the most intense component at very early times. This is supported by the very low values of $\alpha$ (i.e., $<-1$) obtained in the first time intervals when fitting C+BB+PL with all the parameters free (see Figure~\ref{fig09}b \& \ref{fig10}b). Indeed, in Section~\ref{sec:CPL-component} we reported that we could not fit the C+BB+PL model at early times, but only C+BB. Intriguingly, in these time intervals, the values of $\alpha$ were significantly lower than average and closer to the typical index values of the additional PL. This can be explained if we consider the fit resulting from C+BB+PL$_\mathrm{5params}$ as the ``true'' model. With C+BB+PL$_\mathrm{5params}$, the additional PL overpowers the other two components, and when fitting C+BB to the data, the value of $\alpha$ is a hybrid (i.e., $\sim-1.1$) between the real value of the C component (i.e., $\sim-0.7$) and the index of the additional PL that is not included in the fit (i.e., $\sim-1.5$). The same conclusions can be drawn for \grbb to explain the low values of $\alpha$ obtained when fitting C+BB at late times (see Figure~\ref{fig10}b). Indeed, in the C+BB+PL$_\mathrm{5params}$ scenario, the contribution coming from the additional PL is the most intense at late times (see panels (d) of Figure~\ref{fig08}).

The reconstructed photon light curves resulting from C+BB+PL$_\mathrm{5params}$ are presented in panels (d) of Figure~\ref{fig07} and ~\ref{fig08} for GRBs 080916C and 090926A, respectively. For both bursts, these light curves are in very good agreement with the observed count light curves, in particular the LLE light curve between 20 and 100 MeV. Indeed, although we do not use the LAT-LLE data in our fits, the extrapolations of our model to higher energies reproduce very well the observations with a smooth evolution of the additional PL flux with time. The resemblance of the observed 20--100 MeV light curves with the reconstructed one is particularly striking for \grbnosa: not only the intensity peak is well recovered, but the long tails before and after it are also very adequately reproduced. Figure~\ref{fig18} shows the evolution of the energy content of each component with time (panels (a1) and (b1)) as well as their relative contributions to the total energy flux (panels (a2) and (b2)) between 8 keV and 100 MeV in the observer frame. The Band function contribution to the total energy flux dominates at early time while the additional PL contribution is dominant at later time; however, the contribution of the additional PL is also important at very early times during the first instant of the bursts. Because of the narrow shape of the thermal-like component, its contribution to the total energy flux is limited although it may reach few tens of percent in some time intervals in the best case scenario.

A larger sample of bursts will help to better estimate the typical values of $\alpha$ and of the additional PL index. While for the two GRBs studied in this article, as well as for many others, a value of $\alpha$ around $-0.7$ seems to be a good estimate, for others the value of $\alpha$ seems to peak around $-1.2$ \citep[e.g.,][]{Guiriec:2013a}; both values seem to be typical ones for $\alpha$.

The F$_\mathrm{i}^\mathrm{NT}$--E$_\mathrm{peak,i}^\mathrm{NT}$ relation can be used to better constrain the C+BB+PL$_\mathrm{5params}$ model and to even further decrease its complexity. Indeed, in the C+BB+PL$_\mathrm{5params}$ model the values of $\alpha$, ``$\beta$'' (i.e., an exponential cutoff for C), and the index of the additional PL are frozen, therefore, the F$_\mathrm{i}^\mathrm{NT}$--E$_\mathrm{peak,i}^\mathrm{NT}$ relation can be reduced to a correlation between the amplitude of the Band function (or C), A, and its peak energy E$_{peak}$. This relation between A and E$_\mathrm{peak,i}$ can then be determined with an initial time-resolved analysis of the data for each burst with the C+BB+PL$_\mathrm{5params}$ model, and subsequently be used as input to the C+BB+PL$_\mathrm{5params}$ model to tighten even more the fit results. 

Moreover, if the L$_\mathrm{i}^\mathrm{NT}$--E$^\mathrm{rest,NT}_\mathrm{peak,i}$ relation is confirmed and shown to hold for a large sample of bursts, then the relation between A and E$_{peak}$ may be firmly determined beforehand for GRBs with known redshift and would allow us to constrain the C+BB+PL$_\mathrm{5params}$ model from the beginning; therefore, the C+BB+PL$_\mathrm{5params}$ model would then have only 4 independent parameters (i.e., C+BB+PL$_\mathrm{4params}$) like the Band function alone. For GRBs with unknown redshift, z would be an additional parameter (i.e., 5 independent parameters and a constraint between A and E$_\mathrm{peak}$) and it may be possible to evaluate the redshift of the burst from the fits of C+BB+PL$_\mathrm{4params,z}$ to each time-interval. Finally, it could also be possible to fit directly the cosmological parameters to the time-resolved spectra in the case of GRBs with known redshifts (i.e., C+BB+PL$_\mathrm{4params,\Omega_M,\Omega_\Lambda}$). Although preliminary results look promising, a comprehensive study is required to validate or not those relations prior to any possible use; those results will be presented in a follow up article.

\section{Interpretation}
\label{section:interpretation}
There are several theoretical possibilities for the physical origin of each of the three components identified in the spectra of GRB~080916C and GRB~090926A. However the general framework of the discussion is well established: 
(i) GRBs are produced by ultra-relativistic outflows and (ii) the prompt emission has an internal origin.
More precisely, (i) is required to avoid a strong $\gamma$--$\gamma$ annihilation (e.g. \citealt{Baring:1997, Lithwick:2001}). 
A detailed calculation leads to constraints for the Lorentz factor and the radius of the emission site to allow the escape of photons at a given energy $E$ \citep{Granot:2008, Hascoet:2012}. 
For the Band (or C) and BB components, this calculation does not provide a strong constraint on the emission site, that can be anywhere from the photosphere to the deceleration radius. On the other hand, as the PL component is extending at least up to 100 MeV, it must be produced above the photosphere.
Condition (ii) is obtained from the study of the variability of the prompt emission, which cannot be reproduced by the external shock during the phase when the outflow is decelerated by the ambient medium (e.g. \citealt{Sari:1997}). 
This implies that the Band (or C) and BB components must be of internal origin. For the PL, our analysis clearly shows that variability below the one-second timescale is also present (see panels (c) and (d) of Figure~\ref{fig07} \&~\ref{fig08}, red curves), indicating that the origin is at least partially internal. 
On the other hand, the evolution of this component, which becomes brighter at the end of the burst suggests that a link with the deceleration of the outflow should also be considered.

\subsection{Origin of the Band (or C) and BB components}

In the first class of models---dissipative photospheres---
the bulk of the keV--MeV prompt emission is released
at the photosphere of the relativistic outflow \citep{Goodman:1986, Paczynski:1986}, i.e. at a low radius compared to other models. 
The spectrum can have a complex shape, far from a thermal spectrum, due to a sub-photospheric dissipation process that accelerates electrons, allowing Comptonization at high-energies, and, if the outflow is magnetized, synchrotron radiation at low-energies (e.g. \citealt{Rees:2005, Peer:2006, Beloborodov:2010, Vurm:2011}).  In this scenario, the Band (or C) component and the BB component should be interpreted as a single component with a complex shape.
The physics governing the evolution of this shape is related not only to the well-understood physics of the photosphere, but also to the nature of the dissipative sub-photospheric process, which is not well constrained. It is then too early to conclude if this scenario can reproduce the observed spectral evolution (see the discussions in \citealt{Vurm:2013, Asano:2013, Gill:2014}).

In the second class of models the photosphere is not strongly affected by sub-photospheric dissipation. The spectrum of the photosphere then remains thermal, with a Planck spectrum slightly modified at low-energy due to the peculiar geometry of the photosphere in a relativistic outflow
\citep{Beloborodov:2011,Lundman:2013,Deng:2014}. This provides an obvious interpretation of the BB component in our analysis. However, as pointed out in previous cases (e.g. \citealt{Guiriec:2011a, Guiriec:2013a}), the photospheric emission in our two bursts appears to be less hot and bright than the prediction of the standard fireball model where the acceleration of the ultra-relativistic outflow is purely thermal (\citealt{Meszaros:2000, Daigne:2002,Nakar:2005}). This implies that the outflow is strongly magnetized close to the central source \citep{Zhang:2009, Hascoet:2013}. This important conclusion of previous analysis of bright \textit{Fermi} bursts remains unchanged with the three-components spectral analysis of GRBs~080916C and 090926A. On the other hand, our analysis tends to show that the BB component is always detected in bright bursts where an accurate spectral analysis is possible, whereas GRB~080916C was often presented in previous studies as a good example of burst without photospheric emission
(e.g \citealt{Abdo:2009:GRB080916C,Zhang:2009,Zhang:2011c}). 
If this is also true for the whole GRB population where a multi-component spectral analysis is not possible, this implies that all GRBs are associated to magnetized outflows, with initially only 1 to 10\% of the energy in thermal form \citep{Hascoet:2013}, allowing for a  residual photospheric emission at the observed level, as in the two cases presented here.

In this scenario, the non-thermal emission corresponding to the Band (or C) component is produced in the optically thin regime, above the photosphere. It is associated to the emission of relativistic electrons, accelerated either in internal shocks (\citealt{Rees:1994, Kobayashi:1997, Daigne:1998}) or due to magnetic reconnection (e.g. \citealt{Lyutikov:2003, Zhang:2011b}). The first possibility would occur if the conversion of the magnetic energy into kinetic energy is efficient during the acceleration of the outflow, leading to a low magnetization $\sigma\lesssim 0.1$ at large radius. The second possibility corresponds to the opposite case, where the acceleration is inefficient and the magnetization is still high at large radius, $\sigma\gtrsim 1$.
In both cases, the dominant radiative process should be synchrotron, as the synchrotron self-Compton (SSC) predicts an additional strong component either in the optical-UV-soft X-ray range, or above 100 MeV, in contradiction with observations
\citep{Piran:2009, Bosnjak:2009}. 

Our analysis confirms that when considering only the Band (or C) and BB components, this scenario is in good agreement with observations. In the case of GRB 080916C, the low-energy photon index $\alpha$ of the non-thermal spectrum remains below -1 for almost the whole evolution, which is compatible with the expected fast-cooling regime, when taking into account the effect on the electron cooling of the inverse Compton scatterings in the Klein-Nishina regime \citep{Derishev:2001,Bosnjak:2009, Wang:2009, Daigne:2011}. 
In GRB 090926A, $\alpha$ is larger, usually in the -0.7 -- -1 range. This is of course fully compatible with the synchrotron radiation in slow cooling regime. However, this interpretation is not satisfactory as the radiative efficiency  is very low in this regime, leading to an energy crisis for a bright burst like GRB 090926A. Large photon indices close to -0.7 can however be obtained if electrons are only marginally fast cooling (the cooling break being close to the peak energy, the intermediate slope -1.5 is masked and the slope $-2/3$ is observed (e.g. \citealt{Daigne:2011,Beniamini:2013}) or if the magnetic field in the shocked region is decaying on a timescale smaller than the dynamical timescale but comparable to the electron cooling timescale \citep{Derishev:2007,Lemoine:2013, Uhm:2014, Zhao:2014}. 

Synchrotron radiation in both the internal shock and the reconnection scenario can reproduce the observed lightcurves. In the case of internal shocks, detailed simulations show in addition that the spectral evolution can also be reproduced, including the correlation between the peak energy and the flux \citep{Bosnjak:2014}. 
Such spectral calculations are not available yet for reconnection models.
The generic calculation of synchrotron radiation by \citet{Uhm:2014} shows that the continuous injection of electrons by reconnection events as expected in the ICMART model could lead to the correct spectral slope (see also \citealt{Beniamini:2014}). It remains to test that the addition of the contributions from several simultaneous emission sites in the outflow would not broaden the spectrum too much.
Detailed calculations of the photospheric emission in a variable outflow predicts that evolution is also expected for the BB component. However, due to the weakness of this component, we probably detect only the brightest peaks, which does not allow us to evaluate the real variations of the temperature \citep{Hascoet:2013}. 

Overall, both the spectral parameters and the spectral evolution observed in GRBs~080916C and 090926A are compatible with the scenario where these two bursts are associated with an ultra-relativistic outflow, which is initially highly-magnetized, producing a weak thermal emission at the photosphere (BB component) and where dissipation in the optically thin regime (shocks or reconnection) leads to synchrotron radiation of accelerated electrons (Band component).

Unfortunately, the situation regarding the synchrotron origin of the Band (or C) component becomes more complex in the three-component analysis, as the addition of the PL leads to larger photon index $\alpha$, usually close to $-2/3$, which could be explained by a modified synchrotron radiation as discussed above.
 However, we point out that our analysis shows that a third component, dominant at low ($<10 $ keV) and high ($>$ 100 MeV), is required to better fit the observed GRB spectrum, but that a power-law is only the simplest possible assumption for the shape of this component. The real shape is difficult to constrain, especially because this third component is masked in the intermediate MeV range. From a theoretical point of view, it is difficult to understand which process could produce a single power-law extending over at least five decades in energy. It is more natural (see below) to expect that the spectral shape of the third component is in reality more complex. A small change of the shape at low-energy would allow us to reconcile the three-components analysis with the synchrotron radiation for the dominant component in the MeV range.
 
\subsection{Origin of the PL component}

The third, PL, component is clearly the most puzzling. It has been suggested that it could be entirely of external origin (e.g. \citealt{Abdo:2009:GRB090902B}.) 
However, our analysis shows that it is present in both GRBs since the very beginning, when the deceleration of the outflow is still negligible. In addition, it also shows clearly that time-variability is present, on the 1~s timescale and probably below. This suggests an internal origin, at least at early times. 
One possibility discussed in the litterature is a hadronic component  \citep[see e.g.][]{Abdo:2009:GRB090902B}. However its low efficiency leads to an energy crisis for bright bursts as GRBs~080916C and 090926A \citep{Asano:2009,Asano:2012}. This leaves only electrons as a reasonable  candidate for this emission. 
If the Band (or C) component is due to synchrotron radiation, then the radiating electrons must have initially very large Lorentz factors ($\gtrsim 10^4$) to produce MeV photons, which limits the inverse Compton scatterings due to Klein-Nishina corrections. However, the evolution of the Band (or C) component shows large variations of the spectral peak, allowing it to be less deep in the Klein-Nishina regime when the peak energy is lower, and even to enter the much more efficient Thomson regime. Therefore, inverse Compton emission is expected at high energy and is not expected to peak exactly at the same time as the synchrotron radiation, in agreement with the observed lightcurves of the Band (or C) and PL components in GRB 080916C and 090926A. Calculations made in the context of internal shocks predict a flat spectrum in agreement with the observed index of the PL close to -2 
\citep{Bosnjak:2009,Bosnjak:2014}.

The inverse Compton scenario does not explain why the PL is extending down to the keV range. There are however some mechanisms that could correlate a component in the keV range and one above 100 MeV, leading to a more complex shape than a simple power-law: (i) one possibility is associated to the $\gamma$--$\gamma$ annihilation at high energies. The signature of this process is a time-evolving attenuation leading to a broken power-law at high-energy followed by a strong, close to exponential, cutoff. Our study cannot confirm this signature in GRBs~080916C and 090926A but the fact that a C in replacement of the PL is not rejected by the analysis may indicate that $\gamma$--$\gamma$ annihilation is indeed present in the 100 MeV--10 GeV range. Then a cascade of pairs is produced, which radiate at lower energy by synchrotron radiation \citep[see e.g.][]{Asano:2011}. It should be tested if such an emission could emerge in the keV range ; (ii)
a second possibility is to assume that less relativistic electrons are also present in the flow, either because the acceleration process does not accelerate all of them, or because the dissipation process is weaker in some regions (residual internal shocks with small relative Lorentz factors for instance.) Then, the synchrotron radiation of these electrons peaks in the keV range, and inverse Compton scatterings occur in the Thomson regime, producing high-energy photons. This could explain the third component fitted by a PL, but with a more complex spectrum.
A detailed modeling, which is out the scope of this paper, will be needed to test these ideas. Note that in the inverse Compton scenario for the PL component, the possible cutoff in the PL found in some time bins could be an evidence of the intrinsic curvature of this component close to its peak energy, rather than a signature of the $\gamma$--$\gamma$ annihilation.

Finally, the fact that the PL component becomes dominant at the end of both bursts suggests that a new component, of external origin, is emerging at late times, when deceleration starts. The physical origin would then be the same as for the long-lasting emission revealed by \textit{Fermi}-LAT, which is observed both in GRBs~080916C and GRB 090926A. This origin is still debated, but synchrotron and inverse Compton radiation in the external shock are strong candidates \citep{Zou:2009, Kumar:2009, Ghisellini:2010, Kumar:2010, Lemoine:2013, Beloborodov:2014, Vurm:2014}.
Interestingly, an early temporal break is observed in 
the LAT long-lasting emission of GRB~090926A~\citep{Ackermann:2011:GRB090926A}, which reinforces the scenario where the high-energy emission is initially dominated by a component of internal origin, with a transition to an external origin once the deceleration starts.

\section{Conclusion}
\label{sec:partialConclusion}

We summarize our results as follows:
\begin{itemize}
\item GRB prompt emission spectra are more complex than the shape resulting from the fit to a Band function alone. An adequate option to capture the complex shape of GRB prompt emission is a combination of at least three separate spectral components whose intensity and relative contribution to the total energy flux evolve with time. (i) The three components are not always present in all GRBs; or (ii) in some GRBs certain components strongly overpower the others making their identification more difficult. Figure~\ref{fig06} shows an example of the three-component model (i.e., (1) a non-thermal Band function with a high-energy power law index $<$-3 and which is statistically equivalent to a cutoff power law, (2) a thermal-like component adequately approximated with a black body component, and (3) a non-thermal additional power law with or without cutoff) in GRB~090926A.
\item The new three-component model is composed of (i) a smoothly broken PL with a constant low-energy index of $\sim$-0.7 (or $\sim$-1.2 as for GRB 120323A---see \citet{Guiriec:2013a}) and a high-energy slope compatible with an exponential cutoff based on the quality of our current data (i.e., C), (ii) a thermal-like component adequately approximated with a BB component, and (iii) another PL component with a fixed index of $\sim$-1.5, whose cutoff energy is often challenging to identify. However, the cutoff would be most of the time below 100 MeV. When a cutoff is used, the index of the additional PL gets a bit closer to -1. It is important to note that this paper aims to establish the existence of at least three distinct spectral components, but the spectral shapes of the various components will be explored in greater detail in a following article.
\item In the C+BB+PL model, (i) the C component can be interpreted as synchrotron radiation in an optically thin region above the photosphere, either from internal shocks or magnetic field dissipation regions, (ii) the BB component can be interpreted as the photosphere emission of a magnetized relativistic outflow and (iii) the extra PL extending to high energies likely has an inverse Compton origin of some sort even though its extension to a much lower energy remains a mystery.
\item We succeed in reducing the number of parameters in the three-component model (i.e., C+BB+PL) from 7 down to 5 (i.e., C+BB+PL$_\mathrm{5params}$) still keeping the same quality for the fits; with 5 free parameters C+BB+PL$_\mathrm{5params}$ is statistically competitive with the 4 parameters of the Band function. Additional constraints to the model will be added if the F$_\mathrm{Band,i}$--E$_\mathrm{peak,i}$ and L$_\mathrm{Band,i}$--E$^\mathrm{rest}_\mathrm{peak,i}$ relations are confirmed on a large sample of GRBs; This will result in an even simpler model.
\item The C and BB components are intense at early times in a burst. The additional PL may kick off at very early times and even before the two other components and last much longer at low energies. The intensity peak of the additional PL seems to be often related to sharp and bright structures present in the light curves at all energies from 8 keV to tens of MeV and even GeV. The additional PL is the most intense component at late times at low energies (tens of keV). This additional PL may connect smoothly with the GeV LAT emission observed for thousands of seconds after the prompt phase and contemporaneously with the X-ray and optical afterglow emissions. Because the BB component is narrow, its contribution to the total energy flux between 8 keV and 100 MeV is very limited compared to the two other components (see figure~\ref{fig18}); however, the BB has an important relative contribution to the total emission at its spectral peak around 100--200 keV (see Figure~\ref{fig21}~\&~\ref{fig22} of Appendix~\ref{section:Time-resolved-analysis}).
\item From the data set we used, it is difficult to conclude if the additional PL is a single component. However, it is clear that there is a simultaneous evolution of the low- and high-energy fluxes which seems to propagate all across the spectrum.
\item There is a strong correlation between the energy flux of the non-thermal component---adequately fitted with either Band or C---and its E$_\mathrm{peak}$ within each burst. Taking into account the various components in the fit process reduces the scatter of this relation. When fitted with a PL, the F$_\mathrm{i}^\mathrm{NT}$-E$_\mathrm{peak,i}^\mathrm{NT}$ relations have similar slopes for all GRBs. This relation may be useful to understand the mechanism responsible for the non-thermal prompt emission.
\item In the central engine rest frame, the F$_\mathrm{i}^\mathrm{NT}$--E$_\mathrm{peak,i}^\mathrm{NT}$ relation points towards a universal L$_\mathrm{i}^\mathrm{NT}$--E$^\mathrm{rest,NT}_\mathrm{peak,i}$ relation. If confirmed, this relation could be used to estimate GRB redshifts as well as to constrain the cosmological parameters.
\end{itemize}

Beyond the purpose of the current article, this analysis using multiple spectral components may have a broader impact on other analyses:
\begin{itemize}
\item Since the cutoff in the additional PL seems to evolve with time---and most of the time bellow 100 MeV---and since it is sometimes hidden by other components, it may indicate that this cutoff is a property of the emission mechanism(s) producing the additional PL and not the result of $\gamma$--$\gamma$ opacity. If it is the case, it is not correct to estimate the bulk Lorentz factor of the jet based on the additional PL cutoff energy.
\item The presence of multiple components evolving independently with time and exhibiting highly variable relative fluxes may strongly bias the conclusion resulting from spectral lag analysis. Indeed, lags may be identified between high intensity structures belonging to different components.
\item Such multi-component spectral analysis may help in the context of analysis conducted to constrain Lorentz invariance. Indeed, such analysis considers that photons are emitted from the same region and at the same time. By identifying the various components we may improve the association between low- and high-energy photons, as it would be the case e.g., for \grbnosa. Based on the count light curves of Figure~\ref{fig07}a, we may be tempted to consider the association of the peak above 20 MeV around 6\,s with peaks at lower energy happening at earlier (or later) times, especially if we do not have the knowledge of the low-energy light curve below 20 keV. With the multiple component analysis, we see that the peak at high-energy is directly correlated to the one at low-energy. Therefore we can increase the accuracy of the Lorentz invariance measurement by reducing the uncertainties on the time dispersion.
\end{itemize}

\section{Acknowledgments}


The Fermi-LAT Collaboration acknowledges generous ongoing support from a number of agencies and institutes that have supported both the development and the operation of the LAT as well as scientific data analysis. These include the National Aeronautics and Space Administration and the Department of Energy in the United States; the Commissariat {\`a} l'Energie Atomique and the Centre National de la Recherche Scientifique/Institut National de Physique Nucl{\'e}aire et de Physique des Particules in France; the Agenzia Spaziale Italiana and the Istituto Nazionale di Fisica Nucleare in Italy; the Ministry of Education, Culture, Sports, Science and Technology (MEXT), High Energy Accelerator Research Organization (KEK), and Japan Aerospace Exploration Agency (JAXA) in Japan; and the K. A. Wallenberg Foundation, the Swedish Research Council, and the Swedish National Space Board in Sweden.

Additional support for science analysis during the operations phase is gratefully acknowledged from the Istituto Nazionale di Astrofisica in Italy and the Centre National d'{\'E}tudes Spatiales in France.

To complete this project, S.G. was supported by the NASA Postdoctoral Program (NPP) at the NASA/Goddard Space Flight Center, administered by Oak Ridge Associated Universities through a contract with NASA as well as by the NASA grants NNH11ZDA001N and NNH13ZDA001N awarded to S.G. during the cycles 5 and 7 of the NASA Fermi Guest Investigator Program.

\newpage

\begin{appendix}

\section{Cross-calibration}
\label{section:crosscalibration}

\subsection{Time-Integrated Spectral Analysis}

We find that a normalization correction is not required between the selected NaI and BGO detectors in both GRBs for all fitted models, i.e., all EAC factors are compatible with unity within 1--$\sigma$ uncertainties corresponding to $\sim$5\% and $\sim$10\% for GRBs 080916C and 090926A, respectively (see Tables~\ref{tab01} \&~\ref{tab02}). When we fit the GBM and LAT data of GRB~080916C simultaneously, the three LAT data sets (i.e., LLE, LAT-Front and LAT-Back) are consistent with the GBM ones for all the tested models except B+PL (see Table~\ref{tab01}). For GRB~090926C, the GBM and LAT data are only compatible for the  B+C, B+BB+B2 or C+BB+B2 fits, but large corrections are required with all other models (see Table~\ref{tab02}). In all cases, however, where EACs are implemented, their uncertainties are large, making it difficult to precisely quantify the intercalibration of the two instruments. In addition we note that the inclusion of EACs does not affect significantly the values of the spectral parameters of both GRBs in any of the models used (see Tables~\ref{tab01} \&~\ref{tab02}). We decided, therefore, to not consider any EAC corrections in the time-integrated spectral analysis. 

\subsection{Coarse-Time Spectral Analysis}

\hspace{-0.3 cm}\grbnosa:
\vspace{0.3 cm}

We find that all GBM detectors are consistent with each other within 5 to 10\% uncertainties for all tested models and in all intervals (see Table~\ref{tab03}). Simultaneous fits of GBM and LLE data are consistent with fits of GBM data alone even in the second time interval, where the high-energy emission is most intense (see Table~\ref{tab03}). However, since the intensities of the BGO and LLE signals are too weak to constrain with enough accuracy the EAC factors between GBM and LLE data, a detailed study of EAC factors here is meaningless (we also note that we found a small effect of EAC factors on the Cstat values).

We conclude that all GBM detectors and LLE data have a consistent relative calibration and we note that the spectral parameters are similar with or without EACs. Therefore, in the following we will only describe spectra resulting from fits that do not include any EAC.

\vspace{0.3 cm}
\hspace{-0.3 cm}\grbnosb:
\vspace{0.3 cm}

In the case of \grbnosb, we used both BGO detectors and determined that they are consistent within $\sim$10\% to $\sim$20\% uncertainties (see Table~\ref{tab04}). Beyond T$_\mathrm{0}$+11s, however, the signal in the BGO detectors becomes weak and it is difficult to obtain good constraints on their EAC factors. All the NaI detectors are also consistent with each other within 10--20\% uncertainties (see Table~\ref{tab04}). We identified discrepancies of the order of a few percent between the NaIs and the BGOs supported by the change in Cstat values when EAC factors were included in the fits, although we could not quantify these precisely (see Table~\ref{tab04}). Despite the effect on the Cstat values, EAC factors do not significantly impact the spectral parameters resulting from the fits using the various models nor the Cstat variations (i.e., $\Delta$Cstat) from model to model. Therefore, we conclude that, similarly to \grbnosa, we do not need to include EAC factors between the various GBM detectors.

Significant EAC corrections are required between the GBM and the LLE data for models that do not include any additional PL to the Band function (see Table~\ref{tab04}). This is easily explained by the fact that the high-energy emission observed with the LAT is clearly incompatible with an extrapolation of the high-energy PL of the Band function above several tens of MeV. Therefore, to account for these discrepancies, we can either include an EAC factor or an additional component at high-energy; the latter option looks more satisfactory because such strong discrepancies between GBM and LAT are not expected. In addition, since the additional PL to the Band function can be identified from GBM only data, it is quite natural to consider that the discrepancies observed between GBM and LLE data---when no additional PL is used---are not only due to calibration issues but also depend significantly on the accuracy/goodness of the model fitted to the data. For the models that include an additional PL, GBM and LLE data look consistent with EAC factors close to unity; however, the EAC uncertainties are very large making any correction unreliable. We conclude from this analysis that the GBM and LLE data are in agreement but that correction factors between the two data sets cannot be accurately determined; however, the results are not significantly affected, so in the following, we only present the spectra obtained without EAC between the GBM and the LLE data.

\section{Model Comparison Procedure}
\label{section:modelComparison}

In every time intervals of the time-integrated, coarse-time and fine-time spectral analysis, we performed multiple sets of Monte Carlo simulations (i) to compare all the tested models one to one, (ii) to estimate the goodness of fit of every tested model, and (iii) to estimate the trustworthiness of the measured spectral parameters for all the tested models. To generate the synthetic spectra of our simulation sets, we followed exactly the same procedure as already described in \citet{Guiriec:2011a} and in the Appendix of~\citet{Guiriec:2013a}.

In the article, we will use the statistical terminology and call ``best model'' and ``best fit'' the model and the fit, respectively, that have the lowest Cstat value.

To assess which model is a ``better description of the real data'' set when comparing models M$_\mathrm{1}$ and M$_\mathrm{2}$, we performed ten statistical tests addressing the ten following questions (shortened for simplicity). In all tests we assume that both signal and background fluctuations are following Poisson statistics. 

\begin{itemize}
\item Test 1: estimate the odds to obtain a $\Delta$Cstat value between M$_\mathrm{1}$ and M$_\mathrm{2}$ higher than the observed one if the true underlying model is M$_\mathrm{1}$---with real fit parameters" $=>$ This is the typical ``likelihood-ratio test'' or ``log-likelihood statistic test.''

\item Test 2: estimate the odds to obtain the observed parameters of M$_\mathrm{2}$ when fitting M$_\mathrm{2}$ to the data if the true underlying model is M$_\mathrm{1}$---with real fit parameters

\item Test 3: estimate the odds to obtain the observed parameters of M$_\mathrm{1}$ when fitting M$_\mathrm{1}$ to the data if the true underlying model is M$_\mathrm{1}$---with real fit parameters $=>$ This also serves to estimate the trustworthiness of the measured parameters.

\item Test 4: estimate the odds to obtain the specific Cstat value when fitting M$_\mathrm{1}$ to the data if the true underlying model is M$_\mathrm{1}$---with real fit parameters $=>$ This is the typical goodness of fit test.

\item Test 5: estimate the odds to obtain the specific Cstat value when fitting M$_\mathrm{2}$ to the data if the true underlying model is M$_\mathrm{1}$---with real fit parameters

\item Test 6: estimate the odds to obtain a $\Delta$Cstat value between M$_\mathrm{1}$ and M$_\mathrm{2}$ higher than the observed one if the true underlying model is M$_\mathrm{2}$---with real fit $=>$ This is the typical ``likelihood-ratio test'' or ``log-likelihood statistic test.''

\item Test 7: estimate the odds to obtain the observed parameters of M$_\mathrm{2}$ when fitting M$_\mathrm{2}$ to the data if the true underlying model is M$_\mathrm{2}$---with real fit parameters  $=>$ This also estimates the trustworthiness of the measured parameters.

\item Test 8: estimate the odds to obtain the observed parameters of M$_\mathrm{1}$ when fitting M$_\mathrm{1}$ to the data if the true underlying model is M$_\mathrm{2}$---with real fit parameters

\item Test 9: estimate the odds to obtain the specific Cstat value when fitting M$_\mathrm{1}$ to the data if the true underlying model is M$_\mathrm{2}$---with real fit parameters

\item Test 10: estimate the odds to obtain the specific Cstat value when fitting M$_\mathrm{2}$ to the data if the true underlying model is M$_\mathrm{2}$---with real fit parameters $=>$ This is the typical goodness of fit test.
\end{itemize}
 
We consider that M$_\mathrm{2}$ is a better description of the data than M$_\mathrm{1}$ if:

\begin{enumerate}
\item Test 1 is conclusive (i.e., p-value$<$$\sim$10$^\mathrm{-5}$---see footnote\footnote{For Tests 1, 4, 5, 6, 9 and 10, p-value is the probability to obtain a value for a specific quantity beyond the actually measured value of the same quantity from the real data (i.e., one-tail probability.) For Tests 2, 3, 7 and 8, p-value is the probability to obtain an absolute difference between the value of a parameter resulting from the fit and the most probable value of the distribution for the same parameter beyond the 1--$\sigma$ confidence region around the most probable value of the distribution (i.e., two-tail probability.)} for the definition of p-value) and if Test 6 is inconclusive (i.e., p-value$>>$10$^\mathrm{-5}$.)

\item Tests 3, 7, 8, 9 and 10 are inconclusive (i.e., p-value$>>$10$^\mathrm{-5}$) and Tests 4 and 5 are conclusive (i.e., p-value$<<<$.) In this case, Test 2 is also usually conclusive.
\end{enumerate}

In addition, if M$_\mathrm{2}$ is a ``better description of the data'' than M$_\mathrm{1}$ because of criterion 1, we will also say that M$_\mathrm{2}$ is ``statistically significantly better'' than M$_\mathrm{1}$.

\bigskip

We note that as of today, the GBM instrument team does not anticipate any evidence for global strong instrumental systematic effects that will not be accounted in the instrument response function yet; however, we cannot discard the fact that possible calibration problems discovered at later time may impact the results of the current analysis as well as all the analysis performed with the traditional Band function alone.

\bigskip
Hereafter we introduce two typical cases:
\begin{center}
1. Comparison of the B+BB and C+BB+PL fits to the GBM data alone in the time interval lasting from T$_\mathrm{0}$+4.3 s to T$_\mathrm{0}$+7.5 s of the coarse-time analysis of GRB 080916C:
\end{center}

In the following, we will consider that B+BB and C+BB+PL correspond, respectively, to the M$_\mathrm{1}$ and M$_\mathrm{2}$ models from the description of the various tests presented earlier in this Appendix. B+BB and C+BB+PL have 6 and 7 free parameters, respectively, but they are not nested models.

The B+BB and C+BB+PL fits to the real data result in Cstat values of 508 and 495, respectively. In this time interval, C+BB is clearly a very bad fit to the data (i.e., extremely high Cstat value) and it is, therefore, not considered here. Indeed, when fitting B+BB to this time interval $\beta$ is $>$-2.28 and clearly incompatible with the exponential cutoff of a C component.

We first present the results of the typical log-likelihood statistic test using B+BB with the parameter values resulting from the B+BB fit to the real data as the null-hypothesis (i.e., Test 1.) Despite the fact that C+BB+PL has one more free parameter than B+BB, the $\Delta$Cstat$_\mathrm{M_1--M_2}$ distribution (i.e., Cstat$_\mathrm{M_1}$-Cstat$_\mathrm{M_2}$) resulting from the B+BB and C+BB+PL fits to the 10$^\mathrm{-5}$ synthetic spectra has only negative values; this is possible because those two models are not nested. Since the two compared models are not nested, the $\Delta$Cstat$_\mathrm{M_1--M_2}$ distribution does not respect the Wilk's theorem (i.e., the $\Delta$Cstat$_\mathrm{M_1--M_2}$ distribution does not converge to a $\chi^\mathrm{2}$ distribution with 7-6$=$1 degree of freedom.) Among the 10$^\mathrm{5}$ synthetic spectra, we did not encounter a single case for which $\Delta$Cstat$_\mathrm{M_1--M_2}$ is larger than or equal to the $\Delta$Cstat$_\mathrm{M_1--M_2}$ value resulting from  the B+BB and C+BB+PL fits to the real data (i.e., the maximum value of the $\Delta$Cstat$_\mathrm{M_1--M_2}$ distribution resulting from the simulation is $<<$ 508-495=13.) Moreover, the $\Delta$Cstat$_\mathrm{M_1--M_2}$ distribution has a very steep slope so it is extremely unlikely to get a $\Delta$Cstat$_\mathrm{M_1--M_2}$ value even close to the observed one (i.e., 13). We can, therefore, confidently state that Test 1 is conclusive.

We now present the  results of the log-likelihood statistic test but using C+BB+PL with the parameter values resulting from the C+BB+PL fit to the real data as the null-hypothesis (i.e., Test 6.) The resulting $\Delta$Cstat$_\mathrm{M_1--M_2}$ distribution has only positive values and the observed $\Delta$Cstat$_\mathrm{M_1--M_2}$ (i.e., 13) is well within the 1--$\sigma$ of the distribution and very close to its most probable value. We can, therefore, confidently state that Test 6 is inconclusive.

Based on our previous descriptions of the test results, we can confidently conclude in this case that C+BB+PL is a ``better description of the data'' than B+BB and that in addition C+BB+PL is ``statistically significantly better'' than B+BB.

\begin{center}
2. Comparison of the C+BB and C+BB+PL fits to the GBM data alone in the fine-time spectra of GRB 080916C:
\end{center}

Here we describe the overall results, but the significance depends on the time interval. However, we always observed the same trend, which increases the significance of the overall results compared to the significance obtained for each time-interval individually.

C+BB and C+BB+PL are nested-like models and C+BB+PL has two more free parameters than C+BB; in the following, we will consider that C+BB and C+BB+PL correspond, respectively, to the M$_\mathrm{1}$ and M$_\mathrm{2}$ models from the description of the various tests presented earlier in this Appendix.

In the fine time intervals, the Cstat values resulting from the C+BB and C+BB+PL fits to the real data are usually very similar and Test 1 is typically inconclusive. Similarly, if the true underlying model is C+BB+PL with the parameters resulting from the C+BB+PL fit to the real data then we would also expect the value of $\Delta$Cstat$_\mathrm{M_1--M_2}$ computed from the C+BB and C+BB+PL fits to the real data; therefore, Test 6 is also typically inconclusive.

We can often conclude that (i) if the true underlying model was C+BB with the parameter values resulting from the C+BB fit to the real data, then the measured Cstat values obtained when fitting C+BB (i.e., Test 4---goodness of fit) and C+BB+PL (i.e., Test 5) to the real data are much higher than expected, and (ii) if the true underlying model was C+BB+PL with the parameter values resulting from the C+BB+PL fit to the real data, then the measured Cstat values obtained when fitting C+BB (i.e., Test 9) and C+BB+PL (i.e., Test 10---goodness of fit) to the real data are typical to what we would expect (i.e., within the 1--sigma confidence region of the Cstat distributions.) We can then conclude that Tests 4 and 5 are conclusive while Tests 9 and 10 are inconclusive.

Finally, we also usually observe that (i) if the true underlying model is C+BB, it is likely that we obtain the measured parameters resulting from the C+BB fit to the real data (i.e., Test 3 inconclusive) but it is unlikely that we obtain the measured parameters resulting from the C+BB+PL fit to the real data (i.e., Test 2 conclusive.) Conversely, if the true underlying model is C+BB+PL, it is likely that we obtain the measured parameters resulting from the C+BB+PL and C+BB fits to the real data (i.e., Tests 7 and 8 inconclusive, respectively.)

Therefore, Tests 1 and 6 are inconclusive, but Tests 3, 7, 8, 9 and 10 are inconclusive and Tests 4 and 5 are conclusive so we can conclude that C+BB+PL is overall a "better description of the data."

Let us consider the specific case of the time interval lasting from T$_\mathrm{0}$+8.5 s to T$_\mathrm{0}$+9.5 s in GRB 080916C (see Table \ref{tab05} of Appendix~\ref{section:Spectral-analysis-results}.) The C+BB and C+BB+PL fits to the real data result in the same Cstat values (i.e., 526 for 472 and 470 dof for C+BB and C+BB+PL, respectively.) The result of Tests 1 and 6 are clearly inconclusive because C+BB and C+BB+PL are nested models. It is, therefore, not possible to conclude from Tests 1 and 6 solely that C+BB+PL is a better description of the data than C+BB. However, Test 4 indicates that if the true underlying model was C+BB, it would be extremely unlikely (i.e., $<$10$^\mathrm{-5}$) to get a Cstat value beyond 526 when fitting C+BB to the data; indeed, the Cstat distribution peaks narrowly around 485 when fitting C+BB to the synthetic spectra. The same conclusion is drawn for Test 5. In addition, if C+BB was the true underlying model, it would be very likely that we measure the spectral parameter values resulting from the C+BB fit to the real data (i.e., Test 3 inconclusive), but it would be extremely unlikely that we measure the spectral parameter values resulting from the C+BB+PL fit to the real data (i.e., Test 2 conclusive.) This is particularly true for the value of $\alpha$ of C and the value of the amplitude of the additional PL of the C+BB+PL model; indeed, the values obtained when fitting C+BB+PL to the real data are $\sim$-0.60 and $\gg$0 for $\alpha$ and the amplitude respectively, the values of $\alpha$ and of the amplitude resulting from the C+BB+PL fits to the C+BB synthetic spectra peak $\sim$-1.4 and $\sim$0, respectively. None of the 10$^\mathrm{5}$ synthetic spectra based on the C+BB null-hypothesis result in parameters values compatible with the observed ones when fitting to C+BB+PL. Conversely, if the true underlying model was C+BB+PL, it would be very likely to obtain the observed Cstat values resulting from the C+BB and C+BB+PL fits to the real data (i.e., 526---Tests 6, 9 and 10 are inconclusive.) It would also be very likely to measured the spectral parameters resulting from the C+BB and C+BB+PL fits to the real data and this is especially true for $\alpha$ and the amplitude of the additional PL (i.e., Tests 7 and 8 are conclusive). We can then conclude that C+BB+PL is a "better description of the data than C+BB."

\section{Spectral Analysis Results}
\label{section:Spectral-analysis-results}

The values of the parameters resulting from the time-integrated spectral analysis of GRBs~080916C and 090926A as presented in Section~\ref{sec:tisa} are reported in Tables~\ref{tab01} \&~\ref{tab02}, respectively.

The values of the parameters resulting from the coarse-time analysis of GRBs~080916C and 090926A as presented in Section~\ref{sec:ctr} are reported in Tables~\ref{tab03} \&~\ref{tab04}, respectively.

The values of the parameters resulting from the fine-time analysis of GRBs~080916C and 090926A as presented in Section~\ref{sec:ftr} are reported in Tables~\ref{tab05} \&~\ref{tab06}, respectively.

\section{Time Interval Definition for Fine Time-Resolved Spectroscopy}
\label{section:timeBinSelection}

Although several techniques have been proposed in the past to automate the definition of time intervals, none seems to be really adequate to perform the study that we propose here: tracking the spectral evolution of multiple components on the shortest possible time scale---$>$0.1 s---without including light curve regions with strong spectral evolution in a single time-interval as well as in avoiding any possible artificial correlation between energy flux and spectral hardness.

The simplest automated technique consists of using time intervals of constant duration. If the time intervals are chosen to be too large, they may include emission periods with very different spectral properties, making the spectral fits unusable for good interpretation. Conversely, if the defined time intervals are too short, we may not have enough counts in some time intervals to reconstruct adequately the spectra.

Another technique consists of defining the time intervals based on the signal to noise (S/N) ratio. The signal is integrated until it reaches a certain threshold over the background. When this threshold is reached, another time interval is created and so forth. Similarly to the previous technique, this one may also create time intervals including emission periods with very different spectral properties. In addition, the S/N ratio is a quantity that should depend on the spectral shape. Indeed, the number of counts required to reconstruct a soft spectrum may be very different from the number of counts required to reconstruct a hard spectrum and the spectral distribution of the counts is also very important. This is not taken into account in the S/N ratio technique.

\citet{Scargle:1998} proposed to define time intervals as Bayesian blocks. A new time interval is created when the signal, which is integrated over a certain period of time and in a specific energy range, varies significantly based on a criteria provided by the user. This technique has two major issues. The first one concerns the energy band that we are considering to define significant variations of the signal. Indeed, if we choose an energy band that only includes low energy data, it will not be possible to detect signal variations due to a component that will only be intense at high energy. Conversely, if we select a high-energy band, we will not be sensitive to variations in the low-energy regime. Finally, if we select an energy domain that includes both low- and high-energy counts, any variation of the signal at high energy will be hidden by the low-energy variations since the number of counts is usually much larger at low energy than at high energy (i.e., non-thermal spectral shape). The second issue is intrinsic to the technique itself. With the Bayesian blocks, we define time intervals in which the intensity of the signal does not change significantly; however, the spectral distribution may change a lot. Since one of our studies consists of investigating the correlation between the instantaneous flux and the instantaneous hardness (i.e., E$_\mathrm{peak}$) within each burst, then this technique appears to be inappropriate. Indeed, it is not possible to know if two contiguous intervals with similar fluxes have different hardness since contiguous time intervals are devised to have significantly different fluxes.Therefore, the Bayesian block technique may produce artificial correlations.

In \citet{Guiriec:2013a}, we applied an original technique to define the time intervals. To devise intervals with the least spectral evolution, we applied a Bayesian block technique not to the count light curve, but to the hardness ratio evolution computed from light curves in two different energy ranges. This technique appears to be appropriate for the study of GRB 120323A since only two spectral components were identified and that they were not intense in the same energy domains. Therefore, we computed the time-history hardness ratio in using two energy bands in which each component was the most intense. Such a technique is difficult to apply for the current study since we are considering three spectral components whose energy domains in which they are intense overlap.

All the techniques presented here use criteria defined by the user such as energy bands and thresholds. In addition, they are applied to observed count light curves, which are tainted with energy dispersion effects. Indeed,  a photon has only a certain probability to be measured as a count at its true energy and the measured energy can be more or less spread around the true photon energy. While the spectral analysis corrects for this effect, a direct analysis of the count light curve may be significantly biased. Finally, the user usually defines the energy bands in the observer frame while GRBs are known to have large spread in distances. Defining the energy bands in the rest frame is not always possible since very few GRBs have redshift measurements. In addition, the energy in which each spectral component is the most intense does not only depend on the redshift but also on the micro-physics. For instance, the energy of the emitted radiation depends on the magnetic field profile and intensity as well as on the velocity profile of the solid layers within the jet.

Our procedure to define the fine-time intervals of our analysis is described in the following and it is done using Rmfit. We wanted to define time intervals as short as possible---0.1 s being the shortest allowed time duration---to avoid strong spectral evolution within a time interval and to have as many time intervals as possible to track the spectral evolution in the both the rising and decaying parts of the pulses. To do so, we combined the base 0.1 s time bins until

\begin{itemize}

\item we could adequately fit a Band function to the data (i.e., convergence and amplitude at least 2--$\sigma$ above 0),

\item the parameter values of the amplitude, $\alpha$ and E$_\mathrm{peak}$ resulting from the Band fit to the real data were within the 1--$\sigma$ confidence region of the distributions of the amplitude, $\alpha$ and E$_\mathrm{peak}$ values, respectively, resulting from Band fit to the Band synthetic spectra (i.e., see Appendix~\ref{section:modelComparison})

\item the biases on the parameter values for the amplitude, $\alpha$ and E$_\mathrm{peak}$ when fitting a Band function alone to the data was $<$1--$\sigma$ (i.e., the distribution of $\frac{|\mathrm{x_i^{best}-x_i^{max}}|}{\mathrm{|x_i^{un}|}}$---where $\mathrm{x_i^{best}}$ is the best value of the parameter resulting from the Band fit to the real data, $\mathrm{x_i^{max}}$ is the value of $\mathrm{x_i}$ at maximum of the distribution resulting from the Band fit to the Band synthetic spectra and $\mathrm{x_i^{un}}$ is the 1--$\sigma$ uncertainty on $\mathrm{x_i^{best}}$---is compatible with 0 within 1--$\sigma$),

\item the 1--$\sigma$ widths of the distributions of the amplitude, $\alpha$ and E$_\mathrm{peak}$ were smaller than the 1--$\sigma$ uncertainties of the amplitude, $\alpha$ and E$_\mathrm{peak}$ parameters, respectively, resulting from the Band function fit to the real data.

\item the best parameter values for amplitude, $\alpha$ and E$_\mathrm{peak}$ did not vary by more than 10\% when adding a 0.1 s time bin either toward the positive or negative values of time, or by sliding the time interval window by 0.1 s either toward the positive or negative values of time.

\end{itemize}

\section{Fine Time-Resolved Analysis}
\label{section:Time-resolved-analysis}

Figure~\ref{fig19} shows the results of the fine time-resolved analysis of GRB 080916C with a Band function alone and the C+BB+PL model as presented in Section~\ref{sec:ftr}.

Figure~\ref{fig20} shows the results of the fine time-resolved analysis of GRB 090926A with a Band function alone and the C+BB+PL model as presented in Section~\ref{sec:ftr}.

Figure~\ref{fig21} shows the results of the fine time-resolved analysis of GRB 080916C with the C+BB+PL$_{5params}$ model as presented in Section~\ref{sec:newmodel}.

Figure~\ref{fig22} shows the results of the fine time-resolved analysis of GRB 090926A with the C+BB+PL$_{5params}$ model as presented in Section~\ref{sec:newmodel}.

\end{appendix}

\newpage


\newpage

\setcounter{figure}{0}
\renewcommand{\thefigure}{A\arabic{figure}}

\begin{figure*}
\begin{center}

\includegraphics[totalheight=0.95\textheight, clip]{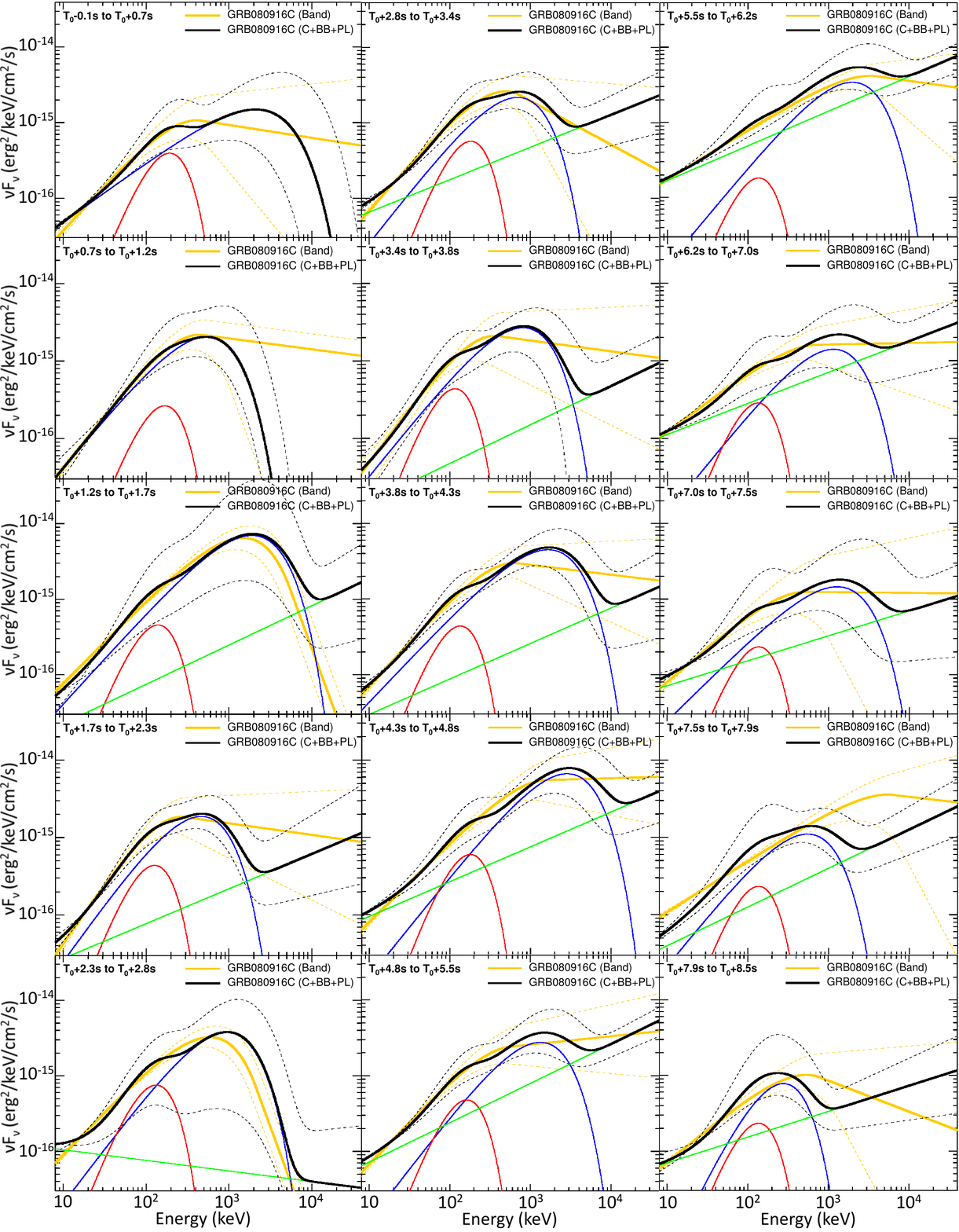}

\end{center}
\end{figure*}

\newpage

\begin{figure*}
\begin{center}

\includegraphics[totalheight=0.78\textheight, clip]{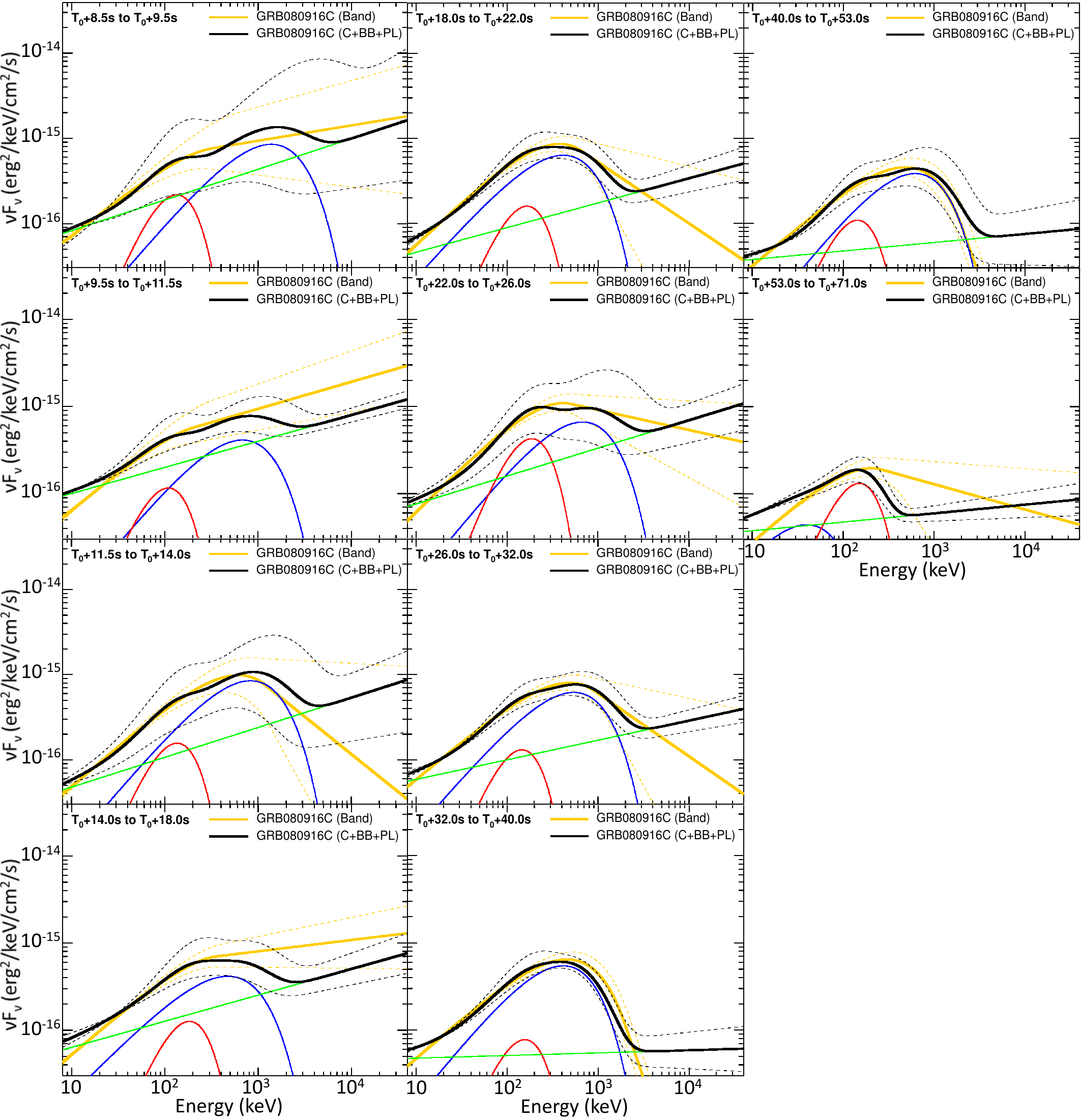}

\caption{\label{fig19}GRB 080916C : $\nu$F$_\nu$ spectra resulting from the fine-time analysis. The solid yellow and black lines correspond to the best Band-only and C+BB+PL fits, respectively. The dashed yellow and black lines correspond to the 1--$\sigma$ confidence regions of the Band-only and C+BB+PL fits, respectively. The solid blue, red and green lines correspond to the cutoff power law, to the BB component and to the additional power law resulting from the best fit with the C+BB+PL model (i.e., solid black line) to the data, respectively.}

\end{center}
\end{figure*}

\newpage

\begin{figure*}
\begin{center}

\includegraphics[totalheight=0.95\textheight, clip]{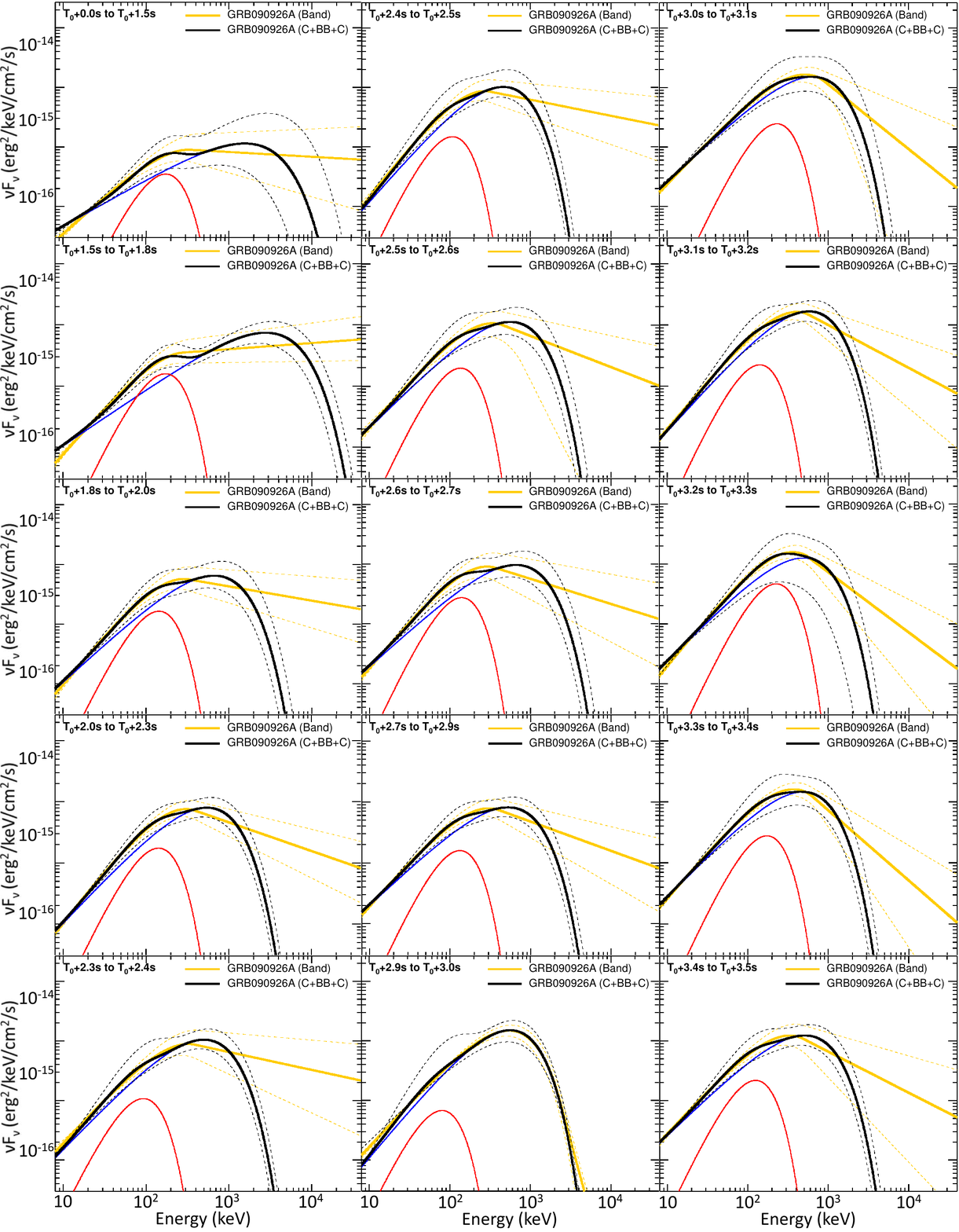}

\end{center}
\end{figure*}

\newpage

\begin{figure*}
\begin{center}

\includegraphics[totalheight=0.95\textheight, clip]{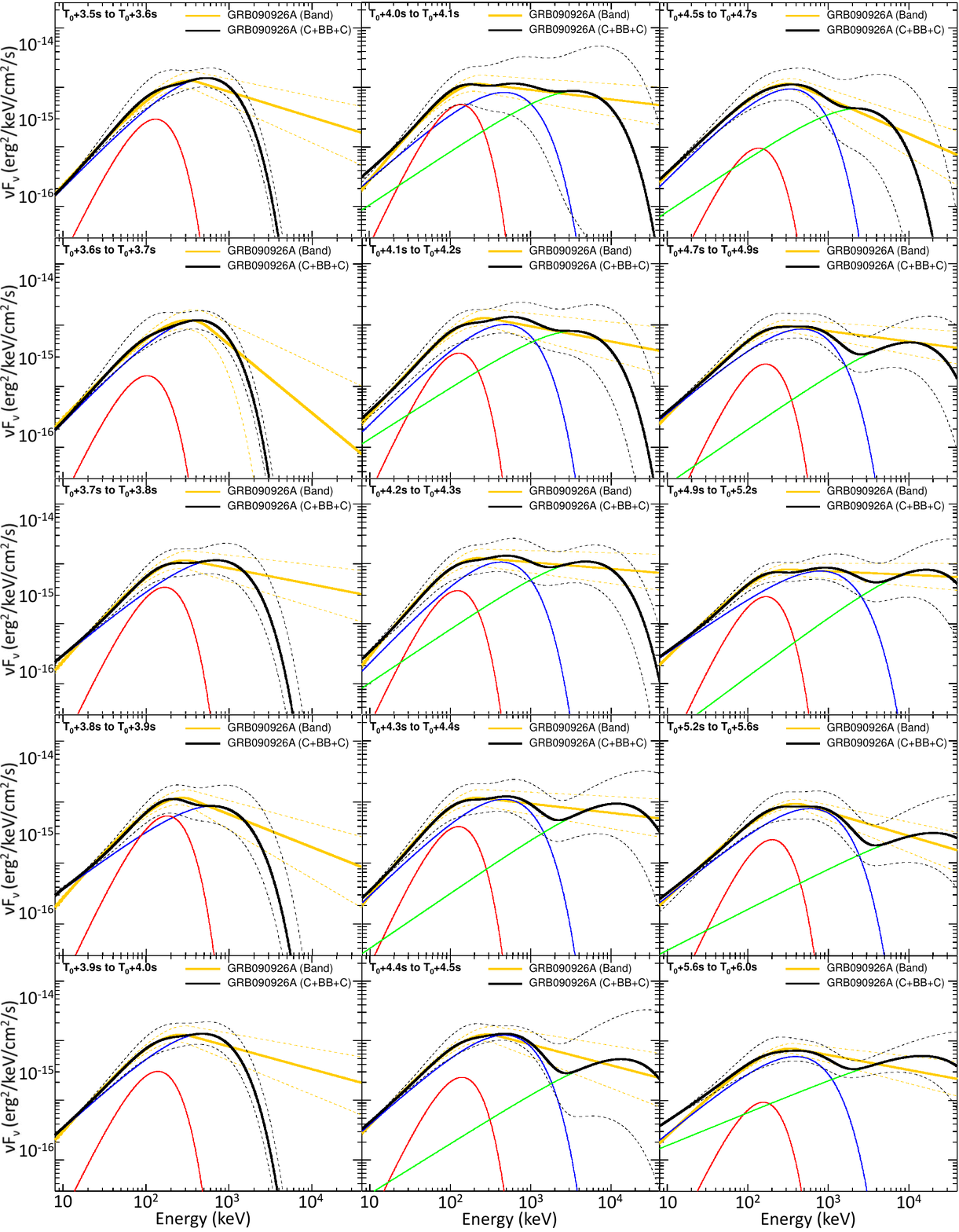}

\end{center}
\end{figure*}

\newpage

\begin{figure*}
\begin{center}

\includegraphics[totalheight=0.95\textheight, clip]{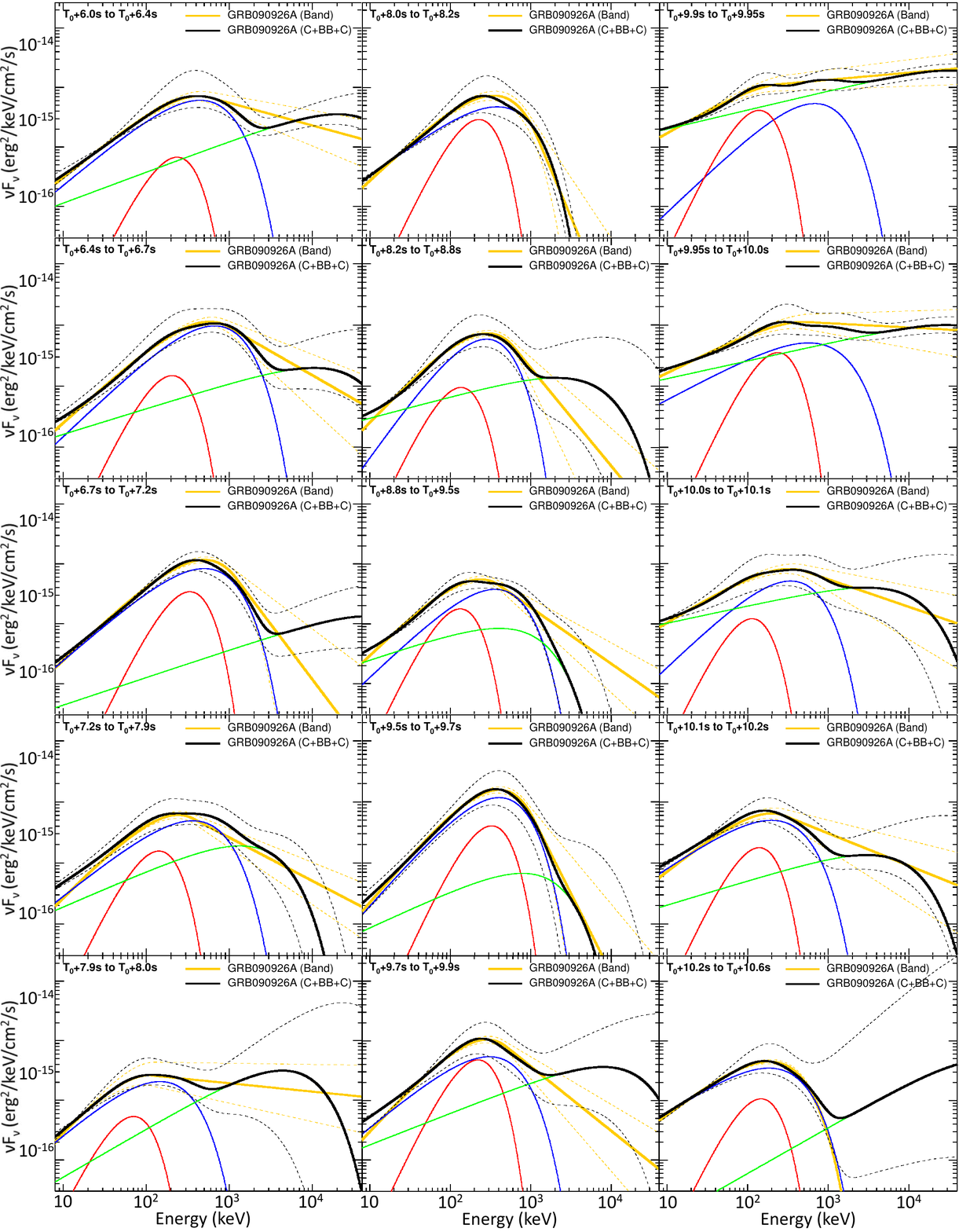}

\end{center}
\end{figure*}

\newpage

\begin{figure*}
\begin{center}

\includegraphics[totalheight=0.78\textheight, clip]{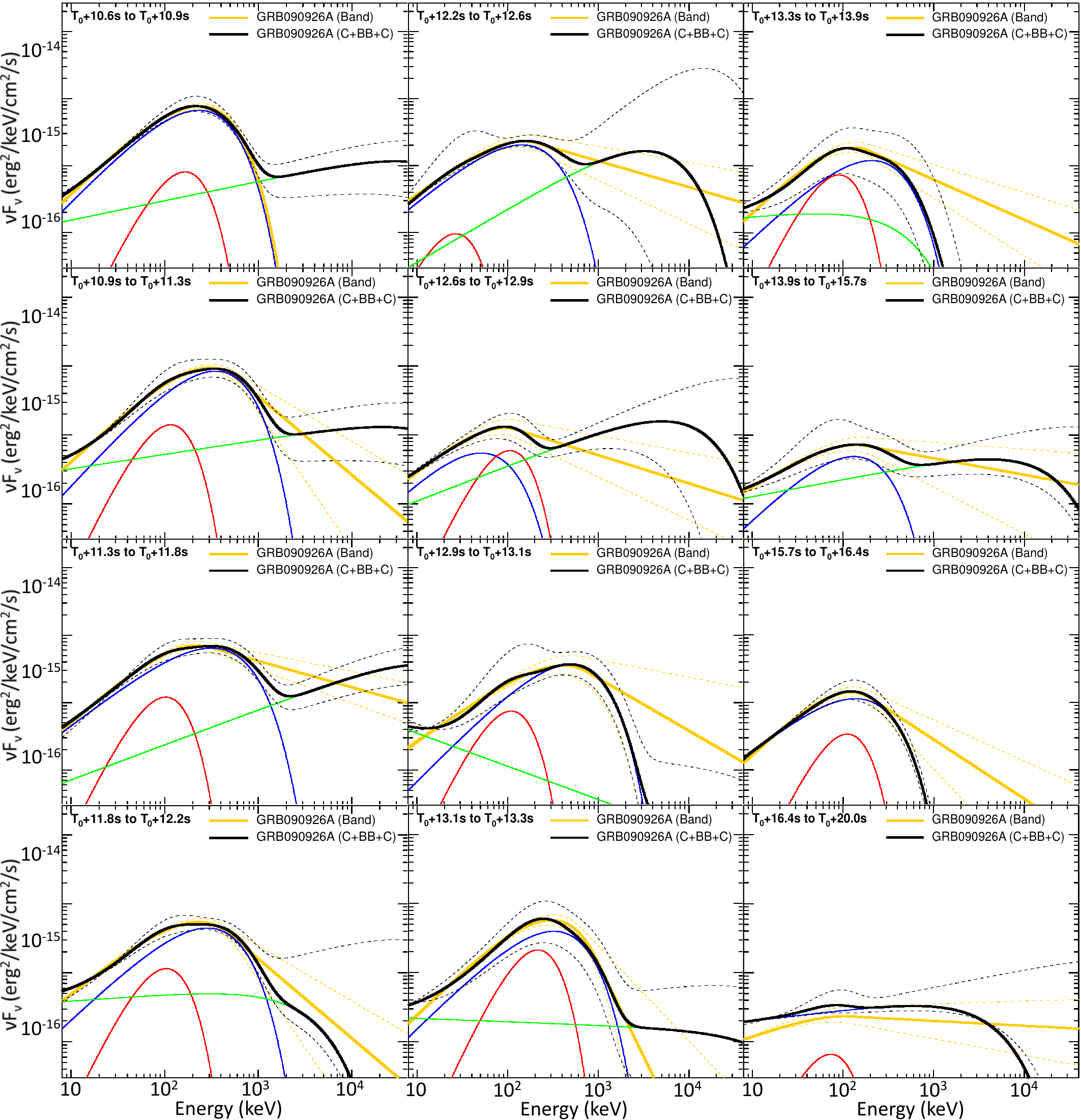}

\caption{\label{fig20}GRB 090926A : $\nu$F$_\nu$ spectra resulting from the fine-time analysis. The solid yellow and black lines correspond to the best Band-only and C+BB+C2 fits, respectively. The dashed yellow and black lines correspond to the 1--$\sigma$ confidence region of the Band-only and C+BB+C2 fits, respectively. The solid blue, red and green lines correspond to the cutoff power law, to the BB component and to the additional cutoff power law resulting from the best fit with the C+BB+C2 model (i.e., solid black line) to the data, respectively.}

\end{center}
\end{figure*}

\newpage

\begin{figure*}
\begin{center}

\includegraphics[totalheight=0.95\textheight, clip]{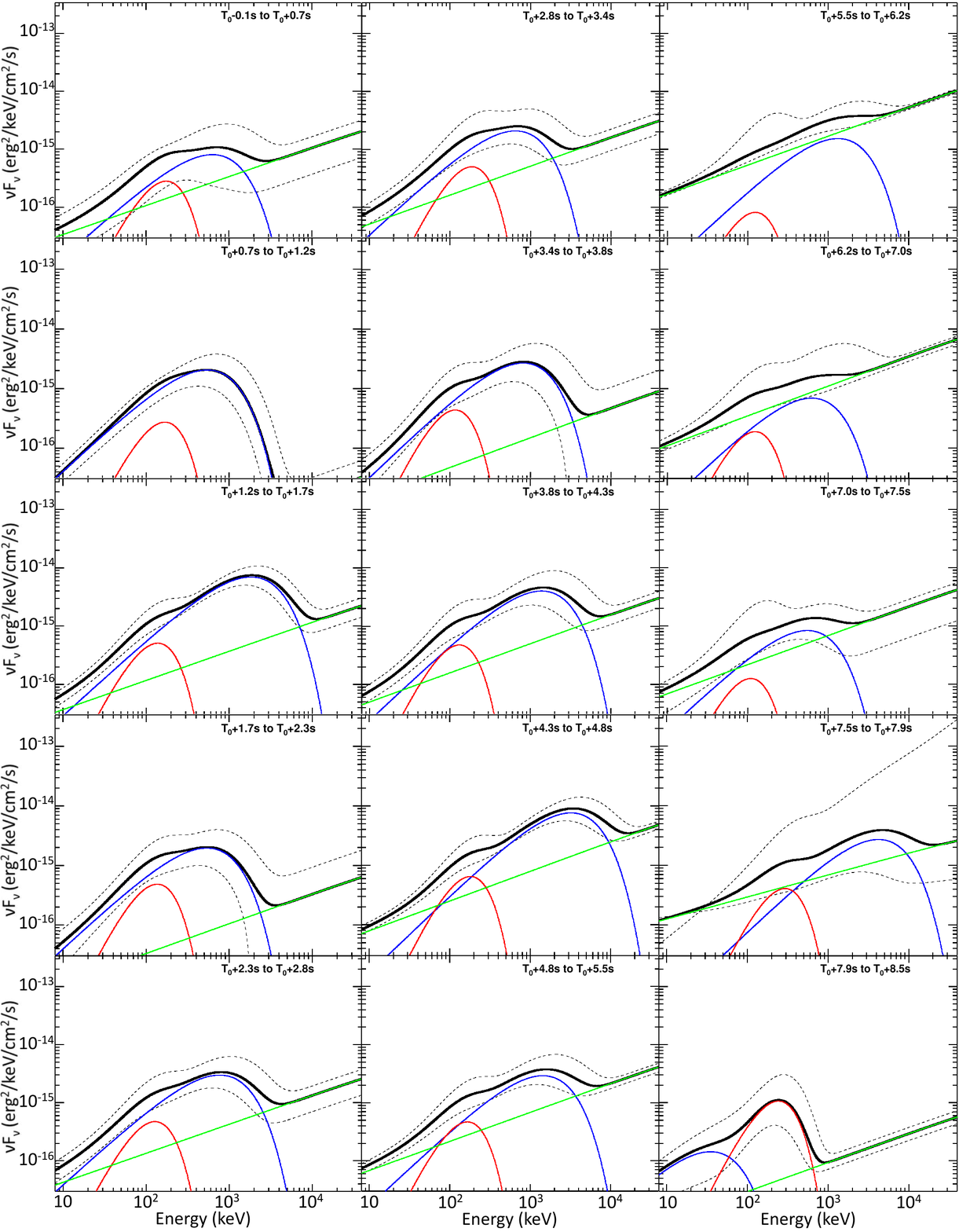}

\end{center}
\end{figure*}

\newpage

\begin{figure*}
\begin{center}

\includegraphics[totalheight=0.78\textheight, clip]{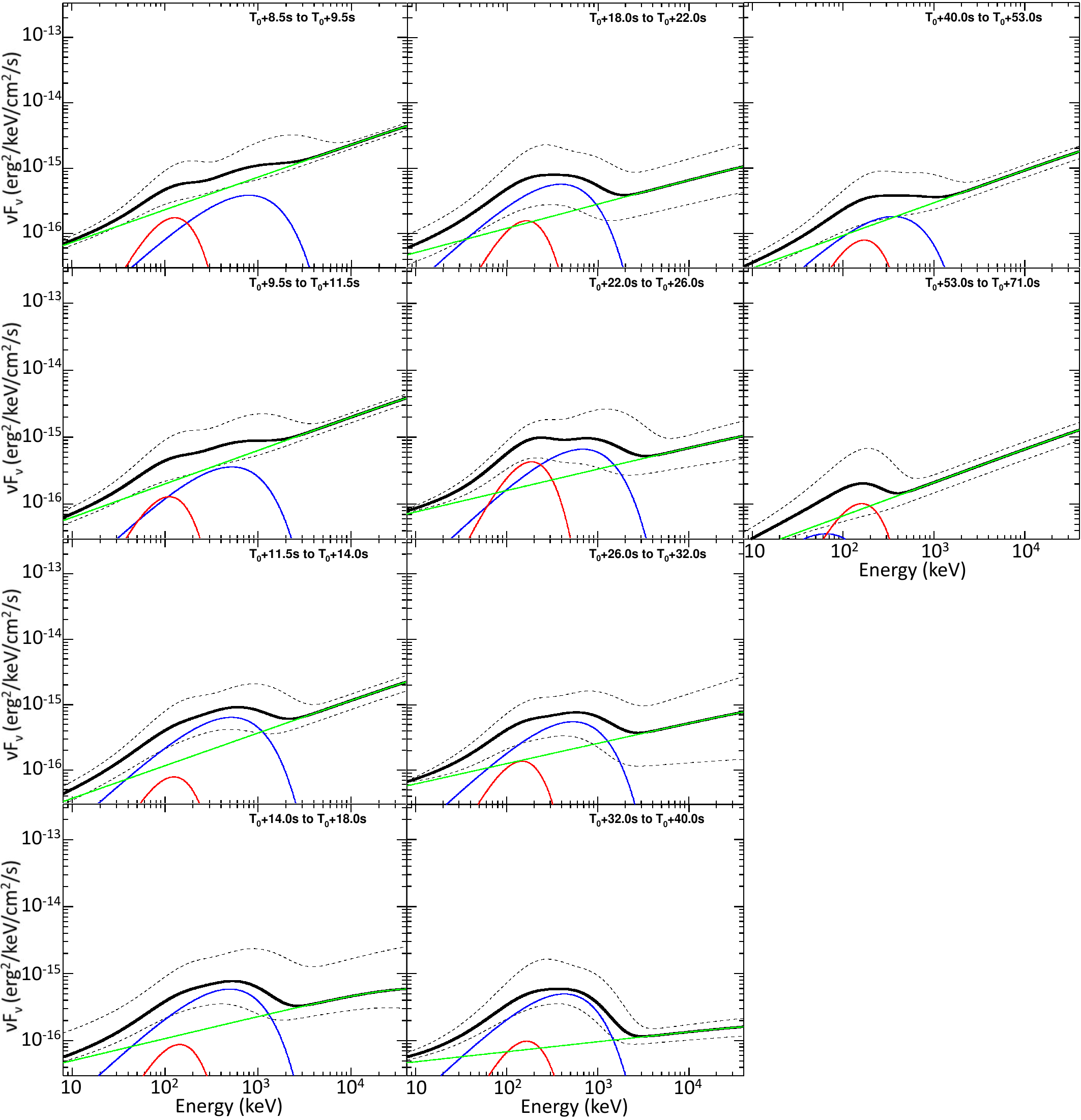}

\caption{\label{fig21}GRB 080916C : $\nu$F$_\nu$ spectra resulting from the fine-time analysis using the C+BB+PL$_\mathrm{5params}$ model (i.e., $\alpha$=-0.7 and index=-1.5). The solid black lines correspond to the best C+BB+PL fits and the dashed ones to the 1--$\sigma$ confidence regions. The solid blue, red and green lines correspond to the cutoff power laws, to the BB components and to the additional power laws resulting from the best fits with the C+BB+PL model.}
\end{center}
\end{figure*}

\clearpage

\begin{figure*}
\begin{center}

\includegraphics[totalheight=0.95\textheight, clip]{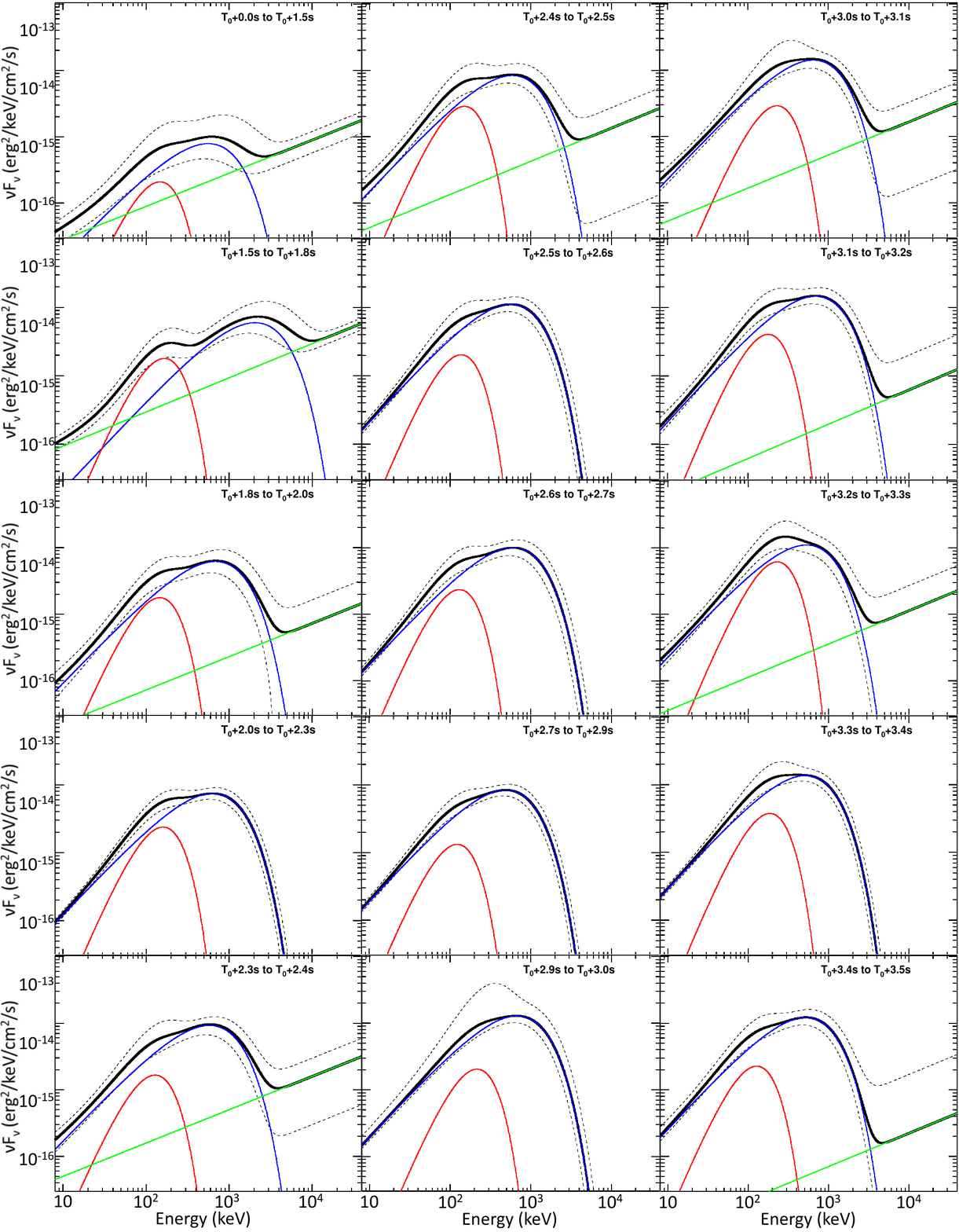}

\end{center}
\end{figure*}

\newpage

\begin{figure*}
\begin{center}

\includegraphics[totalheight=0.95\textheight, clip]{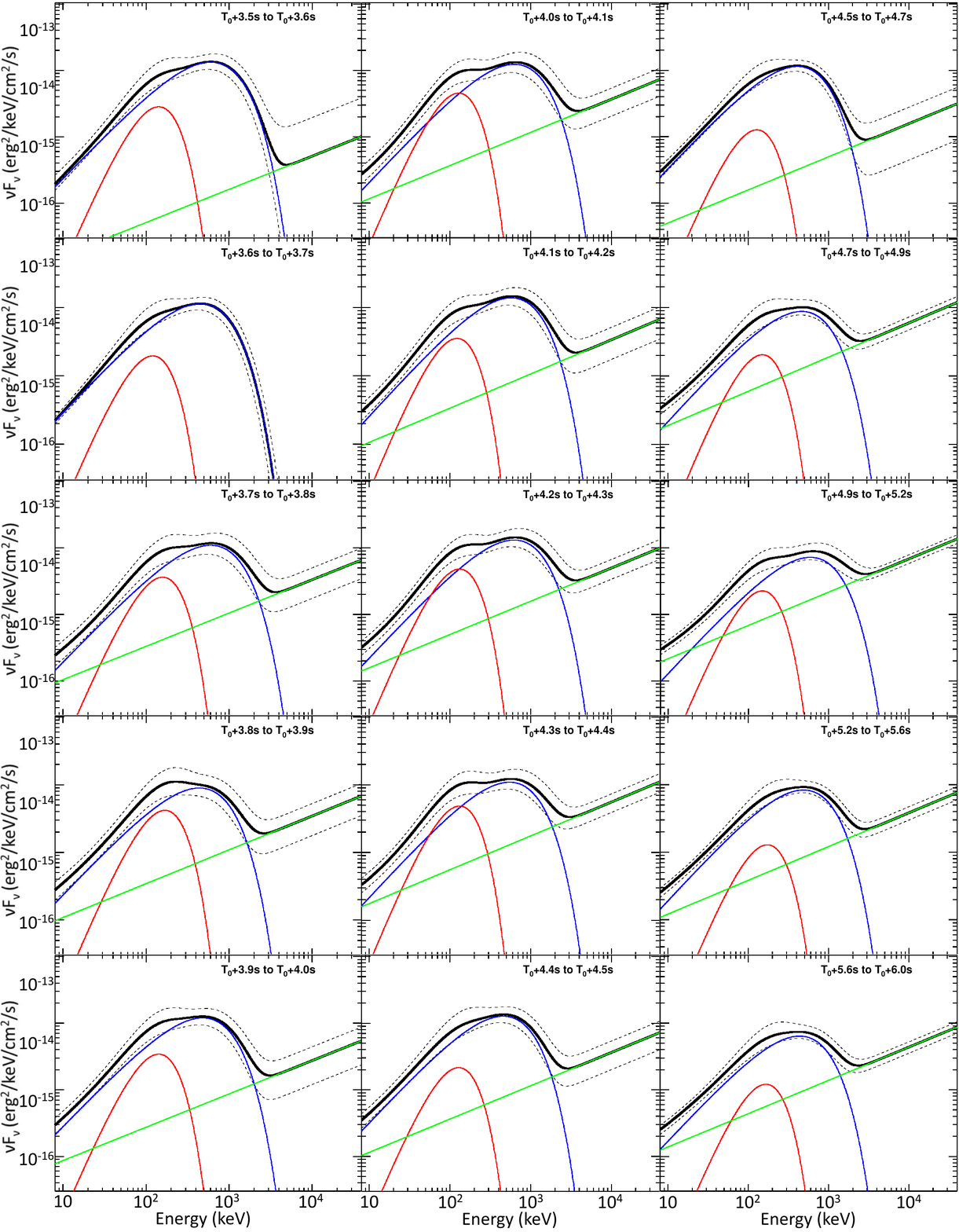}

\end{center}
\end{figure*}

\newpage

\begin{figure*}
\begin{center}

\includegraphics[totalheight=0.95\textheight, clip]{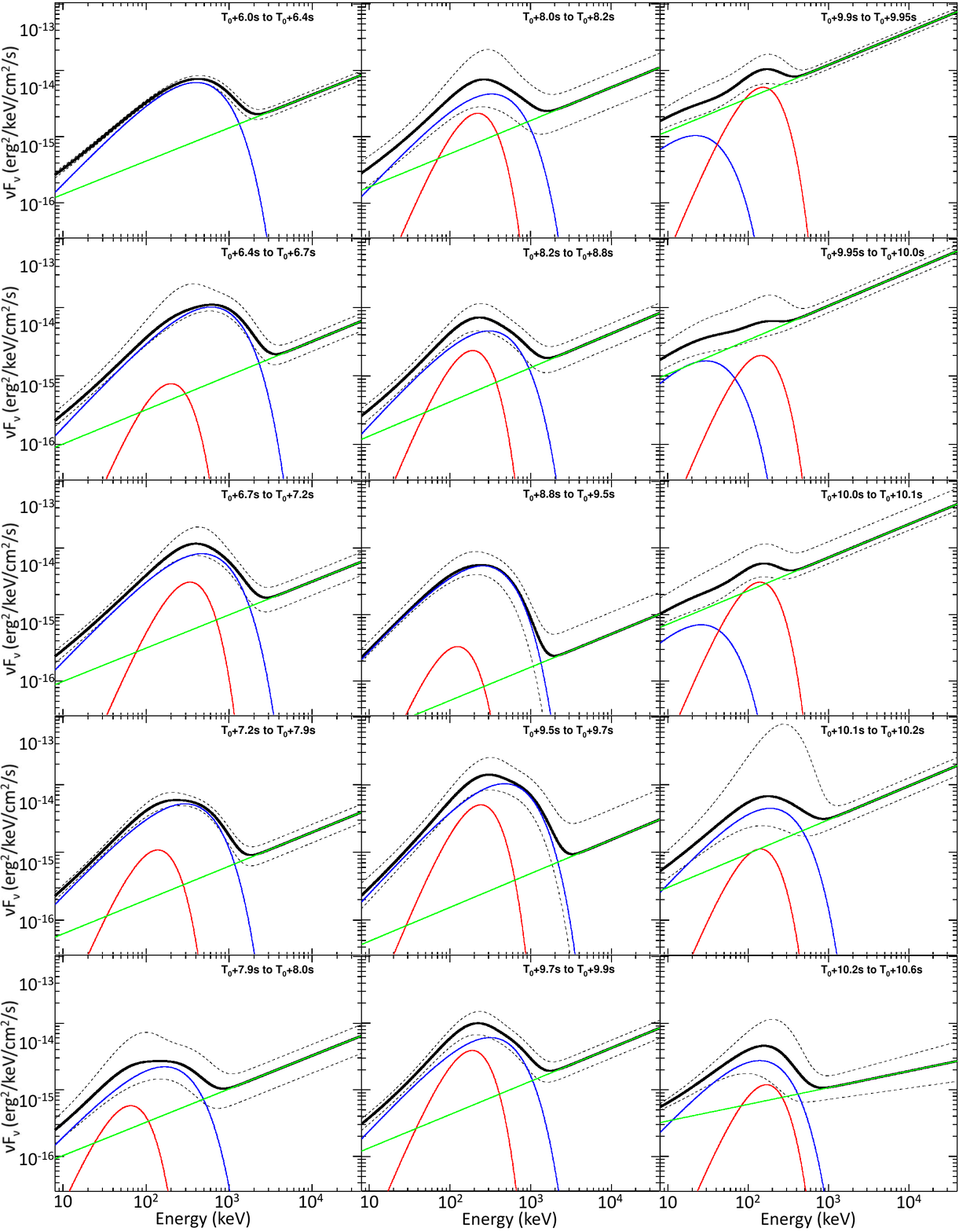}

\end{center}
\end{figure*}

\newpage

\begin{figure*}
\begin{center}

\includegraphics[totalheight=0.78\textheight, clip]{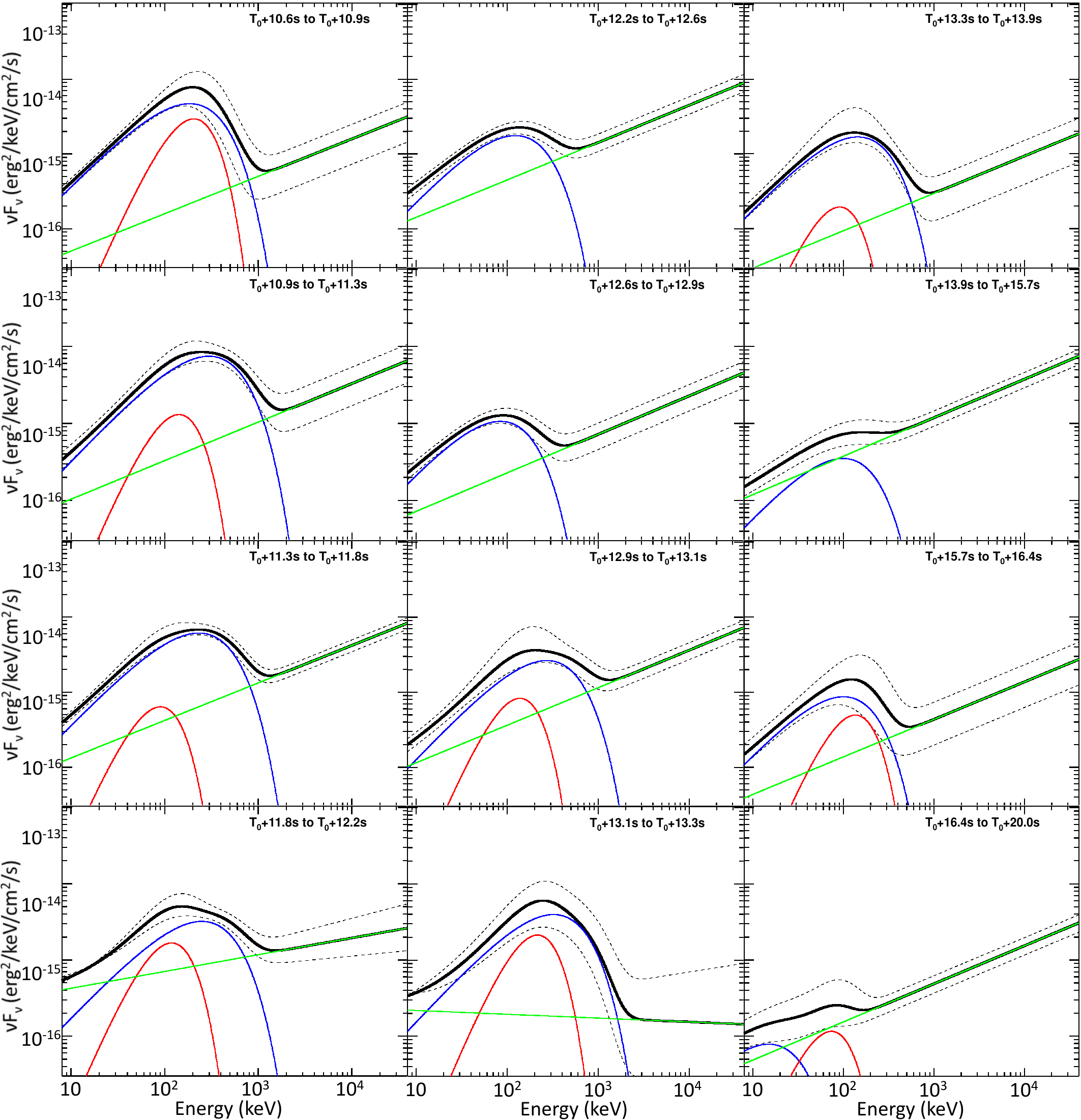}

\caption{\label{fig22}GRB 090926A : $\nu$F$_\nu$ spectra resulting from the fine-time analysis using the C+BB+PL$_\mathrm{5params}$ model (i.e., $\alpha$=-0.7 and index=-1.5). The solid black lines correspond to the best C+BB+PL fits and the dashed ones to the 1--$\sigma$ confidence regions. The solid blue, red and green lines correspond to the cutoff power laws, to the BB components and to the additional power laws resulting from the best fits with the C+BB+PL model.}
\end{center}
\end{figure*}

\clearpage

\setcounter{table}{0}
\renewcommand{\thetable}{A\arabic{table}}

\begin{table*}
\caption{\label{tab01}Parameters of the various tested models resulting from the time-integrated spectral analysis of GRB 080916C from T$_\mathrm{0}$-0.1 s to T$_\mathrm{0}$+71 s with their asymmetrical 1--$\sigma$ uncertainties.}
\begin{center}
{\tiny
\begin{tabular}{|c|c|c|c|c|c|c|c|c|c|c|c|c|c|c|}
\hline
\multicolumn{1}{|c|}{Models} &\multicolumn{3}{|c|}{Standard Model} & \multicolumn{4}{|c|}{Additional Model} & \multicolumn{6}{|c|}{EAC} & \multicolumn{1}{|c|}{Cstat/dof} \\
\hline
     &\multicolumn{3}{|c|}{C or Band} & \multicolumn{1}{|c|}{BB} & \multicolumn{3}{|c|}{ PL, C or Band} & \multicolumn{6}{|c|}{}  & \multicolumn{1}{|c|}{}\\
\hline
\multicolumn{1}{|c|}{Parameters} & \multicolumn{1}{|c|}{E$_{\rm peak}$} & \multicolumn{1}{|c|}{$\alpha$} & \multicolumn{1}{|c|}{$\beta$} & \multicolumn{1}{|c|}{kT} & \multicolumn{1}{|c|}{$\alpha$} & \multicolumn{1}{|c|}{$\beta$} & \multicolumn{1}{|c|}{E$_{\rm 0}$} & \multicolumn{1}{|c|}{n0/b0} & \multicolumn{1}{|c|}{n3/b0} & \multicolumn{1}{|c|}{n4/b0} & \multicolumn{1}{|c|}{L/b0} & \multicolumn{1}{|c|}{F/b0} & \multicolumn{1}{|c|}{B/b0} & \\
\multicolumn{1}{|c|}{} & \multicolumn{1}{|c|}{(keV)} & \multicolumn{1}{|c|}{} & \multicolumn{1}{|c|}{} & \multicolumn{1}{|c|}{(keV)} & \multicolumn{1}{|c|}{} & \multicolumn{1}{|c|}{} & \multicolumn{1}{|c|}{(MeV)} & \multicolumn{1}{|c|}{} & \multicolumn{1}{|c|}{} & \multicolumn{1}{|c|}{} & \multicolumn{1}{|c|}{} & \multicolumn{1}{|c|}{} & \multicolumn{1}{|c|}{} & \\
\hline
 \multicolumn{15}{|c|}{b0+n0+n3+n4}   \\  
 Band & 451 & -1.03 & -2.18 & -- & -- & -- & --  & -- & -- & -- & -- & -- & -- & 666/473 \\
     & $\pm25$ & $\pm0.02$ & $\pm0.08$ & -- & -- & --  & -- & -- & -- & -- & --  & -- & -- &  \\
 B+BB & 1321 & -1.30 & -2.41 & 45.0 & -- & -- & --  & -- & -- & -- & -- & --  & -- & 592/471 \\
\hline
\end{tabular}
}
\end{center}
\end{table*}

\newpage

\begin{table*}
\caption{\label{tab02}Parameters of the various tested models resulting from the time-integrated spectral analysis of GRB 090926A from T$_\mathrm{0}$ s to T$_\mathrm{0}$+20 s with their asymmetrical 1--$\sigma$ uncertainties.}
\begin{center}
{\tiny
\begin{tabular}{|c|c|c|c|c|c|c|c|c|c|c|c|c|c|c|c|}
\hline
\multicolumn{1}{|c|}{Models} &\multicolumn{3}{|c|}{Standard Model} & \multicolumn{4}{|c|}{Additional Model} & \multicolumn{7}{|c|}{EAC} & \multicolumn{1}{|c|}{Cstat/dof} \\
\hline
     &\multicolumn{3}{|c|}{C or Band} & \multicolumn{1}{|c|}{BB} & \multicolumn{3}{|c|}{ PL, C or Band} & \multicolumn{7}{|c|}{}  & \multicolumn{1}{|c|}{}\\
\hline
\multicolumn{1}{|c|}{Parameters} & \multicolumn{1}{|c|}{E$_{\rm peak}$} & \multicolumn{1}{|c|}{$\alpha$} & \multicolumn{1}{|c|}{$\beta$} & \multicolumn{1}{|c|}{kT} & \multicolumn{1}{|c|}{$\alpha$} & \multicolumn{1}{|c|}{$\beta$} & \multicolumn{1}{|c|}{E$_{\rm 0}$} & \multicolumn{1}{|c|}{b1/b0} & \multicolumn{1}{|c|}{n3/b0} & \multicolumn{1}{|c|}{n6/b0} & \multicolumn{1}{|c|}{n7/b0} & \multicolumn{1}{|c|}{L/b0} & \multicolumn{1}{|c|}{F/b0} & \multicolumn{1}{|c|}{B/b0} & \\
\multicolumn{1}{|c|}{} & \multicolumn{1}{|c|}{(keV)} & \multicolumn{1}{|c|}{} & \multicolumn{1}{|c|}{} & \multicolumn{1}{|c|}{(keV)} & \multicolumn{1}{|c|}{} & \multicolumn{1}{|c|}{} & \multicolumn{1}{|c|}{(MeV)} & \multicolumn{1}{|c|}{} & \multicolumn{1}{|c|}{} & \multicolumn{1}{|c|}{} & \multicolumn{1}{|c|}{} & \multicolumn{1}{|c|}{} & \multicolumn{1}{|c|}{} & \multicolumn{1}{|c|}{} & \\
\hline
 \multicolumn{16}{|c|}{b0+b1+n3+n6+n7 with b0$>$756.48 keV and b1$>$609.51 keV}   \\  
 Band & 294 & -0.75 & -2.44 & -- & -- & -- & --  & -- & -- & -- & -- & -- & -- & -- & 1138/566 \\
     & $\pm5$ & $\pm0.01$ & $\pm0.03$ & -- & -- & --  & -- & -- & -- & -- & -- & --  & -- & -- &  \\
 B+BB & 450 & -1.01 & -2.58 & 39.4 & -- & -- & --  & -- & -- & -- & -- & -- & -- & -- & 999/564 \\
\hline
\end{tabular}
}
\end{center}
\end{table*}

\newpage

\begin{table*}
\caption{\label{tab03}Parameters of the various tested models resulting from the coarse-time spectral analysis of GRB 080916C with their asymmetrical 1--$\sigma$ uncertainties.}
\begin{center}
{\tiny
\begin{tabular}{|c|c|c|c|c|c|c|c|c|c|c|c|c|c|c|c|c|}
\hline
\multicolumn{2}{|c|}{Time from T$_\mathrm{0}$} & \multicolumn{1}{|c|}{Models} &\multicolumn{3}{|c|}{Standard Model} & \multicolumn{3}{|c|}{Additional Model} & \multicolumn{4}{|c|}{EAC} & \multicolumn{1}{|c|}{Cstat/dof} \\
\hline
 \multicolumn{2}{|c|}{} &    &\multicolumn{3}{|c|}{C or Band} & \multicolumn{1}{|c|}{BB} & \multicolumn{2}{|c|}{ PL or C} & \multicolumn{4}{|c|}{}  & \multicolumn{1}{|c|}{}\\
\hline
\multicolumn{1}{|c|}{T$_\mathrm{start}$} & \multicolumn{1}{|c|}{T$_\mathrm{stop}$} & \multicolumn{1}{|c|}{Parameters} & \multicolumn{1}{|c|}{E$_{\rm peak}$} & \multicolumn{1}{|c|}{$\alpha$} & \multicolumn{1}{|c|}{$\beta$} & \multicolumn{1}{|c|}{kT} & \multicolumn{1}{|c|}{$\alpha$} & \multicolumn{1}{|c|}{E$_{\rm 0}$} & \multicolumn{1}{|c|}{n0/b0} & \multicolumn{1}{|c|}{n3/b0} & \multicolumn{1}{|c|}{n4/b0} & \multicolumn{1}{|c|}{L/b0} & \\
\multicolumn{1}{|c|}{(s)} & \multicolumn{1}{|c|}{(s)} & \multicolumn{1}{|c|}{} & \multicolumn{1}{|c|}{(keV)} & \multicolumn{1}{|c|}{} & \multicolumn{1}{|c|}{} & \multicolumn{1}{|c|}{(keV)} & \multicolumn{1}{|c|}{} & \multicolumn{1}{|c|}{(MeV)} & \multicolumn{1}{|c|}{} & \multicolumn{1}{|c|}{} &  \multicolumn{1}{|c|}{} & \multicolumn{1}{|c|}{} & \\
\hline
\multicolumn{14}{|c|}{b0+n0+n3+n4}   \\
-0.1 & +4.3 & Band & 454$\pm38$ & -0.69$\pm0.04$ & -2.25$^\mathrm{+0.12}_\mathrm{-0.16}$ & -- & -- & -- & --  & -- & -- & -- & 505/473 \\
 & & B+BB & 1192$^\mathrm{+214}_\mathrm{-184}$ & -0.97$\pm0.04$ & $<$-5 & 39.8$\pm2.6$ & -- & -- & --  & -- & -- & -- & 477/471 \\
& & C+BB & 1189$^\mathrm{+215}_\mathrm{-182}$ & -0.97$\pm0.04$ & -- & 39.8$\pm2.7$ & -- & -- & --  & -- & -- & -- & 477/472 \\
\end{tabular}
}
\end{center}
\end{table*}

\newpage

\begin{table*}
\begin{center}
{\tiny
\begin{tabular}{|c|c|c|c|c|c|c|c|c|c|c|c|c|c|c|c|c|}
\hline
\multicolumn{2}{|c|}{Time from T$_\mathrm{0}$} & \multicolumn{1}{|c|}{Models} &\multicolumn{3}{|c|}{Standard Model} & \multicolumn{3}{|c|}{Additional Model} & \multicolumn{4}{|c|}{EAC} & \multicolumn{1}{|c|}{Cstat/dof} \\
\hline
 \multicolumn{2}{|c|}{} &    &\multicolumn{3}{|c|}{C or Band} & \multicolumn{1}{|c|}{BB} & \multicolumn{2}{|c|}{ PL or C} & \multicolumn{4}{|c|}{}  & \multicolumn{1}{|c|}{}\\
\hline
\multicolumn{1}{|c|}{T$_\mathrm{start}$} & \multicolumn{1}{|c|}{T$_\mathrm{stop}$} & \multicolumn{1}{|c|}{Parameters} & \multicolumn{1}{|c|}{E$_{\rm peak}$} & \multicolumn{1}{|c|}{$\alpha$} & \multicolumn{1}{|c|}{$\beta$} & \multicolumn{1}{|c|}{kT} & \multicolumn{1}{|c|}{$\alpha$} & \multicolumn{1}{|c|}{E$_{\rm 0}$} & \multicolumn{1}{|c|}{n0/b0} & \multicolumn{1}{|c|}{n3/b0} & \multicolumn{1}{|c|}{n4/b0} & \multicolumn{1}{|c|}{L/b0} & \\
\multicolumn{1}{|c|}{(s)} & \multicolumn{1}{|c|}{(s)} & \multicolumn{1}{|c|}{} & \multicolumn{1}{|c|}{(keV)} & \multicolumn{1}{|c|}{} & \multicolumn{1}{|c|}{} & \multicolumn{1}{|c|}{(keV)} & \multicolumn{1}{|c|}{} & \multicolumn{1}{|c|}{(MeV)} & \multicolumn{1}{|c|}{} & \multicolumn{1}{|c|}{} &  \multicolumn{1}{|c|}{} & \multicolumn{1}{|c|}{} & \\
\hline
\multicolumn{14}{|c|}{b0+n0+n3+n4}   \\  
+7.5 & +71.0 & Band & 413$\pm26$ & -1.07$\pm0.02$ & -2.29$^\mathrm{+0.11}_\mathrm{-0.17}$ & -- & -- & -- & --  & -- & -- & -- & 618/473 \\
& & B+BB & 967$^\mathrm{+326}_\mathrm{-210}$ & -1.34$\pm0.04$ & -2.32$^\mathrm{+0.17}_\mathrm{-0.29}$ & 45.5$\pm2.0$ & -- & -- & --  & -- & -- & -- & 569/471 \\
& & B+PL & 263$\pm25$ & -0.40$\pm0.19$ & -2.16$^\mathrm{+0.09}_\mathrm{-0.14}$ & -- & -1.84$^\mathrm{+0.09}_\mathrm{-0.16}$ & -- & --  & -- & -- & -- & 579/471 \\
\hline
\end{tabular}
}
\end{center}
\end{table*}

\newpage

\begin{table*}
\caption{\label{tab04}Parameters of the various tested models resulting from the coarse-time spectral analysis of GRB 090926A with their asymmetrical 1--$\sigma$ uncertainties.}
\begin{center}
{\tiny
\begin{adjustbox}{angle=90}
\begin{tabular}{|c|c|c|c|c|c|c|c|c|c|c|c|c|c|c|c|c|c|}
\hline
\multicolumn{2}{|c|}{Time from T$_\mathrm{0}$} & \multicolumn{1}{|c|}{Models} &\multicolumn{3}{|c|}{Standard Model} & \multicolumn{3}{|c|}{Additional Model} & \multicolumn{5}{|c|}{EAC} & \multicolumn{1}{|c|}{Cstat/dof} \\
\hline
 \multicolumn{2}{|c|}{} &    &\multicolumn{3}{|c|}{C or Band} & \multicolumn{1}{|c|}{BB} & \multicolumn{2}{|c|}{ PL or C} & \multicolumn{5}{|c|}{}  & \multicolumn{1}{|c|}{}\\
\hline
\multicolumn{1}{|c|}{T$_\mathrm{start}$} & \multicolumn{1}{|c|}{T$_\mathrm{stop}$} & \multicolumn{1}{|c|}{Parameters} & \multicolumn{1}{|c|}{E$_{\rm peak}$} & \multicolumn{1}{|c|}{$\alpha$} & \multicolumn{1}{|c|}{$\beta$} & \multicolumn{1}{|c|}{kT} & \multicolumn{1}{|c|}{$\alpha$} & \multicolumn{1}{|c|}{E$_{\rm 0}$} & \multicolumn{1}{|c|}{b1/b0} & \multicolumn{1}{|c|}{n3/b0} & \multicolumn{1}{|c|}{n6/b0} & \multicolumn{1}{|c|}{n7/b0} & \multicolumn{1}{|c|}{L/b0} & \\
\multicolumn{1}{|c|}{(s)} & \multicolumn{1}{|c|}{(s)} & \multicolumn{1}{|c|}{} & \multicolumn{1}{|c|}{(keV)} & \multicolumn{1}{|c|}{} & \multicolumn{1}{|c|}{} & \multicolumn{1}{|c|}{(keV)} & \multicolumn{1}{|c|}{} & \multicolumn{1}{|c|}{(MeV)} & \multicolumn{1}{|c|}{} & \multicolumn{1}{|c|}{} &  \multicolumn{1}{|c|}{} &  \multicolumn{1}{|c|}{} & \multicolumn{1}{|c|}{} & \\
\hline
\multicolumn{15}{|c|}{b0+b1+n3+n6+n7 with b0>756.48 keV and b1>609.51 keV}   \\ 
0.0 & +5.0 & Band & 304$\pm7$ & -0.45$\pm0.02$ & -2.32$\pm0.03$ & -- & -- & --  & -- & -- & -- & -- & -- &  762/566 \\ 
 & & B+BB & 529$\pm23$ & -0.74$\pm0.03$ & -2.68$\pm0.09$ & 36.4$\pm1.2$ & -- & -- & --  & -- & -- & -- & -- & 695/564 \\
\multicolumn{15}{|c|}{b0+b1+n3+n6+n7 with b0>756.48 keV and b1>609.51 keV + EAC}   \\
\end{tabular}
\end{adjustbox}
}
\end{center}
\end{table*}

\newpage

\begin{table*}
\begin{center}
{\tiny
\begin{adjustbox}{angle=90}
\begin{tabular}{|c|c|c|c|c|c|c|c|c|c|c|c|c|c|c|c|c|c|}
\hline
\multicolumn{2}{|c|}{Time from T$_\mathrm{0}$} & \multicolumn{1}{|c|}{Models} &\multicolumn{3}{|c|}{Standard Model} & \multicolumn{3}{|c|}{Additional Model} & \multicolumn{5}{|c|}{EAC} & \multicolumn{1}{|c|}{Cstat/dof} \\
\hline
 \multicolumn{2}{|c|}{} &    &\multicolumn{3}{|c|}{C or Band} & \multicolumn{1}{|c|}{BB} & \multicolumn{2}{|c|}{ PL or C} & \multicolumn{5}{|c|}{}  & \multicolumn{1}{|c|}{}\\
\hline
\multicolumn{1}{|c|}{T$_\mathrm{start}$} & \multicolumn{1}{|c|}{T$_\mathrm{stop}$} & \multicolumn{1}{|c|}{Parameters} & \multicolumn{1}{|c|}{E$_{\rm peak}$} & \multicolumn{1}{|c|}{$\alpha$} & \multicolumn{1}{|c|}{$\beta$} & \multicolumn{1}{|c|}{kT} & \multicolumn{1}{|c|}{$\alpha$} & \multicolumn{1}{|c|}{E$_{\rm 0}$} & \multicolumn{1}{|c|}{b1/b0} & \multicolumn{1}{|c|}{n3/b0} & \multicolumn{1}{|c|}{n6/b0} & \multicolumn{1}{|c|}{n7/b0} & \multicolumn{1}{|c|}{L/b0} & \\
\multicolumn{1}{|c|}{(s)} & \multicolumn{1}{|c|}{(s)} & \multicolumn{1}{|c|}{} & \multicolumn{1}{|c|}{(keV)} & \multicolumn{1}{|c|}{} & \multicolumn{1}{|c|}{} & \multicolumn{1}{|c|}{(keV)} & \multicolumn{1}{|c|}{} & \multicolumn{1}{|c|}{(MeV)} & \multicolumn{1}{|c|}{} & \multicolumn{1}{|c|}{} &  \multicolumn{1}{|c|}{} &  \multicolumn{1}{|c|}{} & \multicolumn{1}{|c|}{} & \\
\hline
\multicolumn{15}{|c|}{b0+b1+n3+n6+n7 with b0>756.48 keV and b1>609.51 keV}\\ 
 +9.0 & +11.0 & Band & 254$\pm7$ & -0.81$\pm0.02$ & -2.79$\pm0.13$ & -- & -- & --  & -- & -- & -- & -- & -- &  640/566 \\
 & & B+BB & 331$\pm36$ & -1.16$\pm0.05$ & -2.46$^\mathrm{+0.11}_\mathrm{-0.16}$ & 43.4$\pm1.7$ & -- & -- & --  & -- & -- & -- & -- & 569/564 \\
 & & B+PL & 211$\pm7$ & -0.21$\pm0.08$ & -2.96$^\mathrm{+0.15}_\mathrm{-0.20}$ & -- & -1.82$\pm0.05$ & -- & --  & -- & -- & -- & -- & 560/564 \\
\end{tabular}
\end{adjustbox}}
\end{center}
\end{table*}

\newpage

\begin{table*}
\begin{center}
{\tiny
\begin{adjustbox}{angle=90}
\begin{tabular}{|c|c|c|c|c|c|c|c|c|c|c|c|c|c|c|c|c|c|}
\hline
\multicolumn{2}{|c|}{Time from T$_\mathrm{0}$} & \multicolumn{1}{|c|}{Models} &\multicolumn{3}{|c|}{Standard Model} & \multicolumn{3}{|c|}{Additional Model} & \multicolumn{5}{|c|}{EAC} & \multicolumn{1}{|c|}{Cstat/dof} \\
\hline
 \multicolumn{2}{|c|}{} &    &\multicolumn{3}{|c|}{C or Band} & \multicolumn{1}{|c|}{BB} & \multicolumn{2}{|c|}{ PL or C} & \multicolumn{5}{|c|}{}  & \multicolumn{1}{|c|}{}\\
\hline
\multicolumn{1}{|c|}{T$_\mathrm{start}$} & \multicolumn{1}{|c|}{T$_\mathrm{stop}$} & \multicolumn{1}{|c|}{Parameters} & \multicolumn{1}{|c|}{E$_{\rm peak}$} & \multicolumn{1}{|c|}{$\alpha$} & \multicolumn{1}{|c|}{$\beta$} & \multicolumn{1}{|c|}{kT} & \multicolumn{1}{|c|}{$\alpha$} & \multicolumn{1}{|c|}{E$_{\rm 0}$} & \multicolumn{1}{|c|}{b1/b0} & \multicolumn{1}{|c|}{n3/b0} & \multicolumn{1}{|c|}{n6/b0} & \multicolumn{1}{|c|}{n7/b0} & \multicolumn{1}{|c|}{L/b0} & \\
\multicolumn{1}{|c|}{(s)} & \multicolumn{1}{|c|}{(s)} & \multicolumn{1}{|c|}{} & \multicolumn{1}{|c|}{(keV)} & \multicolumn{1}{|c|}{} & \multicolumn{1}{|c|}{} & \multicolumn{1}{|c|}{(keV)} & \multicolumn{1}{|c|}{} & \multicolumn{1}{|c|}{(MeV)} & \multicolumn{1}{|c|}{} & \multicolumn{1}{|c|}{} &  \multicolumn{1}{|c|}{} &  \multicolumn{1}{|c|}{} & \multicolumn{1}{|c|}{} & \\
\hline
\multicolumn{15}{|c|}{b0+b1+n3+n6+n7 with b0>756.48 keV and b1>609.51 keV}   \\  
 +13.0 & +20.0 & Band & 161$\pm12$ & -1.10$\pm0.04$ & -2.33$\pm0.15$ & -- & -- & --  & -- & -- & -- & -- & -- &  706/566 \\
 & & B+BB & 241$^\mathrm{+134}_\mathrm{-60}$ & -1.39$\pm0.10$ & -2.25$^\mathrm{+0.15}_\mathrm{-0.77}$ & 27.2$\pm2.5$ & -- & -- & --  & -- & -- & -- & -- & 696/564 \\
 & & B+PL & 136$\pm11$ & -0.53$^\mathrm{+0.24}_\mathrm{-0.29}$ & -2.34$^\mathrm{+0.13}_\mathrm{-0.25}$ & -- & -1.97$^\mathrm{+0.13}_\mathrm{-0.58}$ & -- & --  & -- & -- & -- & -- & 697/564 \\
\hline
\end{tabular}
\end{adjustbox}}
\end{center}
\end{table*}

\newpage

\begin{table*}
\caption{\label{tab05}Parameters of the various tested models resulting from the fine-time spectral analysis of GRB 080916C with their asymmetrical 1--$\sigma$ uncertainties.}
\begin{center}
{\tiny
\begin{tabular}{|c|c|c|c|c|c|c|c|c|c|c|c|c|c|c|c|}
\hline
\multicolumn{2}{|c|}{Time from T$_\mathrm{0}$} & \multicolumn{1}{|c|}{Models} &\multicolumn{3}{|c|}{Standard Model} & \multicolumn{2}{|c|}{Additional Model} & \multicolumn{1}{|c|}{Cstat/dof} \\
\hline
 \multicolumn{2}{|c|}{} &    &\multicolumn{3}{|c|}{C or Band} & \multicolumn{1}{|c|}{BB} & \multicolumn{1}{|c|}{ PL} & \multicolumn{1}{|c|}{}\\
\hline
\multicolumn{1}{|c|}{T$_\mathrm{start}$} & \multicolumn{1}{|c|}{T$_\mathrm{stop}$} & \multicolumn{1}{|c|}{Parameters} & \multicolumn{1}{|c|}{E$_{\rm peak}$} & \multicolumn{1}{|c|}{$\alpha$} & \multicolumn{1}{|c|}{$\beta$} & \multicolumn{1}{|c|}{kT} & \multicolumn{1}{|c|}{$\alpha$} & \\
\multicolumn{1}{|c|}{(s)} & \multicolumn{1}{|c|}{(s)} & \multicolumn{1}{|c|}{} & \multicolumn{1}{|c|}{(keV)} & \multicolumn{1}{|c|}{} & \multicolumn{1}{|c|}{} & \multicolumn{1}{|c|}{(keV)} & \multicolumn{1}{|c|}{} & \\
\hline
\multicolumn{9}{|c|}{ }   \\ 
-0.1 & +0.7 & Band & 408$^\mathrm{+162}_\mathrm{-110}$ & -0.77$^\mathrm{+0.17}_\mathrm{-0.13}$ & -2.17$^\mathrm{+0.29}_\mathrm{-1.14}$ & -- & -- & 492/473 \\
& & C+BB & 2107$^\mathrm{+2130}_\mathrm{-1130}$ & -1.21$^\mathrm{+0.13}_\mathrm{-0.11}$ & -- & 49.2$\pm8.4$ & -- & 488/472 \\
& & C+BB+PL$_\mathrm{5params}$ & 632$^\mathrm{+238}_\mathrm{-155}$ & -0.7 (fix) & -- & 44.8$^\mathrm{+20.5}_\mathrm{-11.4}$ & -1.5 (fix) & 492/472 \\
\end{tabular}
}
\end{center}
\end{table*}

\newpage

\begin{table*}
\begin{center}
{\tiny
\begin{tabular}{|c|c|c|c|c|c|c|c|c|c|c|c|c|c|c|c|}
\hline
\multicolumn{2}{|c|}{Time from T$_\mathrm{0}$} & \multicolumn{1}{|c|}{Models} &\multicolumn{3}{|c|}{Standard Model} & \multicolumn{2}{|c|}{Additional Model} & \multicolumn{1}{|c|}{Cstat/dof} \\
\hline
 \multicolumn{2}{|c|}{} &    &\multicolumn{3}{|c|}{C or Band} & \multicolumn{1}{|c|}{BB} & \multicolumn{1}{|c|}{ PL} & \multicolumn{1}{|c|}{}\\
\hline
\multicolumn{1}{|c|}{T$_\mathrm{start}$} & \multicolumn{1}{|c|}{T$_\mathrm{stop}$} & \multicolumn{1}{|c|}{Parameters} & \multicolumn{1}{|c|}{E$_{\rm peak}$} & \multicolumn{1}{|c|}{$\alpha$} & \multicolumn{1}{|c|}{$\beta$} & \multicolumn{1}{|c|}{kT} & \multicolumn{1}{|c|}{$\alpha$} & \\
\multicolumn{1}{|c|}{(s)} & \multicolumn{1}{|c|}{(s)} & \multicolumn{1}{|c|}{} & \multicolumn{1}{|c|}{(keV)} & \multicolumn{1}{|c|}{} & \multicolumn{1}{|c|}{} & \multicolumn{1}{|c|}{(keV)} & \multicolumn{1}{|c|}{} & \\
\hline
\multicolumn{9}{|c|}{ }   \\ 
+7.0 & +7.5 & Band & 509$^\mathrm{+579}_\mathrm{-192}$ & -1.07$\pm0.14$ & $<$-1.75 & -- & -- & 467/473 \\
& & C+BB & 1788$^\mathrm{+1960}_\mathrm{-1010}$ & -1.28$\pm0.12$ & -- & 39.0$^\mathrm{+17.5}_\mathrm{-15.3}$ & -- & 465/472 \\
& & C+BB+PL & 1274$^\mathrm{+1220}_\mathrm{-571}$ & -0.94$^\mathrm{+0.19}_\mathrm{-0.14}$ & -- & 34.7$^\mathrm{+10.3}_\mathrm{-8.5}$ & -1.67$^\mathrm{+0.12}_\mathrm{-0.24}$ & 465/470 \\
\hline
\end{tabular}
}
\end{center}
\end{table*}

\newpage

\begin{table*}
\caption{\label{tab06}Parameters of the various tested models resulting from the fine-time spectral analysis of GRB 090926A with their asymmetrical 1--$\sigma$ uncertainties.}
\begin{center}
{\tiny
\begin{tabular}{|c|c|c|c|c|c|c|c|c|c|c|c|c|c|c|c|c|}
\hline
\multicolumn{2}{|c|}{Time from T$_\mathrm{0}$} & \multicolumn{1}{|c|}{Models} &\multicolumn{3}{|c|}{Standard Model} & \multicolumn{3}{|c|}{Additional Model} & \multicolumn{1}{|c|}{Cstat/dof} \\
\hline
 \multicolumn{2}{|c|}{} &    &\multicolumn{3}{|c|}{C or Band} & \multicolumn{1}{|c|}{BB} & \multicolumn{2}{|c|}{ PL or C} & \multicolumn{1}{|c|}{}\\
\hline
\multicolumn{1}{|c|}{T$_\mathrm{start}$} & \multicolumn{1}{|c|}{T$_\mathrm{stop}$} & \multicolumn{1}{|c|}{Parameters} & \multicolumn{1}{|c|}{E$_{\rm peak}$} & \multicolumn{1}{|c|}{$\alpha$} & \multicolumn{1}{|c|}{$\beta$} & \multicolumn{1}{|c|}{kT} & \multicolumn{1}{|c|}{$\alpha$} & \multicolumn{1}{|c|}{E$_\mathrm{0}$} & \\
\multicolumn{1}{|c|}{(s)} & \multicolumn{1}{|c|}{(s)} & \multicolumn{1}{|c|}{} & \multicolumn{1}{|c|}{(keV)} & \multicolumn{1}{|c|}{} & \multicolumn{1}{|c|}{} & \multicolumn{1}{|c|}{(keV)} & \multicolumn{1}{|c|}{} & \multicolumn{1}{|c|}{(MeV)} & \\
\hline
\multicolumn{10}{|c|}{ }   \\ 
+0.0 & +1.5 & Band & 326$^\mathrm{+112}_\mathrm{-59}$ & -0.71$\pm0.13$ & -2.08$^\mathrm{+0.14}_\mathrm{-0.32}$ & -- & -- & -- & 609/566 \\
& & C+BB & 1551$^\mathrm{+1320}_\mathrm{-790}$ & -1.21$^\mathrm{+0.17}_\mathrm{-0.10}$ & -- & 44.0$\pm5.5$ & -- & -- & 605/565 \\
& & C+BB+PL$_\mathrm{5params}$ & 561$^\mathrm{+229}_\mathrm{-126}$ & -0.7 (fix) & -- & 37.6$\pm7.0$ & -1.5 (fix) & -- & 607/565 \\
\end{tabular}
}
\end{center}
\end{table*}

\newpage

\begin{table*}
\begin{center}
{\tiny
\begin{tabular}{|c|c|c|c|c|c|c|c|c|c|c|c|c|c|c|c|c|}
\hline
\multicolumn{2}{|c|}{Time from T$_\mathrm{0}$} & \multicolumn{1}{|c|}{Models} &\multicolumn{3}{|c|}{Standard Model} & \multicolumn{3}{|c|}{Additional Model} & \multicolumn{1}{|c|}{Cstat/dof} \\
\hline
 \multicolumn{2}{|c|}{} &    &\multicolumn{3}{|c|}{C or Band} & \multicolumn{1}{|c|}{BB} & \multicolumn{2}{|c|}{ PL or C} & \multicolumn{1}{|c|}{}\\
\hline
\multicolumn{1}{|c|}{T$_\mathrm{start}$} & \multicolumn{1}{|c|}{T$_\mathrm{stop}$} & \multicolumn{1}{|c|}{Parameters} & \multicolumn{1}{|c|}{E$_{\rm peak}$} & \multicolumn{1}{|c|}{$\alpha$} & \multicolumn{1}{|c|}{$\beta$} & \multicolumn{1}{|c|}{kT} & \multicolumn{1}{|c|}{$\alpha$} & \multicolumn{1}{|c|}{E$_\mathrm{0}$} & \\
\multicolumn{1}{|c|}{(s)} & \multicolumn{1}{|c|}{(s)} & \multicolumn{1}{|c|}{} & \multicolumn{1}{|c|}{(keV)} & \multicolumn{1}{|c|}{} & \multicolumn{1}{|c|}{} & \multicolumn{1}{|c|}{(keV)} & \multicolumn{1}{|c|}{} & \multicolumn{1}{|c|}{(MeV)} & \\
\hline
\multicolumn{10}{|c|}{ }   \\
+3.5 & +3.6 & Band & 328$^\mathrm{+41}_\mathrm{-35}$ & -0.39$\pm0.09$ & -2.43$^\mathrm{+0.14}_\mathrm{-0.20}$ & -- & -- & -- & 577/566 \\
& & C+BB & 536$^\mathrm{+70}_\mathrm{-55}$ & -0.58$\pm0.10$ & -- & 33.3$^\mathrm{+6.0}_\mathrm{-5.1}$ & -- & -- & 575/565 \\
& & C+BB & 598$^\mathrm{+57}_\mathrm{-50}$ & -0.7 (fix) & -- & 35.5$^\mathrm{+4.4}_\mathrm{-3.6}$ & -- & -- & 576/566 \\
\end{tabular}
}
\end{center}
\end{table*}

\newpage

\begin{table*}
\begin{center}
{\tiny
\begin{tabular}{|c|c|c|c|c|c|c|c|c|c|c|c|c|c|c|c|c|}
\hline
\multicolumn{2}{|c|}{Time from T$_\mathrm{0}$} & \multicolumn{1}{|c|}{Models} &\multicolumn{3}{|c|}{Standard Model} & \multicolumn{3}{|c|}{Additional Model} & \multicolumn{1}{|c|}{Cstat/dof} \\
\hline
 \multicolumn{2}{|c|}{} &    &\multicolumn{3}{|c|}{C or Band} & \multicolumn{1}{|c|}{BB} & \multicolumn{2}{|c|}{ PL or C} & \multicolumn{1}{|c|}{}\\
\hline
\multicolumn{1}{|c|}{T$_\mathrm{start}$} & \multicolumn{1}{|c|}{T$_\mathrm{stop}$} & \multicolumn{1}{|c|}{Parameters} & \multicolumn{1}{|c|}{E$_{\rm peak}$} & \multicolumn{1}{|c|}{$\alpha$} & \multicolumn{1}{|c|}{$\beta$} & \multicolumn{1}{|c|}{kT} & \multicolumn{1}{|c|}{$\alpha$} & \multicolumn{1}{|c|}{E$_\mathrm{0}$} & \\
\multicolumn{1}{|c|}{(s)} & \multicolumn{1}{|c|}{(s)} & \multicolumn{1}{|c|}{} & \multicolumn{1}{|c|}{(keV)} & \multicolumn{1}{|c|}{} & \multicolumn{1}{|c|}{} & \multicolumn{1}{|c|}{(keV)} & \multicolumn{1}{|c|}{} & \multicolumn{1}{|c|}{(MeV)} & \\
\hline
\multicolumn{10}{|c|}{ }   \\
+5.2 & +5.6 & Band & 390$\pm30$ & -0.68$\pm0.04$ & -2.39$^\mathrm{+0.11}_\mathrm{-0.14}$ & -- & -- & -- & 609/566 \\
& & C+BB & 709$^\mathrm{+154}_\mathrm{-109}$ & -1.01$\pm0.08$ & -- & 51.5$\pm24.5$ & -- & -- & 615/565 \\
& & C+BB+PL & 592$^\mathrm{+125}_\mathrm{-95}$ & -0.90$^\mathrm{+0.25}_\mathrm{-0.13}$ & -- & 50.6$^\mathrm{+5.7}_\mathrm{8.3}$ & -1.44$^\mathrm{+0.55}_\mathrm{-0.16}$ & -- & 603/563 \\
\end{tabular}
}
\end{center}
\end{table*}

\newpage

\begin{table*}
\begin{center}
{\tiny
\begin{tabular}{|c|c|c|c|c|c|c|c|c|c|c|c|c|c|c|c|c|}
\hline
\multicolumn{2}{|c|}{Time from T$_\mathrm{0}$} & \multicolumn{1}{|c|}{Models} &\multicolumn{3}{|c|}{Standard Model} & \multicolumn{3}{|c|}{Additional Model} & \multicolumn{1}{|c|}{Cstat/dof} \\
\hline
 \multicolumn{2}{|c|}{} &    &\multicolumn{3}{|c|}{C or Band} & \multicolumn{1}{|c|}{BB} & \multicolumn{2}{|c|}{ PL or C} & \multicolumn{1}{|c|}{}\\
\hline
\multicolumn{1}{|c|}{T$_\mathrm{start}$} & \multicolumn{1}{|c|}{T$_\mathrm{stop}$} & \multicolumn{1}{|c|}{Parameters} & \multicolumn{1}{|c|}{E$_{\rm peak}$} & \multicolumn{1}{|c|}{$\alpha$} & \multicolumn{1}{|c|}{$\beta$} & \multicolumn{1}{|c|}{kT} & \multicolumn{1}{|c|}{$\alpha$} & \multicolumn{1}{|c|}{E$_\mathrm{0}$} & \\
\multicolumn{1}{|c|}{(s)} & \multicolumn{1}{|c|}{(s)} & \multicolumn{1}{|c|}{} & \multicolumn{1}{|c|}{(keV)} & \multicolumn{1}{|c|}{} & \multicolumn{1}{|c|}{} & \multicolumn{1}{|c|}{(keV)} & \multicolumn{1}{|c|}{} & \multicolumn{1}{|c|}{(MeV)} & \\
\hline
\multicolumn{10}{|c|}{ }   \\
+9.7 & +9.9 & Band & 285$\pm13$ & -0.51$\pm0.05$ & -3.08$^\mathrm{+0.28}_\mathrm{-0.47}$ & -- & -- & -- & 635/566 \\
& & C+BB & 409$^\mathrm{+82}_\mathrm{-55}$ & -1.00$\pm0.12$ & -- & 54.9$\pm3.0$ & -- & -- & 624/565 \\
& & C+BB+PL & 355$^\mathrm{+75}_\mathrm{-58}$ & -0.88$^\mathrm{+0.40}_\mathrm{-0.18}$ & -- & 55.4$^\mathrm{+3.4}_\mathrm{-4.3}$ & -1.51$^\mathrm{+0.44}_\mathrm{-0.17}$ & -- & 607/589 \\
\end{tabular}
}
\end{center}
\end{table*}

\newpage

\begin{table*}
\begin{center}
{\tiny
\begin{tabular}{|c|c|c|c|c|c|c|c|c|c|c|c|c|c|c|c|c|}
\hline
\multicolumn{2}{|c|}{Time from T$_\mathrm{0}$} & \multicolumn{1}{|c|}{Models} &\multicolumn{3}{|c|}{Standard Model} & \multicolumn{3}{|c|}{Additional Model} & \multicolumn{1}{|c|}{Cstat/dof} \\
\hline
 \multicolumn{2}{|c|}{} &    &\multicolumn{3}{|c|}{C or Band} & \multicolumn{1}{|c|}{BB} & \multicolumn{2}{|c|}{ PL or C} & \multicolumn{1}{|c|}{}\\
\hline
\multicolumn{1}{|c|}{T$_\mathrm{start}$} & \multicolumn{1}{|c|}{T$_\mathrm{stop}$} & \multicolumn{1}{|c|}{Parameters} & \multicolumn{1}{|c|}{E$_{\rm peak}$} & \multicolumn{1}{|c|}{$\alpha$} & \multicolumn{1}{|c|}{$\beta$} & \multicolumn{1}{|c|}{kT} & \multicolumn{1}{|c|}{$\alpha$} & \multicolumn{1}{|c|}{E$_\mathrm{0}$} & \\
\multicolumn{1}{|c|}{(s)} & \multicolumn{1}{|c|}{(s)} & \multicolumn{1}{|c|}{} & \multicolumn{1}{|c|}{(keV)} & \multicolumn{1}{|c|}{} & \multicolumn{1}{|c|}{} & \multicolumn{1}{|c|}{(keV)} & \multicolumn{1}{|c|}{} & \multicolumn{1}{|c|}{(MeV)} & \\
\hline
\multicolumn{10}{|c|}{ }   \\
+11.8 & +12.2 & Band & 230$\pm12$ & -0.89$\pm0.04$ & -3.15$^\mathrm{+0.35}_\mathrm{-0.83}$  & -- & -- & -- & 516/566 \\
& & C+BB & 300$^\mathrm{+36}_\mathrm{-28}$ & -1.09$\pm0.07$ & -- & 31.0$\pm3.4$ & -- & -- & 509/565 \\
& & C+PL & 227$\pm14$ & -0.75$\pm0.14$ & -- & -- & -1.79$^\mathrm{+0.20}_\mathrm{-0.30}$ & -- & 516/565 \\
\hline
\end{tabular}
}
\end{center}
\end{table*}


\newpage

\begin{thebibliography}{}


\bibitem[\protect\astroncite{{Abdo} et al.}{2009a}]{Abdo:2009:GRB080916C}
Abdo, A.~A., Ackermann, M., Arimoto, M., et al.: 2009a,
\newblock{Science}, 323, 1688

\bibitem[\protect\astroncite{{Abdo} et al.}{2009b}]{Abdo:2009:GRB090902B}
Abdo, A.~A., Ackermann, M., Ajello, M., et al.: 2009b,
\newblock{\apjl}, 706, L138

\bibitem[\protect\astroncite{{Ackermann} et al.}{2010}]{Ackermann:2010:GRB090510}
Ackermann, M., Asano, K., Atwood, W.~B., et al.: 2010,
\newblock{\apj}, 716, 1178

\bibitem[\protect\astroncite{{Ackermann} et al.}{2011}]{Ackermann:2011:GRB090926A}
Ackermann, M., Ajello, M., Asano, K., et al.: 2011,
\newblock{\apj}, 729, 114


\bibitem[\protect\astroncite{{Ackermann} et al.}{2012b}]{Ackermann:2012b}
Ackermann, M., Ajello, M., Albert, A., et al.: 2012,
\newblock{\apjs}, 203, 4

\bibitem[\protect\astroncite{{Ackermann} et al.}{2013}]{Ackermann:2013}
Ackermann, M., Ajello, M., Asano, K., et al.: 2013,
\newblock{\apjs}, 209, 34 

\bibitem[\protect\astroncite{{Ackermann} et al.}{2014}]{Ackermann:2014}
Ackermann, M., Ajello, M., Asano, K., et al.: 2014,
\newblock{Science}, 343, 42

\bibitem[\protect\astroncite{{Ajello} et al.}{2014}]{Ajello:2014}
Ajello, M., Albert, A., Allafort, A., et al.: 2014,
\newblock{\apj}, 789, 20 

\bibitem[\protect\astroncite{{Asano} et al.}{2009}]{Asano:2009}
Asano, K., Inoue, S., \& M{\'e}sz{\'a}ros, P.: 2009,
\newblock{\apj}, 699, 953

\bibitem[\protect\astroncite{{Asano} \& {M{\'e}sz{\'a}ros}}{2011}]{Asano:2011}
Asano, K., \& M{\'e}sz{\'a}ros, P.: 2011,
\newblock{\apj}, 739, 103

\bibitem[\protect\astroncite{{Asano} \& {M{\'e}sz{\'a}ros}}{2012}]{Asano:2012}
Asano, K., \& M{\'e}sz{\'a}ros, P.: 2012,
\newblock{\apj}, 757, 115

\bibitem[\protect\astroncite{{Asano} \& {M{\'e}sz{\'a}ros}}{2013}]{Asano:2013}
Asano, K., \& M{\'e}sz{\'a}ros, P.: 2013,
\newblock{JCAP}, 9, 8 

\bibitem[\protect\astroncite{{Atwood} et al.}{2009}]{Atwood:2009}
Atwood, W.~B., Abdo, A.~A., Ackermann, M., et al.: 2009,
\newblock{\apj}, 697, 1071 

\bibitem[\protect\astroncite{{Axelsson} et al.}{2012}]{Axelsson:2012:GRB110721A}
Axelsson, M., Baldini, L., Barbiellini, G., et al.: 2012,
\newblock{\apjl}, 757, L31 

\bibitem[\protect\astroncite{{Band} et al.}{1993}]{Band:1993}
Band, D., Matteson, J., Ford, L., et al.: 1993,
\newblock{\apj}, 413, 281

\bibitem[\protect\astroncite{{Baring} \& {Harding}}{1997}]{Baring:1997}
Baring, M.~G., \& Harding, A.~K.: 1997,
\newblock{\apj}, 491, 663 

\bibitem[\protect\astroncite{{Beloborodov}}{2010}]{Beloborodov:2010}
Beloborodov, A.~M.: 2010, 
\newblock{\mnras}, 407, 1033 

\bibitem[\protect\astroncite{{Beloborodov}}{2011}]{Beloborodov:2011}
Beloborodov, A.~M.: 2011, 
\newblock{\apj}, 737, 68

\bibitem[\protect\astroncite{{Beloborodov} et al.}{2014}]{Beloborodov:2014}
Beloborodov, A.~M., Hasco{\"e}t, R., \& Vurm, I.: 2014,
\newblock{\apj}, 788, 36

\bibitem[\protect\astroncite{{Beniamini} \& {Piran}}{2013}]{Beniamini:2013}
Beniamini, P., \& Piran, T.: 2013,
\newblock{\apj}, 769, 69

\bibitem[\protect\astroncite{{Beniamini} \& {Piran}}{2014}]{Beniamini:2014}
Beniamini, P., \& Piran, T.: 2014,
\newblock{arXiv:1402.4113}

\bibitem[\protect\astroncite{{Bissaldi}}{2009}]{Bissaldi:2009}
Bissaldi, E.: 2009,
\newblock{GRB Coordinates Network}, 9933, 1 

\bibitem[\protect\astroncite{{Borgonovo} \& {Ryde}}{2001}]{Borgonovo:2001}
Borgonovo, L., \& Ryde, F.: 2001,
\newblock{\apj}, 548, 770

\bibitem[\protect\astroncite{{Bo{\v s}njak} et al.}{2009}]{Bosnjak:2009}
Bo{\v s}njak, {\v Z}., Daigne, F., \& Dubus, G.: 2009,
\newblock{\aap}, 498, 677

\bibitem[\protect\astroncite{{Bo{\v s}njak} \& {Daigne}}{2014}]{Bosnjak:2014}
Bo{\v s}njak, Z., \& Daigne, F.: 2014,
\newblock{arXiv:1404.4577}

\bibitem[\protect\astroncite{{Burgess} et al.}{2011}]{Burgess:2011}
Burgess, J.~M., Preece, R.~D., Baring, M.~G., et al.: 2011,
\newblock{\apj}, 741, 24 

\bibitem[\protect\astroncite{{Cavallo} \& {Rees}}{1978}]{Cavallo:1978}
Cavallo, G., \& Rees, M.~J.: 1978,
\newblock{\mnras}, 183, 359

\bibitem[\protect\astroncite{{Clemens} et al.}{2008}]{Clemens:2008}
Clemens, C., Rossi, A., Greiner, J., \& McBreen, S.: 2008,
\newblock{GRB Coordinates Network}, 8257, 1 

\bibitem[\protect\astroncite{{Crider} et al.}{1997}]{Crider:1997}
Crider, A., Liang, E.~P., Smith, I.~A., et al.: 1997,
\newblock{\apjl}, 479, L39

\bibitem[\protect\astroncite{{Daigne} \& {Mochkovitch}}{1998}]{Daigne:1998}
Daigne, F., \& Mochkovitch, R.: 1998,
\newblock{\mnras}, 296, 275

\bibitem[\protect\astroncite{{Daigne} \& {Mochkovitch}}{2002}]{Daigne:2002}
Daigne, F., \& Mochkovitch, R.: 2002,
\newblock{\mnras}, 336, 1271

\bibitem[\protect\astroncite{{Daigne} et al.}{2011}]{Daigne:2011}
Daigne, F., Bo{\v s}njak, {\v Z}., \& Dubus, G.: 2011,
\newblock{\aap}, 526, A110

\bibitem[\protect\astroncite{{Deng} \& {Zhang}}{2014}]{Deng:2014}
Deng, W., \& Zhang, B.: 2014,
\newblock{\apj}, 785, 112

\bibitem[\protect\astroncite{{Derishev} et al.}{2001}]{Derishev:2001}
Derishev, E.~V., Kocharovsky, V.~V., \& Kocharovsky, V.~V.: 2001,
\newblock{\aap}, 372, 1071

\bibitem[\protect\astroncite{{Derishev}}{2007}]{Derishev:2007}
Derishev, E.~V.: 2007,
\newblock{\apss}, 309, 157

\bibitem[\protect\astroncite{{Fryer} et al.}{1999}]{Fryer:1999}
Fryer, C.~L., Woosley, S.~E., Herant, M., \& Davies, M.~B.: 1999,
\newblock{\apj}, 520, 650

\bibitem[\protect\astroncite{{Gehrels}}{1997}]{Gehrels:1997}
Gehrels, N.: 1997,
\newblock{Nuovo Cimento B Serie}, 112, 11

\bibitem[\protect\astroncite{{Ghirlanda} et al.}{2003}]{Ghirlanda:2003}
Ghirlanda, G., Celotti, A., \& Ghisellini, G.: 2003,
\newblock{\aap}, 406, 879

\bibitem[\protect\astroncite{{Ghirlanda} et al.}{2010}]{Ghirlanda:2010}
Ghirlanda, G., Ghisellini, G., \& Nava, L.: 2010a,
\newblock{\aap}, 510, L7 

\bibitem[\protect\astroncite{{Ghirlanda} et al.}{2010}]{Ghirlanda:2010b} 
Ghirlanda, G., Nava, L., \& Ghisellini, G.: 2010b,
\newblock{\aap}, 511, A43

\bibitem[\protect\astroncite{{Ghirlanda} et al.}{2011a}]{Ghirlanda:2011a}
Ghirlanda, G., Ghisellini, G., Nava, L., \& Burlon, D.: 2011a,
\newblock{\mnras}, 410, L47

\bibitem[\protect\astroncite{{Ghirlanda} et al.}{2011b}]{Ghirlanda:2011b}
Ghirlanda, G., Ghisellini, G., \& Nava, L.: 2011b,
\newblock{\mnras}, 418, L109

\bibitem[\protect\astroncite{{Ghisellini} et al.}{2000}]{Ghisellini:2000}
Ghisellini, G., Celotti, A., \& Lazzati, D.: 2000,
\newblock{\mnras}, 313, L1

\bibitem[\protect\astroncite{{Ghisellini} et al.}{2010}]{Ghisellini:2010}
Ghisellini, G., Ghirlanda, G., Nava, L., \& Celotti, A.: 2010,
\newblock{\mnras}, 403, 926

\bibitem[\protect\astroncite{{Gill} \& {Thompson}}{2014}]{Gill:2014}
Gill, R., \& Thompson, C.: 2014,
\newblock{arXiv:1406.4774}

\bibitem[\protect\astroncite{{Goldstein} \& {van der Horst}}{2008}]{Goldstein:2008}
Goldstein, A., \& van der Horst, A.: 2008,
\newblock{GRB Coordinates Network}, 8245, 1 

\bibitem[\protect\astroncite{{Golenetskii} et al.}{1983}]{Golenetskii:1983}
Golenetskii, S.~V., Mazets, E.~P., Aptekar, R.~L., \& Ilinskii, V.~N.: 1983,
\newblock{\nat}, 306, 451

\bibitem[\protect\astroncite{{Gonz{\'a}lez} et al.}{2003}]{Gonzalez:2003}
Gonz{\'a}lez, M.~M., Dingus, B.~L., Kaneko, Y., et al.: 2003,
\newblock{\nat}, 424, 749

\bibitem[\protect\astroncite{{Gonz{\'a}lez} et al.}{2012}]{Gonzalez:2012}
Gonz{\'a}lez, M.~M., Sacahui, J.~R., Ramirez, J.~L., Patricelli, B., \& Kaneko, Y.: 2012,
\newblock{\apj}, 755, 140 

\bibitem[\protect\astroncite{{Goodman}}{1986}]{Goodman:1986}
Goodman, J.: 1986,
\newblock{\apjl}, 308, L47 

\bibitem[\protect\astroncite{{Granot} et al.}{2008}]{Granot:2008}
Granot, J., Cohen-Tanugi, J., \& do Couto e Silva, E.: 2008,
\newblock{\apj}, 677, 92

\bibitem[\protect\astroncite{{Greiner} et al.}{1995}]{Greiner:1995}
Greiner, J., Sommer, M., Bade, N., et al.: 1995,
\newblock{\aap}, 302, 121

\bibitem[\protect\astroncite{{Greiner} et al.}{2009}]{Greiner:2009}
Greiner, J., Clemens, C., Kr{\"u}hler, T., et al.: 2009,
\newblock{\aap}, 498, 89 

\bibitem[\protect\astroncite{{Gruber} et al.}{2014}]{Gruber:2014}
Gruber, D., Goldstein, A., Weller von Ahlefeld, V., et al.: 2014,
\newblock{\apjs}, 211, 12

\bibitem[\protect\astroncite{{Guiriec} et al.}{2010}]{Guiriec:2010}
Guiriec, S., Briggs, M.~S., Connaughton, V., et al.: 2010,
\newblock{\apj}, 725, 225 

\bibitem[\protect\astroncite{{Guiriec} et al.}{2011a}]{Guiriec:2011a}
Guiriec, S., Connaughton, V., Briggs, M.~S., et al.: 2011a,
\newblock{\apjl}, 727, L33

\bibitem[\protect\astroncite{{Guiriec} et al.}{2011b}]{Guiriec:2011b}
Guiriec, S.: 2011b,
\newblock{AAS/HEAD}, 12, 01.04

\bibitem[\protect\astroncite{{Guiriec} et al.}{2013}]{Guiriec:2013a}
Guiriec, S., Daigne, F., Hasco{\"e}t, R., et al.: 2013,
\newblock{\apj}, 770, 32 


\bibitem[\protect\astroncite{{Hasco{\"e}t} et al.}{2012}]{Hascoet:2012}
Hasco{\"e}t, R., Daigne, F., Mochkovitch, R., \& Vennin, V.: 2012,
\newblock{\mnras}, 421, 525 

\bibitem[\protect\astroncite{{Hasco{\"e}t} et al.}{2013}]{Hascoet:2013}
Hasco{\"e}t, R., Daigne, F., \& Mochkovitch, R.: 2013,
\newblock{\aap}, 551, A124

\bibitem[\protect\astroncite{{Kobayashi} et al.}{1997}]{Kobayashi:1997}
Kobayashi, S., Piran, T., \& Sari, R.: 1997,
\newblock{\apj}, 490, 92

\bibitem[\protect\astroncite{{Kouveliotou} et al.}{1993}]{Kouveliotou:1993}
Kouveliotou, C., Meegan, C.~A., Fishman, G.~J., et al.: 1993,
\newblock{\apjl}, 413, L101 

\bibitem[\protect\astroncite{{Kumar} \& {Barniol Duran}}{2009}]{Kumar:2009}
Kumar, P., \& Barniol Duran, R.: 2009,
\newblock{\mnras}, 400, L75

\bibitem[\protect\astroncite{{Kumar} \& {Barniol Duran}}{2010}]{Kumar:2010}
Kumar, P., \& Barniol Duran, R.: 2010,
\newblock{\mnras}, 409, 226

\bibitem[\protect\astroncite{{Lazzati} et al.}{2011}]{Lazzati:2011}
Lazzati, D., Morsony, B.~J., \& Begelman, M.~C.: 2011,
\newblock{\apj}, 732, 34

\bibitem[\protect\astroncite{{Lemoine}}{2013}]{Lemoine:2013}
Lemoine, M.: 2013,
\newblock{\mnras}, 428, 845

\bibitem[\protect\astroncite{{Liang} et al.}{2004}]{Liang:2004}
Liang, E.~W., Dai, Z.~G., \& Wu, X.~F.: 2004
\newblock{\apjl}, 606, L29

\bibitem[\protect\astroncite{{Lithwick} \& {Sari}}{2001}]{Lithwick:2001}
Lithwick, Y., \& Sari, R.: 2001,
\newblock{\apj}, 555, 540

\bibitem[\protect\astroncite{{Lu} et al.}{2012}]{Lu:2012}
Lu, R.-J., Wei, J.-J., Liang, E.-W., et al.: 2012,
\newblock{\apj}, 756, 112 

\bibitem[\protect\astroncite{{Lundman} et al.}{2013}]{Lundman:2013}
Lundman, C., Pe'er, A., \& Ryde, F.: 2013,
\newblock{\mnras}, 428, 2430 

\bibitem[\protect\astroncite{{Lyutikov} \& {Blandford}}{2003}]{Lyutikov:2003}
Lyutikov, M., \& Blandford, R.: 2003,
\newblock{arXiv:astro-ph/0312347}

\bibitem[\protect\astroncite{{MacFadyen} \& {Woosley}}{1999}]{MacFadyen:1999}
MacFadyen, A.~I., \& Woosley, S.~E.: 1999,
\newblock{\apj}, 524, 262

\bibitem[\protect\astroncite{{Malesani} et al.}{2009}]{Malesani:2009}
Malesani, D., Goldoni, P., Fynbo, J.~P.~U., et al.: 2009,
\newblock{GRB Coordinates Network}, 9942, 1

\bibitem[\protect\astroncite{{Meegan} et al.}{2009}]{Meegan:2009}
Meegan, C., Lichti, G., Bhat, P.~N., et al.: 2009,
\newblock{\apj}, 702, 791

\bibitem[\protect\astroncite{{M{\'e}sz{\'a}ros} \& {Rees}}{1993}]{Meszaros:1993}
M{\'e}sz{\'a}ros, P., \& Rees, M.~J.: 1993,
\newblock{\apjl}, 418, L59

\bibitem[\protect\astroncite{{M{\'e}sz{\'a}ros} \& {Rees}}{2000}]{Meszaros:2000}
M{\'e}sz{\'a}ros, P., \& Rees, M.~J.: 2000,
\newblock{\apj}, 530, 292

\bibitem[\protect\astroncite{{M{\'e}sz{\'a}ros}}{2002}]{Meszaros:2002}
M{\'e}sz{\'a}ros, P.: 2002,
\newblock{\araa}, 40, 137

\bibitem[\protect\astroncite{{Nakar} et al.}{2005}]{Nakar:2005}
Nakar, E., Piran, T., \& Sari, R.: 2005,
\newblock{\apj}, 635, 516 

\bibitem[\protect\astroncite{{Paczynski}}{1986}]{Paczynski:1986}
Paczynski, B.: 1986,
\newblock{\apjl}, 308, L43

\bibitem[\protect\astroncite{{Pe'er} et al.}{2006}]{Peer:2006}
Pe'er, A., M{\'e}sz{\'a}ros, P., \& Rees, M.~J.: 2006,
\newblock{\apj}, 642, 995

\bibitem[\protect\astroncite{{Pe'er} et al.}{2012}]{Peer:2012}
Pe'er, A., Zhang, B.-B., Ryde, F., et al.: 2012,
\newblock{\mnras}, 420, 468 

\bibitem[\protect\astroncite{{Preece} et al.}{1998}]{Preece:1998}
Preece, R.~D., Briggs, M.~S., Mallozzi, R.~S., et al.: 1998,
\newblock{\apjl}, 506, L23

\bibitem[\protect\astroncite{{Piran} et al.}{2009}]{Piran:2009}
Piran, T., Sari, R., \& Zou, Y.-C.: 2009,
\newblock{\mnras}, 393, 1107

\bibitem[\protect\astroncite{{Rees} \& {M{\'e}sz{\'a}ros}}{1992}]{Rees:1992}
Rees, M.~J., \& M{\'e}sz{\'a}ros, P.: 1992,
\newblock{\mnras}, 258, 41P

\bibitem[\protect\astroncite{{Rees} \& {M{\'e}sz{\'a}ros}}{1994}]{Rees:1994}
Rees, M.~J., \& M{\'e}sz{\'a}ros, P.: 1994,
\newblock{\apjl}, 430, L93

\bibitem[\protect\astroncite{{Rees} \& {M{\'e}sz{\'a}ros}}{2005}]{Rees:2005}
Rees, M.~J., \& M{\'e}sz{\'a}ros, P.: 2005,
\newblock{\apj}, 628, 847

\bibitem[\protect\astroncite{{Rosswog}}{2003}]{Rosswog:2003}
Rosswog, S.: 2003,
Gamma-Ray Burst and Afterglow Astronomy 2001: A Workshop Celebrating the First Year of the HETE Mission, 662, 220

\bibitem[\protect\astroncite{{Ryde}}{2004}]{Ryde:2004}
Ryde, F.: 2004,
\newblock{\apj}, 614, 827

\bibitem[\protect\astroncite{{Ryde} et al.}{2010}]{Ryde:2010}
Ryde, F., Axelsson, M., Zhang, B.~B., et al.: 2010,
\newblock{\apjl}, 709, L172

\bibitem[\protect\astroncite{{Sari} \& {Piran}}{1997}]{Sari:1997}
Sari, R., \& Piran, T.: 1997,
\newblock{\apj}, 485, 270

\bibitem[\protect\astroncite{{Scargle}}{1998}]{Scargle:1998}
Scargle, J.~D.: 1998,
\newblock{\apj}, 504, 405

\bibitem[\protect\astroncite{{Shemi} \& {Piran}}{1990}]{Shemi:1990}
Shemi, A., \& Piran, T.: 1990,
\newblock{\apjl}, 365, L55

\bibitem[\protect\astroncite{{Tajima} et al.}{2008}]{Tajima:2008}
Tajima, H., Bregeon, J., Chiang, J., \& Thayer, G.: 2008,
\newblock{GRB Coordinates Network}, 8246, 1 

\bibitem[\protect\astroncite{{Thompson}}{2006}]{Thompson:2006}
Thompson, C.: 2006,
\newblock{\apj}, 651, 333

\bibitem[\protect\astroncite{{Uehara} et al.}{2009}]{Uehara:2009}
Uehara, T., Takahashi, H., \& McEnery, J.: 2009,
\newblock{GRB Coordinates Network}, 9934, 1 

\bibitem[\protect\astroncite{{Uhm} \& {Zhang}}{2014}]{Uhm:2014}
Uhm, Z.~L., \& Zhang, B.: 2014,
\newblock{Nature Physics}, 10, 351

\bibitem[\protect\astroncite{{Vasileiou}}{2013}]{Vasileiou:2013}
Vasileiou, V.: 2013,
\newblock{Astroparticle Physics}, 48, 61

\bibitem[\protect\astroncite{{Vetere} et al.}{2009}]{Vetere:2009}
Vetere, L., Evans, P.~A., \& Goad, M.~R.: 2009,
\newblock{GRB Coordinates Network}, 9936, 1 

\bibitem[\protect\astroncite{{von Kienlin} et al.}{2014}]{vonKienlin:2014}
von Kienlin, A., Meegan, C.~A., Paciesas, W.~S., et al.: 2014,
\newblock{arXiv:1401.5080}

\bibitem[\protect\astroncite{{Vurm} et al.}{2011}]{Vurm:2011}
Vurm, I., Beloborodov, A.~M., \& Poutanen, J.: 2011,
\newblock{\apj}, 738, 77

\bibitem[\protect\astroncite{{Vurm} et al.}{2013}]{Vurm:2013}
Vurm, I., Lyubarsky, Y., \& Piran, T.: 2013,
\newblock{\apj}, 764, 143

\bibitem[\protect\astroncite{{Vurm} et al.}{2014}]{Vurm:2014}
Vurm, I., Hasco{\"e}t, R., \& Beloborodov, A.~M.: 2014,
\newblock{\apjl}, 789, L37

\bibitem[\protect\astroncite{{Wang} et al.}{2009}]{Wang:2009}
Wang, X.-Y., Li, Z., Dai, Z.-G., \& M{\'e}sz{\'a}ros, P.: 2009,
\newblock{\apjl}, 698, L98

\bibitem[\protect\astroncite{{Woosley}}{1993}]{Woosley:1993}
Woosley, S.~E.: 1993,
\newblock{\apj}, 405, 273

\bibitem[\protect\astroncite{{Woosley} \& {Heger}}{2006}]{Woosley:2006}
Woosley, S.~E., \& Heger, A.: 2006,
\newblock{\apj}, 637, 914

\bibitem[\protect\astroncite{{Zhang} \& {Pe'er}}{2009}]{Zhang:2009}
Zhang, B., \& Pe'er, A.: 2009,
\newblock{\apjl}, 700, L65

\bibitem[\protect\astroncite{{Zhang}}{2011}]{Zhang:2011}
Zhang, B.: 2011,
\newblock{Comptes Rendus Physique}, 12, 206

\bibitem[\protect\astroncite{{Zhang} \& {Yan}}{2011b}]{Zhang:2011b}
Zhang, B., \& Yan, H.: 2011b,
\newblock{\apj}, 726, 90

\bibitem[\protect\astroncite{{Zhang} et al.}{2011c}]{Zhang:2011c}
Zhang, B.-B., Zhang, B., Liang, E.-W., et al.: 2011,
\newblock{\apj}, 730, 141

\bibitem[\protect\astroncite{{Zhang} \& {Zhang}}{2014}]{Zhang:2014}
Zhang, B., \& Zhang, B.: 2014,
\newblock{\apj}, 782, 92

\bibitem[\protect\astroncite{{Zhao} et al.}{2014}]{Zhao:2014}
Zhao, X., Li, Z., Liu, X., et al.: 2014,
\newblock{\apj}, 780, 12

\bibitem[\protect\astroncite{{Zou} et al.}{2009}]{Zou:2009}
Zou, Y.-C., Fan, Y.-Z., \& Piran, T.: 2009,
\newblock{\mnras}, 396, 1163

\end{thebibliography}
\end{document}